%% file: main.tex
\PassOptionsToPackage{table}{xcolor}

\documentclass[sigconf]{acmart}

\setcopyright{none}
\renewcommand\footnotetextcopyrightpermission[1]{}
\usepackage{amsmath}
\usepackage{ebproof}
\usepackage{todonotes}
\usepackage{amsfonts}
\usepackage{algorithmic}
\usepackage{tikz}
\usepackage{multirow}
\usepackage{tikz-network}
\usepackage{pgfplots}
\usepackage{fontawesome}
\usepackage{textcomp}
\usepackage{mathpartir}
\usepackage{color}
\usepackage[table]{xcolor}
\DeclareMathAlphabet{\mathpzc}{OT1}{pzc}{m}{it}
\usepackage{cleveref} % must be loaded after hyperref and amsmath
\usepackage{subcaption} %% For complex figures with subfigures/subcaptions
\usepackage{epsfig}
\usepackage{listings}
\usepackage[override]{cmtt}
\usepackage{graphicx}
\usepackage{balance}
\usepackage{xspace}
\usepackage{booktabs}
\usepackage{url}
\usepackage{xparse}
\usepackage{enumitem}
\usepackage[frozencache,cachedir=.]{minted}
\usepackage{tikz}
\usetikzlibrary{chains,fit,shapes,positioning,calc,arrows.meta,shapes.arrows}
\usepackage{xspace}

\newcommand{\inlinecode}[1]{%
  \mintinline[fontsize=\footnotesize{},mathescape, escapeinside=||]{c}{#1}%
}

\lstdefinestyle{customc}{
  belowcaptionskip=1\baselineskip,
  breaklines=true,
  breakatwhitespace, %needed to avoid extra space after lstinline + linebreak
  numbers=left,
  xleftmargin=\parindent,
  language=C,
  columns=flexible,      
  showstringspaces=false,
  basicstyle=\small\sffamily,
  otherkeywords={uint,uintptr_t,let,assert},
  literate={{<-}{{$\leftarrow\,$}}2
            {->}{{$\rightarrow\,$}}2
            {<=}{{$\leq\,$}}2},
  numberstyle=\tiny\rmfamily,
  keywordstyle=\bfseries\color{green!40!black},
  commentstyle=\itshape\color{purple!40!black},
  identifierstyle=\color{blue!80!black},
}
\newenvironment{DIFnomarkup}{}{}
\newcommand{\myparagraph}[1]{\noindent\textbf{#1}.\xspace}

\newcommand{\code}[1]{\lstinline|#1|}
\newcommand{\ignore}[1]{{}}

\definecolor{taintcolor}{rgb}{1.0, 0.65, 0.79}
\definecolor{checkcolor}{rgb}{1.0, 0.7, 0.0}
\newcommand{\vulprevented}{\textcolor{green}{\faCheckCircle\xspace}}
\newcommand{\vulisolated}{\textcolor{green}{\faCheckCircleO\xspace}}

\newcommand{\tbl}[1]{Table~\ref{#1}}
\newcommand{\sect}[1]{Section~\ref{#1}}
\newcommand{\fig}[1]{Figure~\ref{#1}}
\newcommand{\lst}[1]{Listing~\ref{#1}}

\newcommand{\apdx}[1]{Appendix~\ref{#1}}

\newcommand{\etal}{\textit{et al.,}\xspace}
\newcommand{\ie}{\textit{i.e.,}\xspace}
\newcommand{\eg}{\textit{e.g.,}\xspace}

\newcommand{\realbug}{\textcolor{red}{\faBug}}
\newcommand{\complexcode}{\faChainBroken}
\newcommand{\entrypoint}{\faForward}
\newcommand{\rootcause}{\textcolor{orange}{\faWarning}}
\newcommand{\useradded}{\faUserPlus}
\newcommand{\usermods}{\faUser}

\newcommand{\ptr}{\ensuremath{ptr}}
\newcommand{\arr}{\ensuremath{arr}}
\newcommand{\ntarr}{\ensuremath{ntarr}}

\newcommand{\arrT}[1]{\code{_Array_ptr<}$#1$\code{>}}

\newcommand{\ucregion}{\umode region\xspace}
\newcommand{\cregion}{\cmode region\xspace}
\newcommand{\taintt}{\code{t_*}}

\newif\ifsubmit\submitfalse

\ifsubmit
\newcommand{\mwh}[1]{}
\newcommand{\ashe}[1]{}
\newcommand{\dtarditi}[1]{}
\newcommand{\dvh}[1]{}
\newcommand{\leo}[1]{}
\newcommand{\liyi}[1]{}
\newcommand{\yiyun}[1]{}
\newcommand{\review}[1]{}
\newcommand{\aravind}[1]{}
\newcommand{\mz}[1]{}
\newcommand{\lc}[1]{}

\else
\newcommand{\mwh}[1]{\textcolor{red}{Mike: #1}}
\newcommand{\liyi}[1]{\textbf{\textcolor{orange}{Liyi: #1}}}
\newcommand{\yiyun}[1]{\textcolor{cyan}{Yiyun: #1}}
\newcommand{\review}[1]{\textbf{\textcolor{blue}{Review: #1}}}
\newcommand{\aravind}[1]{\textcolor{green}{Aravind: #1}}
\newcommand{\lc}[1]{\textcolor{teal}{Le: #1}}

\usepackage[normalem]{ulem}
\colorlet{MZ}{violet!80!pink}
\newcommand{\mz}[1]{\textbf{\textcolor{MZ}{MZ: #1}}}
\newcommand{\mzs}[1]{{\color{MZ}\sout{#1}}}
\newcommand{\mzu}[1]{{\color{MZ}\uline{#1}}}
\newcommand{\mzr}[1]{{\color{MZ}{#1}}}
\fi

\newcommand{\lang}{\textsc{CoreChkCBox}\xspace}
\newcommand{\elang}{\textsc{CoreC}\xspace}

\newtheorem{defi}{Definition}
\newtheorem{thm}{Theorem}

\newcommand{\checkedc}{\text{Checked C}\xspace }

 \newcommand{\rulelab}[1]{{\small \textsc{#1}}}

\newcommand{\kw}[1]{\ensuremath{\mathtt{#1}}}

\newcommand{\estrlen}[1]{\ensuremath{\kw{strlen}({#1})}}

\newcommand{\tarray}[3]{\tarrayb{({#1},{#2})}{#3}}
\newcommand{\tarrayb}[2]{\ensuremath{[{#1}~{#2}]}}
\newcommand{\tntarray}[3]{\tntarrayb{({#1},{#2})}{#3}}
\newcommand{\tntarrayb}[2]{\tarrayb{#1}{#2}_{nt}}
\newcommand{\tallarrayb}[2]{\ensuremath{[{#1}~{#2}]_{\kappa}}}

\newcommand{\tfun}[3]{\ensuremath{\forall \;#1.\;{#2}\to#3}}
\newcommand{\cupdot}{\mathbin{\mathaccent\cdot\cup}}

\newcommand{\tallarray}[3]{\tallarrayb{({#1},{#2})}{#3}}

\newcommand{\tptr}[2]{\ensuremath{\mathtt{ptr}^{#2}~{#1}}}

\newcommand{\tarrayptr}[4]{{\tptr{\tarray{#1}{#2}{#3}}{#4}}}
\newcommand{\tntarrayptr}[4]{{\tptr{\tntarray{#1}{#2}{#3}}{#4}}}
\newcommand{\tallarrayptr}[4]{{\tptr{\tallarray{#1}{#2}{#3}}{#4}}}

\newcommand{\tstruct}[1]{\ensuremath{\kw{struct}~{#1}}}

\newcommand{\evalue}[2]{\ensuremath{{#1}\!:\!{#2}}}

\newcommand{\emalloc}[2]{\ensuremath{\kw{malloc}({#1},{#2})}}
\newcommand{\emalloctext}{\ensuremath{\kw{malloc}}}
\newcommand{\ecall}[2]{\ensuremath{{#1}({#2})}}
\newcommand{\elcall}[3]{\ensuremath{{(#1,#2)}({#3})}}

\newcommand{\ret}[3]{\ensuremath{\kw{ret}({#1},{#2},{#3})}}

\newcommand{\ecast}[2]{\ensuremath{\kw{(}{#1}\kw{)}{#2}}}
\newcommand{\edyncast}[2]{\ensuremath{\langle{#1}\rangle{#2}}}

\newcommand{\elet}[3]{\ensuremath{\kw{let}~#1\, \texttt{=}\, #2~\kw{in}\;{#3}}}
\newcommand{\elettext}{\ensuremath{\kw{let}}}
\newcommand{\ebinop}[2]{\ensuremath{#1 \plus #2}}
\newcommand{\eassign}[2]{\ensuremath{\texttt{*}~{#1}\,\texttt{=}\, {#2}}}
\newcommand{\elassign}[3]{\ensuremath{\texttt{*}{(#1,#2)}\,\texttt{=}\, {#3}}}

\newcommand{\estar}[1]{\ensuremath{\texttt{*}~{#1}}}
\newcommand{\getstar}[2]{\ensuremath{\texttt{*}({#1},{#2})}}

\newcommand{\eunchecked}[2]{\ensuremath{\kw{unchecked}({#1})\{#2\}}}
\newcommand{\euncheckedtext}{\ensuremath{\kw{unchecked}}}

\newcommand{\echecked}[2]{\ensuremath{\kw{checked}({#1}){\{#2\}}}}
\newcommand{\echeckedtext}{\ensuremath{\kw{checked}}}

\newcommand{\eif}[3]{\ensuremath{\kw{if\;}(#1)\;{#2}\;\kw{else}\;{#3}}}

\newcommand{\ebounds}{\ensuremath{\kw{bounds}}}
\newcommand{\enull}{\ensuremath{\kw{null}}}

\newcommand{\defscope}{\sigma}

\newcommand{\tint}{\ensuremath{\mathtt{int}}}

\newcommand{\heap}{\ensuremath{\mathpzc{H}}}
\newcommand{\heapup}[3]{\ensuremath{\mathpzc{H}}(#1)[#2 \mapsto #3]}
\newcommand{\eret}[3]{\ensuremath{\kw{ret}({#1},{#2},{#3})}}
\newcommand{\erettext}{\ensuremath{\kw{ret}}}

\newcommand{\plus}{\mathbin{\texttt{+}}}

\newcommand{\fv}{\mathit{FV}}

\newcommand{\size}{\mathit{size}}

\newcommand{\cmode}{\textcolor{blue}{\texttt{c}}\xspace}
\newcommand{\umode}{\textcolor{blue}{\texttt{u}}\xspace}
\newcommand{\tmode}{\textcolor{blue}{\texttt{t}}\xspace}

\newcommand{\bvar}{\ensuremath{\beta}}
\newcommand{\mode}{\textit{mode}}
\newcommand{\type}{\textit{type}}
\newcommand{\funptr}{\textit{fun\_t}}

\newcommand{\tgez}{\texttt{ge}\_0}
\newcommand{\teq}[1]{\texttt{eq}\;{#1}}

\newcommand{\systemname}{\textsc{CheckedCBox}\xspace}
\newcommand{\threec}{\textsc{3C}\xspace}
\newcommand{\sourcerewriter}{\textsc{CheckMate}\xspace}
\newcommand{\clang}{\textsc{Clang}\xspace}

\newcommand{\oururl}{\url{https://github.com/REDACTED}}

\newcommand{\numprog}{seven\xspace}
\newcommand{\numcves}{four\xspace}

\definecolor{programs}{gray}{0.1}

\makeatletter
\lst@CCPutMacro
    \lst@ProcessOther {"22}{\lst@ifupquote \textquotedbl
                                     \else \char34\relax \fi}
    \@empty\z@\@empty
\makeatother

\def\BibTeX{{\rm B\kern-.05em{\sc i\kern-.025em b}\kern-.08em
    T\kern-.1667em\lower.7ex\hbox{E}\kern-.125emX}}
\begin{document}
\newif\iftr\trtrue
%% Title information
\iftr
\title{\systemname: Type Directed Program Partitioning with \checkedc for Incremental Spatial Memory Safety {\large (Extended Version)}}
\else
\title{\systemname: Type Directed Program Partitioning with \checkedc for Incremental Spatial Memory Safety}
\fi

\author{Liyi Li$^*$, Arunkumar Bhattar$^*$$^\dagger$, Le Chang, Mingwei Zhu, and Aravind Machiry$^\dagger$\\
 University of Maryland $\quad\quad ~^\dagger$Purdue University}

\begin{abstract}
  % Programs written in C account for a significant portion of code
  % currently in use, despite its lack of memory safety which has been
  % the cause of many disastrous vulnerabilities throughout the years.
  % %
  % Checked C is a Microsoft-backed attempt at a solution to this problem
  % which extends C with types and annotations that guarantees spatial
  % memory safety in a backwards compatible manner, while allowing
  % the mix of checked and unchecked code.
  % %
  % However, Checked C's implementation is a complex beast built on top
  % of Clang, while Checked C's specification is over \leo{X} pages of
  % natural language, reducing confidence in the promised guarantees.
  Spatial memory safety violation is still a major issue for C programs.
Checked C is a safe dialect of C and extends it with checked pointer types and annotations that guarantee spatial memory safety in a backward compatible manner, allowing the mix of checked pointers and regular (unchecked) pointer types.
%; and providing the guarantee that the root cause for spatial memory safety errors will be in unchecked code (\emph{noblame therom}).
However, unchecked code vulnerabilities can violate the checked code's spatial safety guarantees.
We present~\systemname{}, which adds a flexible, type directed program partitioning mechanism to Checked C,
by enhancing the~\checkedc type system with~\code{tainted} types that enable flexible partitioning of a program into~\cregion and~\ucregion, such that~\ucregion code does not affect the spatial safety in~\cregion{}.
We formalize our type system and prove the non-crashing and non-exposure properties of a well-typed~\systemname program.
We implemented~\systemname in a configurable manner, which enables us to use existing sandbox mechanisms (\eg WebAssembly) to execute program partitions seamlessly.
Our evaluation on~\numprog{} programs shows that~\systemname{} can effectively and efficiently partition programs by preventing~\numcves{} known vulnerabilities.
\def\thefootnote{*}\footnotetext{These authors contributed equally to this work.}
%  Spatial memory safety violation is still a major issue for C programs.
%  Checked C is a Microsoft-backed attempt at a solution to this problem
%  which extends C with types and annotations that guarantees spatial
%  memory safety in a backwards compatible manner, while allowing
%  the mix of checked and unchecked code; and provide the guarantee that any spatial memory
%  safety errors can be \emph{blamed} on portions of the program
%  labeled \emph{unchecked}.
%  However, in many cases, users tend not to keep the mix of checked and 
%  unchecked code without rewriting all programs to be spatially safe checked code,
%  which causes many memory safety violations.
%  Here, we present CheckedCBox by merging Checked C with program partitioning mechanism
%  to ensure that spatial safety violations in unchecked code does not affect checked code 
%  and can be resecured by the mechanism, which we named non-crashing property.
%  In addition, we also provide non-exposure guarantee that programs in unchecked code
%  can never learn about important pointers used in checked code.
%  With the formalism, we implement the CheckedCBox based on the WebAssembly sandbox library
%  and show that our implementation not only fulfill the safety guarantees above,
%   but also has a relatively comparable overhead with respect to the Checked C compiler.
%  To the best of our knowledge, this is the first work that combines program partitioning mechanism with C-like programming language.

\end{abstract}

\maketitle
  
\input{introduction}

\input{background}

\input{overview}

% \ignore{
% \section{Null-Terminated Array Pointers}

% Key ideas, formalized:
% \begin{itemize}
% \item Can read one past the end: x[N] can be read, but only written if
%   the value being written is 0.
% \item NT arrays are a refinement of normal arrays: Can convert
%   x:(ntarray N t) to (array N t).
% \item NT arrays can expand their length flow sensitively: Can convert
%   x:(ntarray N t) to (ntarray N+1 t) assuming x[N] != 0.
% \item In general unsafe to convert x:(array N t) to (ntarray N-1 t)
%   even if x[N-1] == 0 because aliases of x could eliminate the NULL
%   terminator. Would need some sort of linearity/alias tracking to
%   support this idea.
% \item Question? What is the syntax of doing a malloc on a NT-array?
%   Should we just do malloc(NT-array l h type) as the malloc of a
%   normal array or should we also include something like int [] nt =
%   “abcd”, where the bound size is not specified?
% \end{itemize}

% See also the Checked C specification
% \ignore{
% https://github.com/microsoft/checkedc/tree/master/spec/bounds\_safety}
% }
%\input{typesystem}
\input{formal}
\input{implementation}
\input{evaluation}

\input{limitations}
\input{comparison}

\input{conclusion}
\begin{small}
%\balance
\bibliographystyle{ACM-Reference-Format}
\bibliography{IEEEabrv,paper,sources}
%%%%%%%%%%%%%%%%%%% Appendix %%%%%%%%%%%%%%%%%%%%%%%%%%%%%%%%%%%%
\end{small}

%\iftr
\newpage
\clearpage

\appendix
\input{appendix}

%\fi

%\newpage
%\input{remaining}
\end{document}

%% file: introduction.tex
\setcopyright{none}
\renewcommand\footnotetextcopyrightpermission[1]{}
\pagestyle{plain}

\section{Introduction}\label{sec:intros}

Vulnerabilities due to memory corruption, especially spatial memory corruption, 
are still a major issue for C programs~\cite{cvetrend, microsoftmemsafe, Zeng:2013:SRF:2534766.2534798} 
despite many efforts that tried to prevent them~\cite{song2019sanitizing}.
Several industrial and research efforts, including CCured~\cite{Necula2005},
Softbound~\cite{softbound}, and ASAN~\cite{Serebryany2012},
have investigated ways to better compile C programs with automatic spatial safety enforcement.
These approaches all impose performance overheads deemed too high for deployment use. 
% 
%\liyi{Should we say this? Is the paper about overhead reducing? }
Recently,~\citet{Elliott2018} and~\citet{li22checkedc} introduced and formalized \checkedc, an
open-source extension to C,
to ensure a program’s spatial safety by introducing new pointer types,~\ie checked ($\cmode$) pointer types.
The checked pointers are represented as system-level memory words without ``fattening'' metadata~\cite{duck2016heap}, 
and ensuring backward compatibility,~\ie developers can use checked and regular (unchecked $\umode$) pointers within the same program.
%users are able to split code into checked and unchecked regions and incrementally convert
%C code in the unchecked region to Checked C code in the checked one.
However, as we explain in~\sect{subsec:nosafetyagsintuncheckedcode}, the
unconverted or unchecked ($\umode$) code can violate guarantees provided in $\cmode$
regions.
We need to ensure that~\emph{code executed as part of 
unchecked ($\umode$) regions does not lead to the safety violations in checked ($\cmode$) regions  with the use of program partitioning mechanism \cite{rul2009towards}}.

Existing such mechanisms are not suitable as they are based on process isolation and have high overhead, and are \emph{data-centric} (\sect{subsec:background:programpart}).
%Furthermore, these techniques are hard to engineer to co-exist with~\checkedc.
%\liyi{why? Can we not say this? It seems that the whole reason people should care our work is we did it in \checkedc, but \checkedc itself is the selling product here.... }
But in our case, we want a low-overhead code-centric partitioning, where the $\umode$ region code (or functions) should be isolated (or partitioned) from $\cmode$ one. We also want the technique to co-exist and be compatible with Checked C guarantees such that the partition containing $\cmode$ region code can maintain spatial safety.

Here, we propose a type-directed code-centric program partitioning approach.
Specifically, our system,~\systemname, extends \checkedc's checked and unchecked pointer types---representing safe and unsafe program pieces---with \textbf{tainted} (\taintt) types running on an isolated sandbox mechanism,
forbids the communication between checked and unchecked type entities, and enforces the communication between checked and unchecked types
through the uses of tainted types with additional validity checks. 
%which can be used to mark functions and pointers that need to be isolated from the original program, while checked types follow the standard \checkedc typing rules.
%However, our type-system disallows any unsafe interactions between taint and untainted types.
% 
% Specifically, in our system,~\systemname, extending CheckedC,
% functions and pointers that need to be isolated from the original program can be
% marked with ~\textbf{tainted} (\taintt) types.
% 

The developer starts by marking desired (\ie unchecked, $\umode$) functions and pointers used in functions as tainted. 
Then, \systemname partitions the given program into two partitions (\umode and~\cmode regions) of different privileges:

\begin{itemize}
\item~\umode \emph{region} (low privilege tainted region, extended from the unchecked region in \checkedc): this partition contains tainted types (\ie functions and pointers) and can only access tainted and unchecked pointers.
\item~\cmode \emph{region} or~\emph{safe region} (high privilege untainted or checked region): This partition contains the remaining (untainted) code and data and has complete access to~\cmode region.
The functions in~\cregion can invoke any function in~\ucregion and access all its data but not the other way around, except for call-back functions, which we will discuss later.
\end{itemize}

The $\cmode$ region code is executed as a regular program, while the~\ucregion partition will be executed in an existing sandboxed environment (\eg WASM sandbox), with additional instrumentations to facilitate the communication between code in~$\cmode$ and~$\umode$ regions.

The combination of tainted types and privileged partitions
enforces isolation and provides memory safety without transforming all unchecked C code to \checkedc code,
because unchecked types can stay in \ucregion, and \cregion code can access tainted type entities that are allocated 
in \ucregion.
Although memory isolation prevents direct violations,~\ucregion code can still affect~\cregion through tainted pointers by confused deputy attacks~\cite{rajani2016access, machiry2017boomerang},~\eg by using a valid~\cregion address in a tainted pointer.
Our compiler avoids these attacks by ensuring using dynamic checks that tainted pointers validly point to~\ucregion address space.
Such checks are statically generated by our compiler.

In summary, we make the following three main contributions.

\myparagraph{\systemname Type System, Formalism and Compiler}
We present a type system that integrates tainted types with~\checkedc and provides additional guarentees---the~\emph{non-crashing} and \emph{non-exposure} guarantees,~\ie a well-typed~\systemname program can never crash due to spatial safety violations,
as well as \ucregion code cannot directly observe a checked pointer address.
%no \cregion pointer addresses will be leaked in
%\ucregion code
We extend the \checkedc compiler to support the type system and
formalize it by extending~\checkedc formalism~\cite{li22checkedc} with the non-crashing and non-exposure guarantees.
We formally prove theorems related to the two guarantees and use model-based randomized testing \cite{Pierce:SF4} to certify the simulation relation between the~\systemname semantics and its compiler formalism.
To the best of our knowledge,~\systemname is the first C(-like) language and compiler formalism with the program partitioning mechanism.

\myparagraph{Type-Directed Program Partitioning}
We present a type-directed program partition technique to separate $\cmode$ and $\umode$ code regions and ensure the above guarantees.
%Our type system restricts checked pointers usage only in~\cregion and ensures that~\ucregion can only access tainted pointers.
Our modular design enables us to use existing sandbox techniques to enforce memory and execution isolation, with the implementation of tainted pointers in the \systemname compiler.

\myparagraph{Supporting callbacks to~\cregion with no Checked Pointer Exposure}
Although we disallow access to~\cregion from~\ucregion directly, there can be cases where such access is needed.
Specifically, when~\cregion wants to provide access to certain shared checked data to~\ucregion.
To enable this, we support callback functions in~\cregion that can be invoked from~\ucregion through function pointers.
%However, knowing the address of~\cregion functions in~\ucregion violates the program partition principle of \systemname.
However, knowing the address of~\cregion functions in~\ucregion violates the non-exposure guarantee and leads to other attacks~\cite{hauser2019sleak}.

We handle this by using indirection. Specifically, instead of directly accessing the~\cregion callbacks, the ~\ucregion accesses them using a tainted-typed protected trampoline function, which directs the execution to the appropriate callback function.
In addition, the trampoline function itself is referenced using an opaque index rather than its virtual address, implemented through
existing sandboxing techniques.
% 
%In our \systemname formalism, we formally verified that our formalism is
%\emph{non-exposure}, \ie, no \cregion pointer addresses will be leaked in
%\ucregion code.

\ignore{

-- NEED TO FINISH THIS--
Our third contribution is an added-up feature
to support checked (function) pointer callbacks in unchecked code regions.
When designing a multi-threaded system, users might want 
to provide a third party interface that allows third party developers to create new program features, while keeping these programs in unchecked code regions.
Moreover, they do want to provide them a (function) pointer pointing to checked data fields.

However, accessing a checked pointer in an unchecked region violates the program partition principle of \systemname.
To resolve the conflict, we develop two mechanisms in \systemname 
and maintain a stronger \textit{non-exposure} guarantee on of the non-crashing guarantee; 
that is, no checked pointer addresses can be observed in an unchecked code region.
The first mechanism allows nested checked and unchecked code regions.
Users can context switch between checked and unchecked code regions 
by nested using the keywords $\echeckedtext$ $\euncheckedtext$.
The type system ensures that no checked pointers can be accessed across the context switching.
The second one is that a call to a checked pointer in unchecked code regions 
must be surrounded by a \textit{tainted shell}; 
i.e., a tainted function pointer that points to a checked region possibly holding checked pointers.
In this case, no checked pointer address will be observed in the unchecked code regions.
\mzu{In \systemname, for every checked function, 
we automatically compile a tainted version by surrounding the function without a
tainted shell.}
\mz{What does this mean?}
}
% 

\iffalse
\myparagraph{Formalizing the Type System, Semantics and Compiler}
%
We developed a core formalism named~\lang, which extends
\citet{li22checkedc}~\checkedc formalism with the non-crashing guarantee and other new features below.
We formally prove the~\emph{non-crashing theorem},~\ie 
a well-typed~\lang program can never crash due to spatial safety violations.
We utilize the model-based randomized testing (CITE) to certify the simulation relation between the~\lang semantics and the compiler.
Specifically, we use a conversion tool that converts expressions from~\lang into actual~\checkedc code that can be compiled by the~\checkedc compiler. We create a random program generator 
based on the typing rules of~\lang and ensure that~\lang and~\checkedc compiler are consistent after conversion, both statically and dynamically.  
To the best of our knowledge,~\lang is the first C(-like) language and compiler formalism with the program partitioning mechanism.
%To the best of our knowledge, \systemname is the first work of formalizing C function pointers with security guarantee.
\fi

We evaluated~\systemname~\footnote{Our implementation is available open source at~\oururl{}.} by partioning seven large real-world programs to demonstrate its effectiveness.
Our evaluation shows that~\systemname provides a flexible, low-overhead program partitioning mechanism and guarantees spatial memory safety.

\ignore{
\aravind{Fix the following text after all the sections of the papers are finalized.}
\liyi{the following might be removed if space is needed, since it is usually not quite useful.}
We begin with a review of \checkedc{} \mzr{and introduction of} new features in
\systemname (Section~\ref{sec:overview}), present our main contributions
(Sections~\ref{sec:formal}--\ref{sec:evaluation}), and conclude with a
discussion of related and future work (Sections~\ref{sec:related},
\ref{sec:conclude}). All code and proof artifacts (both for Coq and Redex) can
be found at \url{https://github.com/plum-umd/checkedc}.
}

\ignore{
\noindent
\textbf{Converting C to Checked C.} The safety guarantees of Checked C come with certain restrictions. For instance, as shown below, Checked C programs cannot use address-taken variables in a bounds expression as the bounds relations may not hold because of possible modifications through pointers.
\begin{minted}[xleftmargin=30pt, mathescape, escapeinside=||, fontsize=\footnotesize]{c}
...
array_ptr<int> p : count (n) = NULL;
|\textcolor{red}{\faTimes}|..,&n,.
\end{minted}
Consequently, converting existing C programs to Checked C might require refactoring, e.g., modifying the above program to not use~\inlinecode{&n} expression, which might require considerable effort~\cite{duanrefactoring} depending on the program's complexity. 
Recently, Machiry et al. developed~\threec~\cite{machiry2022c} that tries to automatically convert a program to Checked C by adding appropriate pointer annotations.
However, as described in \threec, complete automated conversion is infeasible and requires the developer to convert some code regions manually.
Although, the backward compatibility of Checked C helps a partially annotated program to enjoy spatial memory safety on those regions using only Checked pointers (i.e., checked or safe regions).

\noindent
\textbf{No safety against unchecked code.} But, the unconverted code regions (or unsafe regions) can affect pointers in safe regions and violate certain assumptions leading to vulnerabilities, as demonstrated by cross-language attacks~\cite{mergendahlcross}.
Although the blameless proof exists~\cite{ruef2019achieving}, it does not state that spatial safety violations cannot happen in Checked regions but rather states that Checked regions~\emph{cannot be blamed for any spatial safety violations}.
Consider the following example:
\begin{minted}[xleftmargin=25pt, mathescape, linenos, escapeinside=||, fontsize=\footnotesize{}]{c}
// Checked code
int func(array_ptr<char> p : count(5)) {
|\textcolor{red}{\faChainBroken}|..p[4]..
}
// unchecked code
...
str = "he";
...
|\textcolor{red}{\faBug}|assume_bounds_cast<char>(str, 5); 
...
char ptr[16];
...
len <- derived from user input
...
|\textcolor{red}{\faBug}| memcpy(ptr, buff, len); // buffer overflow
\end{minted}
Here, the checked function~\inlinecode{func} expected a pointer to a buffer of five elements, but unchecked code violated it and invoked the function with a buffer of 2 elements.
This results in a spatial safety violation (\textcolor{red}{\faChainBroken}) in the Checked region, but of course, the blame or the root cause is in the unchecked region (\textcolor{red}{\faBug}).
Furthermore, since checked and unchecked regions execute in the same address space, spatial memory corruptions in unchecked regions (Line 15) can take down the complete program despite having checked regions.
We need~\emph{an isolation mechanism to ensure that code executed as part of unchecked regions does not violate the safety guarantees in checked regions}.

This mechanism is called program partitioning~\cite{rul2009towards}, and there has been considerable work~\cite{tan2017principles, brumley2004privtrans, bittau2008wedge, lind2017glamdring, liu2017ptrsplit} in the area. Most of these techniques are~\emph{data-centric}~\cite{lind2017glamdring, liu2017ptrsplit}, wherein program data drives the partitioning. E.g., Given sensitive data in a program, the goal is to partition functions into two parts or partitions based on whether a function can access the sensitive data.
The performance overhead of these approaches is dominated by marshaling costs and depends on the usage of sensitive data.
The overhead of state-of-the-art approaches~\cite{lind2017glamdring, liu2017ptrsplit} is prohibitive and varies from 37\%-163\%.
But in our case, we want a low-overhead code-centric partitioning, where the unchecked code (or functions) should be isolated (or partitioned) from checked code. We also want the technique to co-exist and be compatible with Checked C guarantees such that the partition containing checked code should still enjoy its spatial safety.

In this work, we propose a type-directed code-centric program partitioning approach.
Specifically, our system,~\systemname, extends Checked c using~\textbf{tainted} (\taintt) types, which can be used to mark functions and pointers that need to be isolated from the original program.
The tainted types long with Checked C allow annotating pointer along two dimensions, i.e., (i) taintedness: a pointer can be either tainted or not (untainted), and (ii) checkedness: it can be either checked or not.
The checked types follow the standard Checked C typing rules. However, our type-system disallows all interactions between taint and untainted types.
The developer starts by marking desired (unchecked or risky) functions and pointers used in these functions as tainted.
Second,~\systemname partitions the given program into two partitions (\ucregion and~\cregion)  of different privileges:
\begin{itemize}
\item~\ucregion (low privilege tainted region): This partition contains tainted types (i.e., functions and pointers) and can only access tainted pointers.
\item~\cregion (high privilege untainted region): This partition contains the remaining (untainted) code and data and has complete access to~\cregion{}.
The functions in~\cregion can invoke any function in~\ucregion but not the other way around, except for call-back functions, which we will discuss later.
\end{itemize}
Finally, during program execution, the~\ucregion partition will be executed in an existing sandboxed environment (e.g., WASM sandbox), and our compiler will add the necessary instrumentation to facilitate the communication between code in~\cregion{} and~\ucregion{}.

The combination of tainted types and privileged partitions enables us to enforce isolation and provide memory safety without marshaling costs.
As functions in the~\ucregion can only access tainted types,~\cregion functions should use tainted types to pass pointer arguments to~\ucregion functions. 
We avoid marshaling by allocating all tainted pointers (i.e., tainted buffers) in~\ucregion and thus can be accessed in both partitions.
Although memory isolation prevents direct violations,~\ucregion code can still affect~\cregion through tainted pointers by confused deputy attacks.
Our compiler avoids these attacks by ensuring, either statically or through dynamic checks, that tainted pointers can only point to~\ucregion address space.

\begin{listing}[t!]
  \begin{tabular}{c c}
    \begin{minipage}[b]{.22\textwidth}
\begin{minted}{c}
|\entrypoint| // Entry point
void server_loop(int sock_fd) {
 unsigned b_z;
 struct queue *p;
 ...
 while(1) {
  ...
  p = malloc(sizeof(struct queue) * b_z);
  ...
  if (handle_request(sock_fd)) {
    ...  
  }
 }
}

int handle_request(int sock_fd) {
 char buff[MAX_MSG_SIZE];
 int rc = -1;
 size_t r_len;
 r_len = read_msg(sock_fd, buff, 
                  MAX_MSG_SIZE);
 if (r_len > 0 && 
     r_len < MAX_MSG_SIZE) {
  switch(buff[0]) {
   case REQ1: 
    rc = process_req1(buff, r_len); 
    break;
   case REQ2: 
    rc = process_req2(buff, r_len); 
    break;
   ...
  }
 }
 return rc;
}
\end{minted}
    \end{minipage} &
    \begin{minipage}[b]{.22\textwidth}
\begin{minted}{c}
size_t read_msg(int sock_fd, char *msg, 
                size_t sz) {
 size_t nr;
 nr = read(sock_fd, |\rootcause|(void*)msg, sz);
 ...
 |\complexcode| // complex code
 return nr;
}

int process_req1(char *msg, size_t m_l) {
 int rc = -1, i;
 if (m_l > MIN_SIZE) {
   sscanf(msg, "%d", &i);
   |\realbug| msg[i] = ...
 }
 |\complexcode| // complex code
 return rc;
}

int process_req2(char *msg, size_t m_l) {
 size_t i = 0, j = 0;
 int rc = -1;
 char anum[MAX_MSG_SIZE] = {0};
 if (msg) {
  while(i < m_l && j < MAX_MSG_SIZE-1) {
   ...
   if (isalnum(msg[i]))
    anum[j++] = msg[i]; 
   i++;
  }
   rc = process_data(anum);
 }
 return rc;
}
\end{minted}
    \end{minipage} \\
   (a) Original C code & (b) After initial conversion.\\
  \end{tabular}
\caption{(Contrived) Example demonstrating various phases of.}
\label{lst:comb}
\end{listing}

\begin{listing}[t!]
\begin{minted}{c}
|\entrypoint| // Entry point
void server_loop(int sock_fd) {
 ...
 |\textcolor{checkcolor}{\_Array\_ptr}|<struct queue> p : count(b_z) = NULL;
 ...
}

int process_req2(_Array_ptr<char> msg: count(m_l), 
                 size_t m_l) {
 ...
 char anum |\textcolor{checkcolor}{\_Checked}|[MAX_MSG_SIZE] = {0};
 ...
}
\end{minted}
\caption{(Contrived) Example demonstrating various phases of.}
\label{lst:initialconv}
\end{listing}

\begin{listing}[t!]
\begin{minted}{c}
size_t read_msg(int sock_fd, char *msg, 
                size_t sz) |\textcolor{taintcolor}{\_Tainted}| |\useradded| {
 ...
}

int process_req1(char *msg, size_t m_l) |\textcolor{taintcolor}{\_Tainted}| |\useradded| {
 ...
}
\end{minted}
\caption{(Contrived) Example demonstrating various phases of.}
\label{lst:humantaint}
\end{listing}

\begin{listing}[t!]
\begin{minted}{c}
int handle_request(int sock_fd) {
 char buff[MAX_MSG_SIZE] |\textcolor{taintcolor}{\_Tainted} \usermods|;
 ...
}

int process_req2(|\usermods| |\textcolor{taintcolor}{\_T\_Array\_ptr}|<char> msg: count(m_l), 
                 size_t m_l) {
 ...
}
\end{minted}
\caption{(Contrived) Example demonstrating various phases of.}
\label{lst:humanadjust}
\end{listing}
%}
}

%\input{figures/taintedsplit}

%\begin{listing*}[t!]
%\inputminted[linenos, mathescape, escapeinside=||, fontsize=\tiny{}]{c}{examples/originalprogram.c}
%\caption{Simple server example.}
%\label{lst:comb}
%\end{listing*}

%% file: background.tex
\section{Background and Motivation}
\label{sec:bgm}
%\section{Overview and Transcendence}\label{sec:overview}
%This section describes \checkedc{} and new features \systemname{} provides.
Here, we brief \checkedc and the motivation for~\systemname{}.

\subsection{\checkedc}
\label{subsec:checkedc}
\checkedc{} development began in 2015 by Microsoft Research, but it was forked
in late 2021 and is now actively managed by the Secure Software
Development Project (SSDP). Details can be found in a prior
overview~\cite{Elliott2018} and the formalism~\cite{li22checkedc}.

\noindent
\myparagraph{Checked Pointer Types}
\checkedc{} introduces three varieties of \emph{checked pointer}:
\begin{itemize}
\item \code{_Ptr<}$T$\code{>} ($\ptr$) types a pointer that is either null or
  points to a single object of type $T$.
\item \code{_Array_ptr<}$T$\code{>} ($\arr$) types a pointer that is either null
  or points to an array of $T$ objects. The array width is defined
  by a \emph{bounds} expression, discussed below.
\item \code{_NT_Array_ptr<}$T$\code{>} ($\ntarr$) is like
  \code{_Array_ptr<}$T$\code{>} except that the bounds expression
  defines the \emph{minimum} array width---additional objects may
  be available past the upper bound, up to a null terminator.
\end{itemize}
Both $\arr$ and $\ntarr$ pointers have an associated bounds which defines the
range of memory referenced by the pointer.
The three different ways to declare bounds and the corresponding memory range is:
\begin{footnotesize}
\begin{tabular}{ll}
\arrT{|$T$|} \inlinecode{p: count(|$n$|)}
  &
$[\inlinecode{p}, \inlinecode{p}+\inlinecode{sizeof}(T) \times n) $ \\
%$[\inlinecode{p}, )$ \\
\arrT{|$T$|} \inlinecode{p: byte_count(|$b$|)}

  &
    $[\code{p}, \code{p}+b)$ \\

\arrT{|$T$|} \inlinecode{p: bounds(|$x, y$|)}

  &
    $[x, y)    $\\  
\end{tabular}
\end{footnotesize}
The bounds can be declared for $\ntarr$ as well, but the memory range can extend further to the right,
until a~\code{NULL} terminator is reached (\ie \code{NULL} is not within the bounds).

\noindent
\myparagraph{Ensuring Spatial Memory Safety}
The \checkedc compiler instruments loads and stores of checked
pointers to confirm the pointer is non-null, and additionally the access to $\arr$ and $\ntarr$ pointers is within
their specified bounds.
For example, in the code \inlinecode{if (n>0) a[n-1] =} $...$ the write
is via address $\alpha = \inlinecode{a + sizeof(int)}\!\times\!\code{(n-1)}$. 
If the bounds of \code{a} are \code{count(u)}, the
inserted check confirms
$\inlinecode{a} \leq \alpha < \inlinecode{a + sizeof(int)} \!\times\!
\inlinecode{u}$ prior to dereference.
Failed checks throw an exception.
Oftentimes, inserted checks can be optimized away by LLVM resulting in almost no runtime overhead~\cite{duanrefactoring}.

%\liyi{why we need the following? I suggest we can cut. }
%Oftentimes, inserted checks can be optimized away by LLVM~\cite{duanrefactoring}.
%For instance, Dual~\etal~\cite{duanrefactoring} found essentially no
%overhead when running Checked C-converted portions of the FreeBSD kernel.
%Consider the above code to be enclosed in another condition, such as,~\inlinecode{if (n<u) if (n>0) a[n-1] =}.
%In such cases, the inserted check can be
%removed as the outer condition $\inlinecode{n<u}$ already ensures that the
%dereference is within bounds.
%Dual~\etal~\cite{duanrefactoring} found essentially no
%overhead when running Checked C-converted portions of the FreeBSD kernel.

\noindent
\myparagraph{Backward Compatibility}
\checkedc is backward compatible with legacy C as all legacy code will type-check and compile.
However, the compiler adds the aforementioned spatial safety checks to only checked pointers.
The spatial safety guarantee is partial when the code is not fully ported.
Specifically, only code that appears in \emph{checked
  code regions} (\cregion), is guaranteed to be spatially safe.
$\cmode$ regions can be designated at the level of files, functions, or individual code blocks using the
\code{checked} keyword.\footnote{You can also designate~\emph{unchecked} regions (\ucregion) within checked ones.}
 Within $\cmode$ regions, both legacy pointers and
certain unsafe idioms (\eg \emph{variadic} function calls) are disallowed.

\myparagraph{Converting C to \checkedc}
It is not possible to fully automate the conversion of C code to~\checkedc due to the requirement for semantic reasoning and other modifications such as refactoring.
We provide more details on this in~\apdx{app:convertctocc}.

%%
%The safety guarantees of Checked C come with certain restrictions. For instance,
%as shown below, Checked C programs cannot use address-taken variables in a
%bounds expression as the bounds relations may not hold because of possible
%modifications through pointers.
%% 
%\begin{minted}[xleftmargin=30pt, mathescape, escapeinside=||, fontsize=\footnotesize]{c}
%...
%_Array_ptr<int> p : count (n) = NULL;
%|\textcolor{red}{\faTimes}|..,&n,.
%\end{minted}
%% 
%Consequently, converting existing C programs to Checked C might require
%refactoring,~\eg eliminate~\inlinecode{&n} from the program above without
%changing its functionality.
%% 
%This might require considerable effort~\cite{duanrefactoring} depending on the
%program's complexity.
%% 
%Recently, Machiry~\etal developed~\threec~\cite{machiry2022c} that tries to
%automatically convert a program to Checked C by adding appropriate pointer
%annotations.
%However, as described in \threec, completely automated conversion
%is \emph{infeasible}, and it requires the developer to convert some code regions
%manually.  

\subsection{No Safety Against $\umode$ Regions}
\label{subsec:nosafetyagsintuncheckedcode}
\checkedc provides spatial safety guarantees for completely converted programs,~\ie programs that uses~\emph{only} checked types and no regular pointer types.
A partially annotated program can still enjoy spatial safety only if checked pointers do not communicate with any unchecked ones. For instance, in the example below, there are no spatial safety violations in the function~\code{func} as it uses only checked pointers.
However, the other unconverted code regions (or unsafe regions) can affect pointers in safe regions and violate certain assumptions leading to vulnerabilities, as demonstrated by cross-language attacks~\cite{mergendahlcross}.

%The backward compatibility of~\checkedc helps a partially annotated program to enjoy spatial memory safety on those regions using only Checked pointers (\ie checked or safe regions).
%But, these unconverted code regions (or unsafe regions) can affect pointers in safe regions and violate certain assumptions leading to vulnerabilities, as demonstrated by cross-language attacks~\cite{mergendahlcross}. \lc{I'm confused, does the partially checked program "enjoy spatial memory safety" or not?}
Although the blameless proof exists~\cite{ruef2019achieving, li22checkedc} for~\checkedc, it does not state that spatial safety violations cannot happen in $\cmode$ regions but rather states that $\cmode$ regions~\emph{cannot be blamed for any spatial safety violations}.
Consider the following example:
\begin{minted}[xleftmargin=25pt, mathescape, linenos, escapeinside=||, fontsize=\footnotesize{}]{c}
// c region code
int func(array_ptr<char> p : count(5)) {
|\textcolor{red}{\faChainBroken}|..p[4]..
}
// u region code
...
str = "he";
...
func(|\textcolor{red}{\faBug}|assume_bounds_cast<char>(str, 5)); 
\end{minted}
Here, the $\cmode$ region function~\inlinecode{func} expects a pointer to a buffer of five elements, but the $\umode$ region code
invokes the function (Line 9) with a buffer of 2 elements.
This results in a spatial safety violation (\textcolor{red}{\faChainBroken}) in the $\cmode$ region, but of course, the blame or the root cause is in the $\umode$ region (\textcolor{red}{\faBug}).
Furthermore, since $\cmode$ and $\umode$ regions execute in the same address space, spatial memory corruptions (\eg buffer overflow) in $\umode$ regions can take down the complete program despite having $\cmode$ regions.
We need~\emph{an isolation mechanism to ensure that code executed as part of $\umode$  regions does not violate the safety guarantees in $\cmode$ regions}.

\subsection{Program Partitioning}
\label{subsec:background:programpart}
Program partitioning~\cite{rul2009towards} is a well-known technique to divide a program into multiple isolated parts or partitions.
There has been considerable work~\cite{tan2017principles, brumley2004privtrans, bittau2008wedge, lind2017glamdring, liu2017ptrsplit} in the area.
Most of these techniques are~\emph{data-centric}~\cite{lind2017glamdring, liu2017ptrsplit}, wherein program data drives the partitioning.
Specifically, given sensitive data in a program, the goal is to partition functions into two parts or partitions based on whether a function can access the sensitive data.
The performance overhead of these approaches is dominated by marshaling costs and depends on the usage of sensitive data.
The overhead of state-of-the-art approaches, such as Glamdring~\cite{lind2017glamdring} and  PtrSplit~
\cite{liu2017ptrsplit}, is prohibitive and varies from 37\%-163\%.

%% file: overview.tex
\section{Overview}
\label{sec:overview}

\begin{listing}[t!]
  \begin{tabular}{c c}
    \begin{minipage}[b]{.22\textwidth}
\begin{minted}{c}
|\entrypoint| // Entry point
void server_loop(int sock_fd) {
 unsigned b_z;
 struct queue *p;
 ...
 while(1) {
  ...
  p = malloc(sizeof(struct queue) * b_z);
  ...
  if (handle_request(sock_fd)) {
    ...  
  }
 }
}

int handle_request(int sock_fd) {
 char buff[MAX_MSG_SIZE];
 int rc = -1;
 size_t r_len;
 r_len = read_msg(sock_fd, buff, 
                  MAX_MSG_SIZE);
 if (r_len > 0 && 
     r_len < MAX_MSG_SIZE) {
  switch(buff[0]) {
   case REQ1: 
    rc = process_req1(buff, r_len); 
    break;
   case REQ2: 
    rc = process_req2(buff, r_len); 
    break;
   ...
  }
 }
 return rc;
}
\end{minted}
    \end{minipage} &
    \begin{minipage}[b]{.22\textwidth}
\begin{minted}{c}
size_t read_msg(int sock_fd, char *msg, 
                size_t sz) {
 size_t nr;
 nr = read(sock_fd, |\rootcause|(void*)msg, sz);
 ...
 |\complexcode| // complex code
 return nr;
}

int process_req1(char *msg, size_t m_l) {
 int rc = -1, i;
 if (m_l > MIN_SIZE) {
   sscanf(msg, "%d", &i);
   |\realbug| msg[i] = ...
 }
 |\complexcode| // complex code
 return rc;
}

int process_req2(char *msg, size_t m_l) {
 size_t i = 0, j = 0;
 int rc = -1;
 char anum[MAX_MSG_SIZE] = {0};
 if (msg) {
  while(i < m_l && j < MAX_MSG_SIZE-1) {
   ...
   if (isalnum(msg[i]))
    anum[j++] = msg[i]; 
   i++;
  }
   rc = process_data(anum);
 }
 return rc;
}
\end{minted}
    \end{minipage} %\\
   %(a) Original C code & (b) After initial conversion.\\
  \end{tabular}
\caption{C program snippet of a simple network server with an arbitrary memory write vulnerability indicated by~\realbug.}
\label{lst:comb}
\end{listing}

\begin{listing}[t!]
\begin{minted}{c}
|\entrypoint| // Entry point
void server_loop(int sock_fd) {
 ...
 |\textcolor{checkcolor}{\_Array\_ptr}|<struct queue> p : count(b_z) = NULL;
 ...
}

int process_req2(_Array_ptr<char> msg: count(m_l), 
                 size_t m_l) {
 ...
 char anum |\textcolor{checkcolor}{\_Checked}|[MAX_MSG_SIZE] = {0};
 ...
}
\end{minted}
\caption{Pointers in~\lst{lst:comb} annotated (manually or through automated tools like~\threec) with Checked Types.}
\label{lst:initialconv}
\end{listing}

\begin{listing}[t!]
\begin{minted}{c}
size_t read_msg(int sock_fd, char *msg, 
                size_t sz) |\textcolor{taintcolor}{\_Tainted}| |\useradded| {
 ...
}

int process_req1(char *msg, size_t m_l) |\textcolor{taintcolor}{\_Tainted}| |\useradded| {
 ...
}
\end{minted}
\caption{Initial annotations of tainted functions in~\lst{lst:comb}.}
\label{lst:humantaint}
\end{listing}

\begin{listing}[t!]
\begin{minted}{c}
int handle_request(int sock_fd) {
 char buff[MAX_MSG_SIZE] |\textcolor{taintcolor}{\_Tainted} \usermods|;
 ...
}

int process_req2(|\usermods| |\textcolor{taintcolor}{\_T\_Array\_ptr}|<char> msg: count(m_l), 
                 size_t m_l) {
 ...
}
\end{minted}
\caption{Additional tainted pointers annotations to~\lst{lst:comb} according to typing rules.}
\label{lst:humanadjust}
\end{listing}

\begin{listing}[t!]
  \begin{tabular}{c c}
    \begin{minipage}[b]{.22\textwidth}
\begin{minted}{c}
|\entrypoint| // Entry point
void server_loop(int sock_fd) {
 unsigned b_z;
 |\textcolor{checkcolor}{\_Array\_ptr}|<struct queue> p : 
              count(b_z) = NULL;
 ...
 while(1) {
  ...
  p = malloc(sizeof(struct queue) * b_z);
  ...
  if (handle_request(sock_fd)) {
    ...  
  }
 }
}

int handle_request(int sock_fd) {
 char buff[MAX_MSG_SIZE] |\textcolor{taintcolor}{\_Tainted}|;
 int rc = -1;
 size_t r_len;
 r_len = read_msg(sock_fd, buff, 
                  MAX_MSG_SIZE);
 if (r_len > 0 && 
     r_len < MAX_MSG_SIZE) {
  switch(buff[0]) {
   case REQ1: 
    rc = process_req1(buff, r_len); 
    break;
   case REQ2: 
    rc = process_req2(buff, r_len); 
    break;
   ...
  }
 }
 return rc;
}
\end{minted}
    \end{minipage} &
    \begin{minipage}[b]{.22\textwidth}
\begin{minted}{c}
size_t read_msg(int sock_fd, char *msg, 
                size_t sz) _Tainted {
 size_t nr;
 nr = read(sock_fd, (void*)msg, sz);
 ...
 |\complexcode| // complex code
 return nr;
}

int process_req1(_T_Array_ptr<char> msg :
                 count(m_l), size_t m_l) 
                 _Tainted {

 int rc = -1, i;
 if (m_l > MIN_SIZE) {
   sscanf(msg, "%d", &i);
   msg[i] = ...
 }
 // complex code
 return rc;
}

int process_req2(|\textcolor{taintcolor}{\_T\_Array\_ptr}|<char> msg : 
                 count(m_l), 
                 size_t m_l) {
 size_t i = 0, j = 0;
 int rc = -1;
 char anum |\textcolor{checkcolor}{\_Checked}|[MAX_MSG_SIZE] = {0};
 if (msg) {
  while(i < m_l && j < MAX_MSG_SIZE-1) {
   ...
   if (isalnum(msg[i]))
    anum[j++] = msg[i]; 
   i++;
  }
   rc = process_data(anum);
 }
 return rc;
}
\end{minted}
    \end{minipage} %\\
   %(a) Original C code & (b) After initial conversion.\\
  \end{tabular}
\caption{Final annotated program of~\lst{lst:comb} with~\textcolor{taintcolor}{tainted} and~\textcolor{checkcolor}{checked} types. The~\colorbox{taintcolor}{highlighted} functions will be executed in a sandbox.}
\label{lst:final}
\end{listing}

\input{figures/overview}

\Cref{fig:overview} shows the interaction between various components of~\systemname{}.
Given an appropriately annotated source program and a sandbox configuration, \systemname{} creates an executable such that all tainted functions and data reside in a sandbox and the~\cregion (non-sandboxed) is not affected by the code in the sandbox.
Here, we brief \systemname{} from a developer perspective and the details of individual components in the later sections.

%\Cref{fig:overview} provides the compiler structure of \systemname. The \systemname type checker compiles a \systemname program by partitioning the different program regions, i.e., $c$ and $uc$ regions, as well as different pointers, i.e., checked, unchecked and tainted pointers. Any $uc$ region pointers are executed within a sandbox, while checked pointers are executed via the \checkedc compiler within the checked region.

%\Cref{sec:formal} provides the formalism of the different compiler components.
%Converting a C/checked-c program to \systemname is a two-step process that involves identifying unsafe/unchecked regions described in \Cref{subsec:identifyregionstosbx} followed by making the \systemname changes described in \Cref{subsec:moveregionstosbx}.  
\subsection{Running Example}
\label{subsec:identifyregionstosbx}

To explain changes required by the \systemname compiler, \Cref{lst:comb} shows the C code of the redacted version of a simple network server with~\code{server_loop} as its entry point (\entrypoint).
The server runs in a loop and calls~\code{handle_request}, which handles a network request.
The function~\code{handle_request} reads data from the socket through~\code{read_msg} and based on the first byte, either~\code{process_req1} or~\code{process_req2} is called to handle the request.

\myparagraph{The Vulnerability}
There is an arbitrary memory write vulnerability (indicated by~\realbug) in~\code{process_req1} because of using~\code{i} as an index into the array~\code{msg} without any sanity check.
The variable~\code{i} can take any integer value as it is parsed from~\code{msg}, whose content is read from socket in~\code{read_msg} indicated by~\rootcause.

\myparagraph{Goal} 
The developer's goal is to partition the code in~\lst{lst:comb} so that spatial memory vulnerabilities do not affect the other part of the program.
Ideally, the developer can convert the entire code to \checkedc so that we achieve full spatial memory safety.
However, there might be considerable conversion efforts in \checkedc.
For instance,~\code{void*} pointers are not directly supported by \checkedc.
Consequently, the developer needs to convert functions using~\code{void*} pointers into generic versions -- this could be tedious.
To handle this, the developer can do a best-effort conversion and annotate only a few pointers~\eg by using an automated conversion tool such as~\threec{} which annotates few pointers as shown in~\lst{lst:initialconv}.
However, as shown in~\sect{subsec:nosafetyagsintuncheckedcode}, $\umode$ region code can also affect the safety of checked pointers.
%\aravind{Fix the running example code with the annotations according to the implementation.}

\subsection{\systemname Annotations}
\label{subsec:moveregionstosbx}
%\myparagraph{Using tainted types}
To overcome the above difficulty, The developer marks risky functions with unchecked pointers,~\ie~\code{read_msg} and~\code{process_req1} as~\code{tainted} as shown in~\lst{lst:humantaint} indicated by~\useradded by using with \systemname{},
 and results in the partially annotated program with checked and tainted types~\ie~\Cref{lst:initialconv,lst:humantaint} atop of~\lst{lst:comb}.
The initial tainted functions might require other pointers to be marked as tainted according to our typing rules (\sect{sec:typechecking}).
The developer uses our type checker to identify the additional required annotations (\usermods) and adds them as shown in~\lst{lst:humanadjust}.
The resulting well-typed program as shown in~\lst{lst:final} is passed to our source level program partitioner along with certain configuration parameters of the target sandbox.

\subsection{Partitioning}
Our partitioner splits the provided program into two sets of source files with the necessary source changes required to communicate with sandboxed code.
These sets of source files are compiled with the corresponding compilers to get the corresponding object files.
The \cregion object file has the necessary runtime checks enforcing~\systemname{} guarantees.
The~\ucregion object file is produced according to the corresponding sandbox mechanism.
Finally, these two object files are linked along with the necessary sandbox libraries to produce the final executable such that all the tainted functions are executed in a sandbox (\ucregion) and the rest of the functions as regular code (\cregion).

%newusecase.tex

%% file: figures/overview.tex
% figure source: https://drive.google.com/file/d/1vPE62VCFzwawdjjcqKQc-ALavDqaLec4/view?usp=sharing
\begin{figure*}[t]
\includegraphics[width=1.0\linewidth]{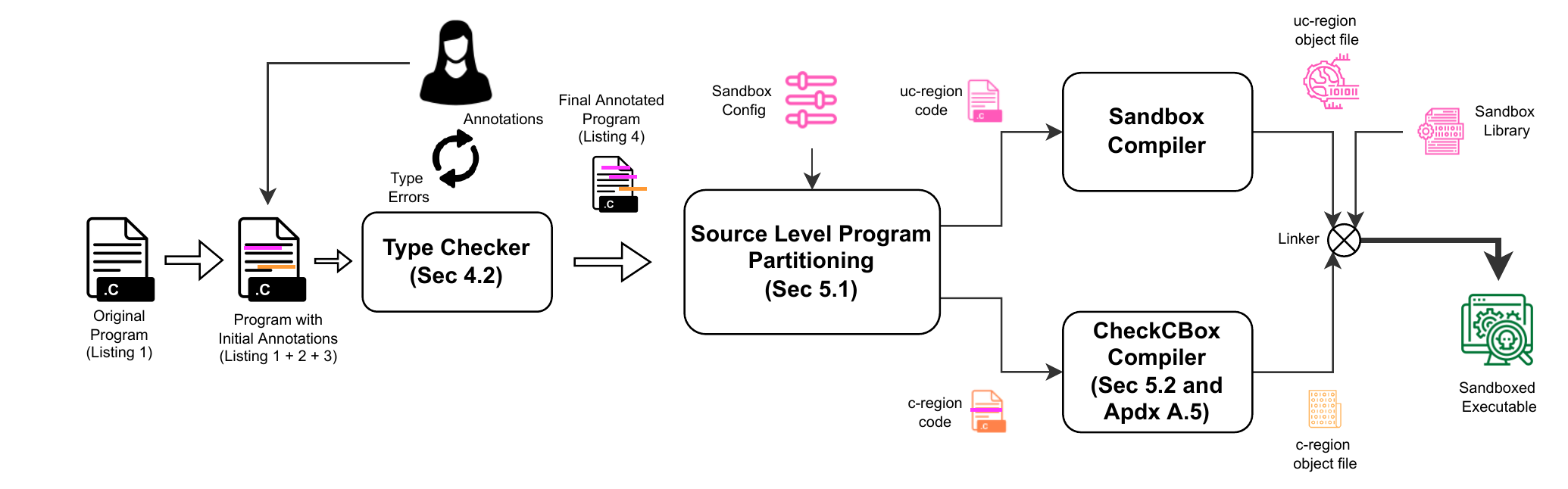}
\caption{Overview of interaction between various phases of~\systemname.}
\label{fig:overview}
\end{figure*}

%% file: formal.tex
\section{\systemname Formalism}
\label{sec:formal}

% \begin{itemize}
% \item Describe the types for checked-C as a graph. \dvh{I don't know what this means.}

% \item Describe the subtyping for checked-C types.
% \begin{itemize}
% \item State that the subtypes in Checked-C are transitive.
% \end{itemize}

% \item Describe the syntax of Checked-C

% \item Describe the semantics of Checked-C

% \item Describe the type system of Checked-C

% \item Describe the progress and preservation theorems, and outline the proofs.

% \item Describe the blame theorem and proof.

% \end{itemize}

% \ignore{
% \begin{figure}
%   \begin{center}
%     \includegraphics[height=6in]{syntax.pdf}
%   \end{center}
%   \caption{\lang: Syntax}
% \end{figure}

% \begin{figure}
%   \begin{center}
%     \includegraphics[height=6in]{types.pdf}
%   \end{center}
%   \caption{\lang: Typing}
% \end{figure}
% }
% \liyi{main text begins here. }

% \review{
% While reading the semantics, I found the fact that S-Def and S-DefNull are
%   applicable non-deterministically if n is 0 a bit confusing. Only when
%   reading the meta-theory section I realized that this is not a concrete issue
%   because well-formed heaps are such that $\mathcal{H}(0)$ is never defined. It
%   might be worth pointing this out early on. }
% \mwh{Done.}

\begin{figure}
  \small \centering
  \[ \hspace*{-1.5em}
\setlength{\arraycolsep}{4pt}
\begin{array}{l}
\begin{array}{ll}
       \text{Variables:}~ x
& \quad\text{Integers:}~n::=\mathbb{Z} 
\end{array}
\\[0.5em]
\begin{array}{llcllcl}
\text{Context Mode:} & m & ::= & \cmode \mid \umode \\
\text{Pointer Mode:} & \xi & ::= & m \mid \tmode \\
\text{Bound:} & b & ::= & n \mid x \plus n \\
              & \bvar & ::= & (b,b) \\
  
     \text{Word Type:}& \tau &::=& \tint\mid \tptr{\omega}{ \xi}
\\
\text{Type Flag:}&\kappa &::=& nt \mid \cdot
\\
\text{Type:}&\omega &::=& \tau \mid \tallarrayb{\bvar}{\tau} \mid \tfun{\overline{x}}{\overline{\tau}}{\tau}
\\
\text{Expression:}& e & ::= & 
\evalue{n}{\tau} \mid x \mid \ebinop{e}{e}\mid \ecast{\tau}{e} \mid \edyncast{\tau}{e}
\mid \ret{x}{\evalue{n}{\tau}}{e}
  \\
&&\mid& \estrlen{x} \mid \emalloc{\xi}{\omega} \mid\estar{e}\mid\eassign{e}{e}  \\
&&\mid& \elet{x}{e}{e} \mid \eif{e}{e}{e} \mid \ecall{e}{\overline{e}}
\\
&&\mid&\eunchecked{\overline{x}}{e}
\mid \echecked{\overline{x}}{e}
\end{array}
    \end{array}
  \]
  \caption{\lang Syntax}
  \label{fig:checkc-syn}
\end{figure}

\ignore{
\begin{figure}[t]
{\small
  \begin{mathpar}

  \inferrule[]
  {}
  {m \vdash \tint}

  \inferrule[]
  {\xi \wedge m\vdash \tau \\ \xi \le m}
  {m \vdash \tptr{\tallarrayb{\bvar}{\tau}}{\xi}}

  \inferrule[]
  {\xi \wedge m \vdash \tau\\ \xi \le m}
  {m \vdash \tptr{\tau}{\xi}}

  \inferrule[]
  {\xi \wedge m \vdash \tau\\ \xi \le m \\\\ \fv(\overline{\tau})\cup\fv(\tau)\subseteq \overline{x}}
  {m \vdash \tptr{(\tfun{\overline{x}}{\overline{\tau}}{\tau}}{\xi})}
  \end{mathpar}
}
{\footnotesize
\[
\begin{array}{l} 
\tmode \wedge \cmode = \umode \qquad \xi \wedge \umode = \umode
\qquad \cmode \wedge m = m 
\qquad  m_1 \wedge m_2 = m_2 \wedge m_1
\\[0.2em]
\xi \le \xi \qquad \tmode \le \xi
\end{array}
\]
}
 \caption{Well-formedness for Types}
\label{fig:wftypes}
\end{figure}
}

%% \dvh{I don't understand the variable grammar.  What is $T$?  What is $\eta$?  I think $\cmode$ and $\umode$ should be in tt font.}
%% \liyi{T and $\eta$ can be moved to the appendix, they are useful only for struct types.}

% \review{
% - Furthermore, inspecting the code also suggests that the expression type (line 155) does not contain constructors for function calls (and I don't see a way to define functions either), conditionals, or strlen, and doesn't distinguish between the two forms of casting. All this contradicts figure 2, and should be clarified
% }
% \liyi{It is in the CheckedC.v file, 392. }
% \mwh{How is this answer helping the reviewer since you've added
%   nothing to the text? Maybe we should add something to an appendix
%   that matches the formalism shown in the paper to definitions in the
%   Coq file?}
% \liyi{The detailed explanation is in the appendix.}

This section describes the formal core model of~\systemname,
named~\lang.
We present its syntax, semantics, type system,
as well as \lang’s meta-theories, including the type soundness, non-exposure, and non-crashing theorems.

%This section describes the formal core model of \systemname, named
%\lang, making precise its syntax, semantics, type system, and compilation. It also
%develops \lang's meta-theories, including the type soundness, non-exposure, non-crashing, and the compiler simulation theorems.

\subsection{Syntax}\label{sec:syntax}
\Cref{fig:checkc-syn} shows the syntax of~\lang.

\myparagraph{Type Syntax}
At a high level, we classify types as word-size value, multi-word
value, or function types. A word-size value can be either an integer or pointer.
Every pointer type ($\tptr{\omega}{\xi}$) includes a
pointer mode annotation ($\xi$, the difference between context and pointer modes
is introduced shortly below) that is either checked (\cmode), tainted (\tmode),
or unchecked (\umode), and a type ($\omega$) denoting the valid value type it is pointed to.
A multi-word value type ($ \tallarrayb{\bvar}{\tau}$) that ranges over 
arrays and null-terminated arrays is constructed by the type of elements in the
array ($\tau$), an array bound ($\bvar$) comprised of an upper and
lower bound on the size of the array ($(b_l,b_h)$), and an array flag ($\kappa$).
Bounds $b$ are
limited to integer literals $n$ and expressions $x + n$.
Whether an array pointer is null terminated or not is determined by annotation
$\kappa$, which is $nt$ for null-terminated arrays, and $\cdot$
otherwise (we elide $\cdot$ when writing types).
An example representation of an array and null-terminated array in~\lang is shown below:
\[\hspace*{-0.5em}
\begin{array}{l}
\begin{array}{rcl}
$\code{_t_Array_ptr<}$\tau$\code{> : count(}$n$\code{)}$
&\Leftrightarrow& \tarrayptr{0}{n}{\tau}{\tmode}
\\[0.2em]
$\code{_NT_Array_ptr<}$\tau$\code{> : count(}$n$\code{)}$
&\Leftrightarrow& \tntarrayptr{0}{n}{\tau}{\cmode}
\end{array}
\end{array}
\]

For simplicity, we write
$\tptr{\tarrayb{b}{\tau}}{\cmode}$ to mean $\tptr{\tarray{0}{b}{\tau}}{\cmode}$,
so the above examples could be rewritten as $\tptr{\tarrayb{n}{\tau}}{\cmode}$ and
$\tptr{\tntarrayb{n}{\tau}}{\cmode}$, respectively.

\myparagraph{Disallowing Unsafe Types}
The well-formedness of these types are presented in~\Cref{app:le}.
It prevents unsafe types from being constructed.
  Consider the type ~\code{_t_Array_ptr<_Ptr<int>>}, which describes a tainted array of checked pointers.
  This is not well-formed in~\lang{} because it potentially exposes the checked pointer addresses in a $\umode$ region when the tainted ($\tmode$) array is used. 
  Nevertheless, we can have a checked array whose elements are tainted pointers\mzs{,}\mzr{:}~\eg~\code{_Array_ptr<_t_Ptr<int>>} is a valid type.
Function types are represented using a dependent function declarations,~\ie $\tfun{\overline{x}}{\overline{\tau}}{\tau}$,
where $\overline{x}$ represents a list of \tint{} type variables that bind type variables appearing in $\overline{\tau}$ and $\tau$.
An example of a function pointer type is shown below:
\[\hspace*{-0.5em}
\begin{array}{l}
$\code{_t_ptr<(int)(_t_NT_Array_ptr<}$\tau_1$\code{> : count(}$n$\code{),}$\\
\qquad\qquad$\code{_t_NT_Array_ptr<}$\tau_2$\code{>: count(}$n$\code{), int}$\;n$\code{)>}$
\\[0.2em]
\Leftrightarrow\; $\tptr{(\tfun{n}{\tint \times \tntarrayptr{0}{n}{\tau_2}{\tmode} \times \tntarrayptr{0}{n}{\tau_1}{\tmode}}{\tint})}{\tmode}$
\end{array}
\]
The function type also has well-formed requirements (\Cref{app:le}), which disallows nesting checked pointers inside tainted pointers.
Furthermore, these requirements also ensure that all variables in $\overline{\tau}$ and $\tau$ are bounded by $\overline{x}$.

\myparagraph{Expressions}
\lang expressions include common expressions such as addition ($\ebinop{e_1}{e_2}$), 
pointer dereference ($\estar{e}$) and assignment ($\eassign{e_1}{e_2}$),
along with expressions that require special handling, such as,
 static casts ($\ecast{\tau}{e}$), dynamic casts ($\edyncast{\tau}{e}$) \footnote{assumed at compile-time and verified at run-time, see \cite{li22checkedc}}, the \texttt{strlen} operation ($\estrlen{x}$),
memory allocations ($\emalloc{\xi}{\omega}$), 
function calls ($\ecall{e}{\overline{e}}$),
unchecked blocks ($\eunchecked{\overline{x}}{e}$), and checked blocks ($\echecked{\overline{x}}{e}$).
For example, a dynamic bounds cast
  { \code{dyn_bounds_cast<_Array_ptr<}$\tau$\code{>>(}$e$\code{,count(}$n$\code{))} }
is formalized as 
{$\edyncast{\tptr{\tarrayb{n}{\tau}}{\cmode}}{e}$} in \lang. 
We denote integer literals $n$ with a type $\tau$ (\ie $\tint$ or $\tptr{\omega}{\xi}$), enabling the use of fixed addresses as pointers.
For example, $\evalue{0}{\tptr{\omega}{\xi}}$ (for any $\xi$ and $\omega$) represents a $\enull$ pointer.
The heap allocation $\emalloc{\xi}{\omega}$ includes a mode flag $\xi$ for allocating memory in different regions, $\cmode$ mode pointer in \cregion or $\umode$ and $\tmode$ mode pointers in \ucregion.
We disallow $\omega$ to be a function type ($\tfun{\overline{x}}{\overline{\tau}}{\tau}$).
The $\echeckedtext$ and $\euncheckedtext$ expressions are used to delimit code regions.
To guarantee the non-exposure property, we extend the \checkedc syntax to include $\echecked{\overline{x}}{e}$ and $\eunchecked{\overline{x}}{e}$ blocks,
where $\overline{x}$ represents all variables that are allowed to communicate between instructions outside and inside of the block $e$, and it cannot contain checked pointers.
$\erettext$ is introduced by the semantics when evaluating a \texttt{let} binding; explained shortly below.

\lang{} aims to be simple enough to work with but powerful enough to
encode realistic \systemname idioms. For example,
loops can be encoded as recursive function calls. \code{struct}s are
not included in~\fig{fig:checkc-syn} for space reasons, but they are
supported as shown in~\cite{li22checkedc}.
C-style \code{union}s have no safe typing in \checkedc, so we omit them.
Although the base syntax of~\lang is similar to that of~\checkedc model~\cite{li22checkedc}, there are considerable enhancements to support tainted types ($t$ in $\xi$), special heap handling (\ie $\emalloc{\xi}{\omega}$), and explicit specification of pointers in the $\echeckedtext$ and $\euncheckedtext$ regions.

\begin{figure}
{\small
$\hspace*{-1.2em}
    \begin{array}{l}
    \begin{array}{lll}
r & ::= & e \mid \enull \mid \ebounds\\
E &::=& \Box \mid \ebinop{E}{e} \mid \ebinop{\evalue{n}{\tau}}{E}\mid \ecast{\tau}{E} \mid \edyncast{\tau}{E} \mid\estar{E}\mid\eassign{E}{e}\\[0.2em]
&&\mid\eassign{\evalue{n}{\tau}}{E}\mid \elet{x}{E}{e}\mid\ret{x}{\evalue{n}{\tau}}{E}\mid \eif{E}{e}{e}\\[0.2em]
&&\mid \ecall{E}{\overline{e}} \mid \ecall{\evalue{n}{\tau}}{\overline{E}} \mid 
\eunchecked{\overline{x}}{E}
\mid \echecked{\overline{x}}{E}

\end{array}
\\ \\
    \end{array} 
$
  \begin{mathpar}
    \inferrule{ m=\mode(E) \\
      e=E[e'] \\
      (\varphi,\heap,e') \longrightarrow (\varphi',\heap',e'')}
    {(\varphi,\heap,e)\longrightarrow_{m} (\varphi',\heap',E[e''])}

    \inferrule{ \umode=\mode(E) \\
      e=E[e'] \\
      \tau=\type(e')}
    {(\varphi,\heap,e)\longrightarrow_{\umode} (\varphi,\heap,E[\evalue{0}{\tau}])}

  \end{mathpar}
}
  \caption{\lang Semantics: Evaluation}
  \label{fig:c-context}
\end{figure}

\subsection{Typing and Semantics}
\label{sec:typechecking}

\begin{DIFnomarkup}
\begin{figure*}[t]
{\small
  \begin{mathpar}
     \inferrule[T-CastPtr]
               {\Gamma;\Theta \vdash_m e : \tau' \\
                 \tau' \sqsubseteq_{\Theta} \tptr{\tau}{\xi}}
               {\Gamma;\Theta \vdash_m \ecast{\tptr{\tau}{\xi}}{e} : \tptr{\tau}{\xi}}
                
    \inferrule[S-Cast]
              {}
              {(\varphi,\heap,\ecast{\tau}{\evalue{n}{\tau'}}) \longrightarrow (\varphi,\heap,\evalue{n}{\varphi(\tau)})}
  \end{mathpar}

  \begin{mathpar}
    \inferrule[T-Def]
              {\xi \leq m \\\Gamma;\Theta \vdash_m e : \tptr{\tau}{\xi}}
              {\Gamma;\Theta \vdash_m \estar{e} : \tau}
\quad
        \inferrule[S-DefC]{\heap(\cmode,n)=\evalue{n_a}{\tau_a} }
    {(\varphi,\heap,\estar{\evalue{n}{\tptr{\tau}{c}}}) \longrightarrow (\varphi,\heap,\evalue{n_a}{\tau})}
\quad
        \inferrule[S-DefT]{\heap(\umode,n)=\evalue{n_a}{\tau_a}
         \\  \emptyset;\heap ; \emptyset \vdash_{\umode}\evalue{n_a}{\tau} }
    {(\varphi,\heap,\estar{\evalue{n}{\tptr{\tau}{\tmode}}}) \longrightarrow (\varphi,\heap,\evalue{n_a}{\tau})}
\quad
    \inferrule[S-DefNull]{}{(\varphi,\heap,\estar{\evalue{0}{\tptr{\omega}{\cmode}}}) \longrightarrow (\varphi,\heap,\enull)}
  \end{mathpar}

  \begin{mathpar}
    \inferrule[T-Checked]
              {\forall x\in\overline{x}\;.\;\neg\cmode(\Gamma(x))\\\neg\cmode(\tau)
                     \\\\\fv(e)\in\overline{x}\\\Gamma;\Theta \vdash_c e : \tau}
              {\Gamma;\Theta \vdash_m \echecked{\overline{x}}{e} : \tau}

    \inferrule[T-Unchecked]
              {\forall x\in\overline{x}\;.\;\neg\cmode(\Gamma(x))\\\neg\cmode(\tau)
                \\\\ \fv(e)\in\overline{x}\\\Gamma;\Theta \vdash_u e : \tau}
              {\Gamma;\Theta \vdash_m \eunchecked{\overline{x}}{e} : \tau}

    \inferrule[S-Unchecked]{}{(\varphi,\heap,\eunchecked{\overline{x}}{\evalue{n}{\tau}}) \longrightarrow (\varphi,\heap,\evalue{n}{\tau})}
  \end{mathpar}

  \begin{mathpar}
\inferrule[T-Fun]
    {\Gamma;\Theta \vdash_m e : \tptr{\tfun{\overline{x}}{\overline{\tau}}{\tau}}{\xi} \\
        \Gamma; \Theta \vdash_m \overline{e} : \overline{\tau'} \\
         \overline{e'}=\{e'|(e',\tint)\in (\overline{e} : \overline{\tau'})\}\\\\
         \forall e'\;.\;e' \in \overline{e'} \Rightarrow e'\in \text{Bound}\\
             \overline{\tau'} \sqsubseteq_{\Theta}
               \overline{\tau}[\overline{e'} / \overline{x}]}
    {\Gamma; \Theta \vdash_m e(\overline{e}) : \tau[\overline{e'} / \overline{x}]}
  \end{mathpar}
  \begin{mathpar}
    \inferrule[S-FunC]{ \Xi(\cmode,n) = \tau\;(\evalue{\overline{x}}{\overline{\tau}})\;(\cmode,e)}
        {(\varphi,\heap,\ecall{\evalue{n}{(\tptr{\tau}{\cmode})}}{{\evalue{\overline{n_a}}{\overline{\tau_a}}}}) \longrightarrow
   (\varphi,\heap, \mathtt{let}\;\overline{x}={\evalue{\overline{n}}{(\overline{\tau}[\overline{n} / \overline{x}])}}\;\mathtt{in}\;\ecast{\tau[\overline{n} / \overline{x}]}{e})}
\quad
    \inferrule[S-FunT]{ \Xi(\umode,n) = \tau\;(\evalue{\overline{x}}{\overline{\tau}})\;(\tmode,e)
                  \\ \emptyset;\heap ; \emptyset \vdash_{\umode}\evalue{n}{\tptr{\tau}{\tmode}}}
        {(\varphi,\heap,\ecall{\evalue{n}{(\tptr{\tau}{\tmode})}}{{\evalue{\overline{n_a}}{\overline{\tau_a}}}}) \longrightarrow
   (\varphi,\heap, \mathtt{let}\;\overline{x}={\evalue{\overline{n}}{(\overline{\tau}[\overline{n} / \overline{x}])}}\;\mathtt{in}\;\ecast{\tau[\overline{n} / \overline{x}]}{e})}

  \end{mathpar}
}
% {\footnotesize
% \begin{center}
% $
% \begin{array}{l}
% \fm(e)\triangleq(\exists x\; n\; \tau. e=x+\evalue{n}{\tau}) \vee (\exists n\;\tau. e = \evalue{n}{\tau})
% \\[0.2em]
% \tau[\overline{e} / \overline{x}]\texttt{(with types }\evalue{\overline{x}}{\overline{\tau}}\texttt{)}\triangleq \forall e_i\in\overline{e}\;x_i\in\overline{x}\;\tau_i\in\overline{\tau}\;.\;\tau_i = \tint \wedge (x_i \in \fv(\tau) \Rightarrow \fm(e_i)) \Rightarrow \tau[e_i / x_i]
% \end{array}
% $
% \end{center}
% }
{\footnotesize
$
\cmode(\tint)=\texttt{false}
\qquad
\cmode(\tptr{\omega}{\cmode})=\texttt{true}
\qquad
\cmode(\tptr{\omega}{\xi})=\texttt{false}\;\;{[\emph{owise}]}
$
}
\caption{Selected typing (\rulelab{T}-$X$) and semantic (\rulelab{S}-$X$) rules. First line is for cast operations, second line is for pointer dereferences, third line is for checked/unchecked blocks, and the rest is for function calls.}
\label{fig:type-system-1}
\end{figure*}
\end{DIFnomarkup}

The \lang type system is a flow-sensitive, gradual type one that generates additional dynamic checks that are inserted in the typing checking stage and executed in the semantic evaluation stage.
Our type checker restricts the usage of tainted and checked pointer types to ensure that tainted pointers do not affect checked types, along with enforcing~\checkedc typing rules~\cite{li22checkedc}.

As partly shown in \Cref{fig:type-system-1} (labeled as \rulelab{T-}$X$),
each typing judgment has the form $\Gamma;\Theta\vdash_m e : \tau$,
which states that in a type environment $\Gamma$ (mapping variables to
their types) and a predicate environment $\Theta$ (mapping integer-typed
variables to Boolean predicates), expression $e$ will have type $\tau$ if evaluated
in context mode $m$, indicating that the code is in $m$ region.
The operational semantics for \lang is defined as a small-step
transition relation with the judgment $ (\varphi,\heap,e)
\longrightarrow_m (\varphi',\heap',r)$, as shown in \Cref{fig:c-context}.
 Here, $\varphi$ is a
\emph{stack} mapping from variables to values $\evalue{n}{\tau}$ and
$\heap$ is a \emph{heap} that is partitioned into two parts ($\cmode$ and $\umode$ heap regions), each of which
maps addresses (integer literals) to values $\evalue{n}{\tau}$.
The complete set of typing rules and special handling of (NT)-arrays are provided in~\Cref{rem-type,sec:rem-semantics}.

We wrote $\heap(m,n)$ to retrieve the $n$-location heap value in the $m$ heap,
and $\heapup{m}{n}{\evalue{n'}{\tau}}$ 
to update location $n$ with the value $\evalue{n'}{\tau}$ in the $m$ heap.
While heap bindings can change, stack bindings are immutable---once
variable $x$ is bound to $\evalue{n}{\tau}$ in $\varphi$, that binding will not
be updated. 
%We can model mutable stack variables as pointers into the mutable heap.
As mentioned, value $\evalue{0}{\tau}$
represents a $\enull$ pointer when $\tau$ is a pointer type.
Correspondingly, $\heap(m,0)$ should always be undefined.
The relation steps to a \emph{result} $r$, which is either an
expression, a $\enull$ or $\ebounds$ failure, represent an expression right
  after the reduction, a null-pointer dereference or out-of-bounds access,
respectively.
Such failures are a \emph{good} outcome; stuck states
(non-value expressions that cannot transition to a result $r$)
characterizing undefined behavior.
%The context mode $m$ (in $\longrightarrow_{m}$) indicates whether the stepped redex within $e$ was in a $\cmode$ or $\umode$ region.

The rules for the main operational semantics
judgment \emph{evaluation} are given at the bottom of
Fig.~\ref{fig:c-context}.
The first rule takes an expression $e$, decomposes
it into an \emph{evaluation context} $E$ and a sub-expression $e'$
(such that replacing the hole $\Box$ in $E$ with $e'$ would yield
$e$), and then evaluates $e'$ according to the \emph{computation}
  relation $(\varphi,\heap,e') \longrightarrow (\varphi,\heap,e'')$,
whose rules are given along with type rules in Fig.~\ref{fig:type-system-1} (labeled as \rulelab{S-}$X$), discussed
shortly.
The $\mode$ function in Fig.~\ref{fig:c-context}
determines the context mode, i.e., region, that the expression $e'$ locates based on the context $E$.
In \Cref{lst:humantaint}, the function call \code{handle_request} is in $\umode$ region since it is inside an unchecked function \code{server_loop}.
The second rule describes the exception handling 
for possible crashing behaviors in $\umode$ regions.
Operations in $\umode$ region can non-deterministically crash
and the \systemname sandbox mechanism recovers
the program to a safe point ($\evalue{0}{\tau}$)
and continues with the existing program state.
Evaluation contexts $E$ define a standard left-to-right evaluation order. 
%(We explain the $\ret{x}{\mu}{e}$ syntax shortly.)
%There are other rules for describing the halts of evaluation to $\enull$ and $\ebounds$ states in \Cref{app:main}.

\myparagraph{Modes, Static Casting, and Subtyping}
In \lang, Context modes $m$ appearing in a type rule determine the code region 
that permits pointer dereferences and value-assignments, which also depends on the pointer modes.
We define a three point lattice $\xi_1 \le \xi_2$ \footnote{In typing rule, the lattice is usually used as $\xi \le m$ as $m$ represents context modes.} to describe such permission, where $\tmode \le \xi$ and $m \le m$.
This means that a $\tmode$ pointer can be dereferenced and value-assigned in any region, while $\cmode$ and $\umode$ pointers can only perform such operations in $\cmode$ and $\umode$ regions, respectively.
%Pointer modes are also useful in determining if a nested pointer has a valid type. For example, in a nested pointer $\tptr{(... \tptr{\tau}{\xi_2} ...)}{\xi_1}$, we require $\xi_2\le \xi_1$ to maintain non-exposure.

\lang also provides static casting operations. As described in rule \textsc{T-CastPtr} in \Cref{fig:type-system-1},
an pointer typed expression of type $\tptr{\tau_1}{\xi_1}$ can be casted to another pointer type \tptr{\tau_2}{\xi_2},
if \tptr{\tau_1}{\xi_1} subtypes ($\sqsubseteq_{\Theta}$) to \tptr{\tau_2}{\xi_2}, i.e., $\tptr{\tau_1}{\xi_1} \sqsubseteq \tptr{\tau_2}{\xi_2}$.
In \lang, except that we can cast a $\tmode$ mode pointer to a $\umode$ mode one, all subtyping relations are between two types with the same mode, meaning that $\xi_1$ and $\xi_2$ above are mostly the same and the above mode lattice ($\le$) has no business with subtyping.

\begin{minted}[xleftmargin=30pt, mathescape, escapeinside=||, fontsize=\footnotesize]{c}
//_Ptr<int> x; _t_Ptr<int> y; int *z;
z = (int *)y; // This is okay.
x = (_Ptr<int>)y; // Not allowed.
\end{minted}

In the above example, a $\tmode$ mode pointer can be cast to $\umode$ mode but casting $\tmode$ mode to $\cmode$ mode is disallowed.
The complete subtyping relation was described in \Cref{app:le}.
Notice that \texttt{let} statements are immutable in \lang, so the following code is not possible, because variables \code{x} and \code{y} must have the same type in \lang.

\begin{minted}[xleftmargin=30pt, mathescape, escapeinside=||, fontsize=\footnotesize]{c}
//_Ptr<int> x;  _t_Ptr<int> y;
x = y; // Not allowed.
\end{minted}

\myparagraph{Pointer Dereference}
The type and semantic rules for pointer dereference (\textsc{T-Def}, \textsc{S-DefC},
\textsc{S-DefT}, \textsc{S-DefNull} in \Cref{fig:type-system-1})
reflect the key \lang feature, where our type checker directs the insertions of dynamic checks executed in the evaluation stage.
The type rule (\textsc{T-Def}) ensures that pointers are used with the right modes in the right region ($\xi \le m$).
With the dynamic checks inserted by the compiler, rule \textsc{S-DefNull} ensure that if a $\enull$ pointer is used,
\lang captures the runtime error.
Type and semantic rules for array types and pointer assignments are given in \Cref{rem-type,sec:rem-semantics}.

Rules \textsc{S-DefC} and \textsc{S-DefT} are for $\cmode$ and $\tmode$ mode pointer dereferences, respectively.
In addition to the no $\enull$ check in $\cmode$ mode pointer dereference,
any dynamic heap access of a tainted ($\tmode$) pointer requires a \textit{verification} ($\emptyset;\heap ; \emptyset \vdash_{\umode}\evalue{n_a}{\tau}$), which refers to that the pointer value $n_a$ is well-defined in $\heap(m,n_a)$ and has right type $\tau$.

\myparagraph{Unchecked and Checked Blocks}
The execution of a $\echeckedtext$ or $\euncheckedtext$ block represents 
the context switching from a $\cmode$ to an $\umode$ region, or vice versa,
with its type and semantic rules given in \Cref{fig:type-system-1}.
In this context switching, to guarantee the checked ($\cmode$) pointer non-exposure property, 
checked pointers are not allowed to go cross different regions, which is guaranteed by the predicates 
$\forall x\in\overline{x}\;.\;\neg\cmode(\Gamma(x))$ and $\neg\cmode(\tau)$,
as well as the check that all free variables in the block content $e$ are in $\overline{x}$.
For example, \code{StringAuth} in \Cref{subsub:gencregion} is a trampoline function that disallows
checked pointers as arguments and return values.
The use of the function in the following \code{_T_StringAuth}, which is in $\umode$ region,
cannot legally acknowledges any checked pointers; otherwise, we might expose a checked pointer address to unsafe code regions.
In \systemname, we actually permits the accesses of checked pointers inside \code{StringAuth},
since the function body of a trampoline function is in $\cmode$ region.
More information is given in \Cref{subsub:gencregion}.

\myparagraph{Dependent Function Pointers} %
Rule \textsc{T-Fun} (\Cref{fig:type-system-1}) states the type judgment for
dependent function pointer application, where we represent the result of
replacing all integer bound variables $\overline{x}$ in the type \(\tau\) with
with bound expressions $\overline{e'}$ by $\tau[\overline{e'} / \overline{x}]$
and write $\overline{\tau}[\overline{e'} / \overline{x}]$ to lift the
substitution to every type in \(\overline{\tau}\).
Given an expression $e$ of function pointer type
($\tptr{\tfun{\overline{x}}{\overline{\tau}}{\tau}}{\xi}$) and arguments
$\overline{e}$ of types $\overline{\tau'}$,
the result of the application will
be of type $\tau[\overline{e'} / \overline{x}]$;
if for each pair of \(\tau'\) and \(\tau''\) in \(\overline{\tau'}\) and
$\overline{\tau}[\overline{e'} / \overline{x}]$, \(\tau'\) is a subtype of
\(\tau''\).
Consider the \code{process_req2} function in
Fig.~\ref{lst:final}, whose parameter type for \code{msg} 
depends on \code{m_1}.
Its function pointer type is 
$\tptr{\tfun{\code{m_1}}{\tint,\tntarray{0}{\code{m_1}}{\texttt{char}}}{\tint}}{\tmode}$.
In \code{handle_request}, the call \code{process_req2(buff, r_len)} binds variable \code{m_1} to \code{r_len}.
After the call returns, \code{m_1}'s scope is ended, so we need to substitute it with \code{r_len} in the final return type
because it might contain \code{m_1}.

\textsc{S-FunC} and \textsc{S-FunT} define the semantics
for $\cmode$ and $\tmode$ mode function pointers, respectively. 
A call to a function pointer $n$ retrieves
 the function definition in $n$'s location in the global function store $\Xi$,
which maps function pointers to
function data $\tau\;(\evalue{\overline{x}}{\overline{\tau}})\;(\xi,e)$, where
$\tau$ is the return type, $(\evalue{\overline{x}}{\overline{\tau}})$
is the parameter list of variables and their types, 
$\xi$ determines the mode of the function, and $e$ is the
function body. 
Similar to \heap, the global function store $\Xi$ is also partitioned into
two parts ($\cmode$ and $\umode$ store regions), each of which
maps addresses (integer literals) to the function data described above.
Rule \textsc{S-FunT} defines the tainted version of function call
with the verification process 
$\emptyset;\heap ; \emptyset \vdash_{\umode}\evalue{n}{\tptr{\tau}{\tmode}}$
makes sure that the function in the global store is well-defined and has the right type.

\subsection{Meta Theories}\label{sec:theorem}

Here, we discuss our main meta-theoretic results for
\lang: type soundness (progress and preservation),
non-exposure, and non-crashing.
These proofs have been conducted in our Coq model.
Type soundness relies on several \emph{well-formedness} given in \cite{li22checkedc} and \Cref{sec:meta}.
The progress theorem below states that a \lang program can always make a move.

\begin{thm}[Progress]\label{thm:progress}

For any \lang program $e$, heap $\heap$, stack
$\varphi$, type environment $\Gamma$, and variable predicate set $\Theta$
that are all are well-formed, consistent
($\Gamma;\Theta\vdash \varphi$ and $\heap \vdash \varphi$) and well
typed ($\Gamma;\Theta\vdash_{\cmode} e : \tau$ for some $\tau$),
one of the following holds:

\begin{itemize}

\item $e$ is a value ($\evalue{n}{\tau}$).

\item there exists $\varphi'$ $\heap'$ $r$, such that $(\varphi,\heap,e) \longrightarrow_m (\varphi',\heap',r)$.

\end{itemize}
\end{thm}
%{\em Proof:} By induction on the typing derivation.

\noindent
There are two forms of preservation regarding the $\cmode$ and $\umode$ regions.
Checked Preservation states that a reduction step preserves both the
type and consistency of the program being reduced, while
unchecked Preservation states that any evaluation happens at $\umode$ region does not affect the $\cmode$ mode heap.

\begin{thm}[Checked Preservation]
For any \lang program $e$, heap $\heap$, stack
$\varphi$, type environment $\Gamma$, and variable predicate set $\Theta$
that are all are well-formed, consistent
($\Gamma;\Theta\vdash \varphi$ and $\heap \vdash \varphi$) and well
typed ($\Gamma;\Theta\vdash_{\cmode} e : \tau$ for some $\tau$), if there exists $\varphi'$,
$\heap'$ and $e'$, such that $(\varphi,\heap,e)
\longrightarrow_{\cmode} (\varphi',\heap',e')$, then $\heap'$ is
$\cmode$ region consistent with $\heap$ ($\heap \triangleright \heap'$) and there exists
$\Gamma'$ and $\tau'$ that are well formed, $\cmode$ region consistent
($\Gamma';\Theta\vdash \varphi'$ and $\heap' \vdash \varphi'$) and
well typed ($\Gamma';\Theta \vdash_{\cmode} e: \tau'$), where
$\tau'\sqsubseteq_{\Theta} \tau$.
\end{thm}
%{\em Proof:} By induction on the typing derivation.
%\smallskip
\begin{thm}[Unchecked Preservation]
For any \lang program $e$, heap $\heap$, stack
$\varphi$, type environment $\Gamma$, and variable predicate set $\Theta$
that are all are well-formed and well
typed ($\Gamma;\Theta\vdash_{\cmode} e : \tau$ for some $\tau$), if there exists $\varphi'$,
$\heap'$ and $e'$, such that $(\varphi,\heap,e)
\longrightarrow_{\umode} (\varphi',\heap',e')$, then $\heap'(\cmode)=\heap(\cmode)$.
\end{thm}

Using the above theorems, we first show the non-exposure theorem,
where code in $\umode$ region cannot observe a valid checked ($\cmode$) pointer address.

\begin{thm}[Non-Exposure]
For any \lang program $e$, heap $\heap$, stack
$\varphi$, type environment $\Gamma$, and variable predicate set $\Theta$
that are all are well-formed and well
typed ($\Gamma;\Theta\vdash_{\cmode} e : \tau$ for some $\tau$), if there exists $\varphi'$,
$\heap'$ and $e'$, such that $(\varphi,\heap,e)
\longrightarrow_{\umode} (\varphi',\heap',e')$ and $e=E[\alpha(x)]$ and $\mode(E)=\umode$,
where $\alpha(x)$ is some expression (not $\echeckedtext$ nor $\euncheckedtext$) containing variable $x$; 
thus, it is not a checked pointer.
\end{thm}

We now state our main result, {\em non-crashing},
which suggests that a well-typed program can never be \emph{stuck} (expression
$e$ is a non-value that cannot take a step\footnote{Note that
  $\ebounds$ and $\enull$ are \emph{not} stuck expressions---they represent a
  program terminated by a failed run-time check. A program that tries to access $\heap{n}$
  but $\heap$ is undefined at $n$ will be stuck, and violates spatial
  safety.}).

% \review{- There appears to be a slight discrepancy between the blame theorem in Coq and the one in the paper: the paper mentions some e', which I believe should be r. Also, the Coq code has a further disjunct m=Unchecked in the conclusion.}
% \liyi{It is a typo. We will add the thing back that we show that either user uses a unchecked mode to evaluate $e$ or $e$ lives in a context that is an unchecked region. This is a bit due to the space limitation. The semantic rules allow users to input the mode $m$ of evaluating an expression, I just forgot to include the $m$ in the result of the proof statement. }

\begin{thm}[Non-Crashing]\label{thm:blame} For any \lang
  program $e$, heap $\heap$, stack
$\varphi$, type environment $\Gamma$, and variable predicate set $\Theta$
that are well-formed and consistent
($\Gamma;\Theta\vdash \varphi$ and $\heap \vdash \varphi$),
if $e$ is well-typed ($\varphi;\Theta\vdash_{\cmode} e :
\tau$ for some $\tau$) and there exists
$\varphi_i$, $\heap_i$, $e_i$, and $m_i$ for $i\in [1,k]$, such that
$(\varphi,\heap,e) \longrightarrow_{m_1} (\varphi_1,\heap_1,e_1)\longrightarrow_{m_2} ...\longrightarrow_{m_k} (\varphi_k,\heap_k,r)$, then $r$ can never be \emph{stuck}.
\end{thm}

%{\em Proof:} By induction on the number of steps of the \checkedc
%evaluation ($\longrightarrow_m^*$), using progress and preservation to
%maintain the invariance of the assumptions.

\ignore{
\subsection{Semantics}\label{sec:semantics}

% The semantics
% gives an independent account of spatial safety in \lang by
% checking pointer bounds based on the annotations carried on types at
% run-time.  While this account makes clear that bounds checking occurs
% as expected, it suggests an implementation that uses fat pointers to
% carry bounds.  We resolve this tension in the subsequent section on
% compilation and show that an implementation faithful to the semantics
% can be obtained without fat pointers.  
% \review{repeat that the stack is immutable at this point?}
% \liyi{Is it? Is the stack immutable? What does the immutable mean? 
%   In a stack, the variable values can be changed? Right?
%   The pointer address itself cannot be changed once it is created, but the stack variable content can be updated?  }
% \mwh{It certainly seems to be immutable: Your create stack frames
%   using let binding, and the let-bound variables will always be bound
%   to the same things. I.e., stack cells are immutable.}

% \review{this raises a fair amount of questions regarding the treatment of the
%   NULL pointer at this stage of the paper... is it modeled as 0, as returned by
%   `malloc`? are dynamic checks inserted by CheckedC to guarantee that no NULL
%   pointer is dereferenced?}
% \mwh{Yes, it is modeled as 0, and the semantics checks for
%   dereferences of 0. }

Here, we discuss the \lang operational semantics (\Cref{fig:semantics}); 
mainly focusing on the new changes on top of \checkedc in \cite{li22checkedc} with function pointers, modes, and function calls.
The other type and semantic rules about (NT)-arrays are given in \cite{li22checkedc} and \Cref{sec:literal-pointer-typing}.

%The typing judgment has the form $\Gamma;\Theta\vdash_m e : \tau$,
%which states that in a type environment $\Gamma$ (mapping variables to
%their types) and a predicate environment $\Theta$ (mapping integer-typed
%variables to Boolean predicates), expression $e$ will have type $\tau$ if evaluated
%in context mode $m$. Key rules for this judgment are given in
%Fig.~\ref{fig:type-system-1},

The operational semantics for \lang is defined as a small-step
transition relation with the judgment $ (\varphi,\heap,e)
\longrightarrow_m (\varphi',\heap',r)$.
 Here, $\varphi$ is a
\emph{stack} mapping from variables to values $\evalue{n}{\tau}$ and
$\heap$ is a \emph{heap} that is partitioned into two parts ($\cmode$ and $\umode$ heaps), each of which
maps addresses (integer literals) to values $\evalue{n}{\tau}$.

\myparagraph{Pointers, Contexts, and Modes}
A $\cmode$ pointer is mapped to a heap location in the $\cmode$ heap, 
while a $\tmode$ and $\umode$ pointer represents a $\umode$ heap location.
We wrote $\heap(m,n)$ to retrieve the $n$-location heap value in the $m$ heap,
and $\heapup{m}{n}{\evalue{n'}{\tau}}$ 
to update location $n$ with the value $\evalue{n'}{\tau}$ in the $m$ heap.
It is worth noting that \systemname is not a fat-pointer system;
thus, in every heap update, the value type annotation remains the same through program executions.
\mz{Does a ``non-fat pointer'' system cause the type preservation of heap
  values?
  The story seems to be that you design the system in a certain way s.t. heap
  value types are unchanged, which allows you to erase the type.
  And, from the type erasure property, we know that we don't need fat pointers.
}
Additionally, for both stack and heap, 
we ensure $\fv(\tau)=\emptyset$ for all the value type annotations $\tau$.

While heap bindings can change, stack bindings are immutable---once
variable $x$ is bound to $\evalue{n}{\tau}$ in $\varphi$, that binding will not
be updated. 
We can model mutable stack variables as pointers into the
mutable heap.
As mentioned, value $\evalue{0}{\tau}$
represents a $\enull$ pointer when $\tau$ is a pointer type.
Correspondingly, $\heap(m,0)$ should always be undefined.
The relation steps to a \emph{result} $r$, which is \mzs{either} \mzr{one of} an
expression, a $\enull$ or $\ebounds$ failure, represent \mzr{an expression right
  after the reduction}, a null-pointer dereference or out-of-bounds access,
respectively.
Such failures are a \emph{good} outcome; stuck states
(non-value expressions that cannot transition to a result $r$)
characterizing undefined behavior.
The context mode $m$ (in $\longrightarrow_{m}$) indicates whether the
stepped redex within $e$ was in a $\cmode$ or $\umode$ region.

The rules for the main operational semantics
judgment \emph{evaluation} are given at the bottom of
Fig.~\ref{fig:c-context}.
The first rule takes an expression $e$, decomposes
it into an \emph{evaluation context} $E$ and a sub-expression $e'$
(such that replacing the hole $\Box$ in $E$ with $e'$ would yield
$e$), and then evaluates $e'$ according to the \emph{computation}
  relation $(\varphi,\heap,e') \longrightarrow (\varphi,\heap,e'')$,
whose rules are given in Fig.~\ref{fig:semantics}, discussed
shortly.
The $\mode$ function  at the bottom of Fig.~\ref{fig:c-context}
determines the context mode that the expression $e'$ locates based on the context $E$.
In \Cref{lst:humantaint}, the function call expression \code{read_msg} has $\umode$ mode since it is inside a tainted function.
The second rule describes the exception handling 
for possible crashing behaviors in unchecked regions.
A $\umode$ mode operation can non-deterministically crash
and the \systemname sandbox mechanism recovers
the program to a safe point ($\evalue{0}{\tau}$)
and continues with the existing program state.
Evaluation contexts $E$ define a standard left-to-right evaluation order. (We explain the
$\ret{x}{\mu}{e}$ syntax shortly.)
%There are other rules for describing the halts of evaluation to $\enull$ and $\ebounds$ states in \Cref{app:main}.

Fig.~\ref{fig:semantics} shows selected rules for the computation relation.
The rules for pointer related operations---\textsc{S-DefC},
\textsc{S-DefT}, \textsc{S-DefNull}, and \textsc{S-Cast}.
The type rule for deference operations is given as rule \rulelab{T-Def} in \Cref{fig:type-system-1}.
The first three define the semantics of deference and assignment operations.
Rule \textsc{S-DefNull} transitions attempted null-pointer
dereferences to $\enull$, whereas \textsc{S-DefC} dereferences a $\cmode$-mode
non-null (single) pointer.
When $\enull$ is returned by the
computation relation, the evaluation relation halts the entire
evaluation with $\enull$ (using a rule not shown in Fig.~\ref{fig:c-context}); it
does likewise when $\ebounds$ is returned (see \Cref{sec:rem-semantics}).
%\textsc{S-AssignArrC} assigns to an array as long as 0 (the point of
%dereference) is within the bounds designated by the pointer's annotation
%and strictly less than the upper bound. 
\textsc{S-DefT} is similar to \textsc{S-DefC} for tainted pointers.
Any dynamic heap access of a tainted pointer requires a \textit{verification}.
Performing such a verification equates to performing a literal type check for a pointer constant in \Cref{fig:const-type}.
We explain this shortly below for \emph{constant validity checks}.
For now, the verification step, e.g. $\emptyset;\heap ; \emptyset \vdash_{\umode}\evalue{n_a}{\tau}$ in \textsc{S-DefC},
refers to that the value $n_a$ is well-defined in $\heap(m,n_a)$ and has type $\tau$, if $\tau$ is a pointer.
Static casts of a literal $n\!:\!\tau'$ to a type $\tau$ are handled
by \textsc{S-Cast}. In a type-correct program, such casts are
confirmed safe by the type system no matter
if the target is a $\tmode$ or $\cmode$ pointer. To evaluate a cast, the rule
updates the type annotation on $n$. Before doing so, it must
``evaluate'' any variables that occur in $\tau$ according to their
bindings in $\varphi$. For example, if $\tau$ was
$\tarrayptr{0}{x+3}{\tint}{\cmode}$, then $\varphi(\tau)$ would
produce $\tarrayptr{0}{5}{\tint}{\cmode}$ if $\varphi(x) = 2$.
%The full formalism, including \kw{struct}
%and null-terminated bound widening pointer operations, is given in \Cref{app:main}.

%\footnote{This approach is that of the PLT Redex model of \lang; the Coq
%development uses a slightly simpler syntax to achieve the same
%effect.}
% \review{the special case raises questions, e.g. why is this syntax-driven and
%   not type-driven? }
% \liyi{This describes the semantic transition rules. We are using context evaluation framework to define the transition rules as the $E$ definition in Fig.3. like $\frac{x \Rightarrow y}{x+z \Rightarrow y + z}$, I don't know how type-driven can help us define translation rules.  }
% \mwh{Don't follow the above. I don't see this ``context transition
%   rule'' anywhere, and I'm not sure how it would fire, if we had it.}
% \liyi{The comment seems to confuse the meaning of the text about the if-then-else rules. Making the rule specific will help. }

\begin{DIFnomarkup}
 \begin{figure}[t]
 {\small

 \begin{mathpar}
   \inferrule
       {}
       {\Theta;\heap;\sigma \vdash_m n : \tint}

   \inferrule
       {}
       {\Theta;\heap;\sigma \vdash_m 0 : \tptr{\omega}{\xi}}

   \inferrule
       {(m = \cmode \Rightarrow \xi \neq \cmode) \\\\ (m=\umode \Rightarrow \xi = \umode)}
       {\Theta;\heap;\sigma \vdash_{\cmode} n : \tptr{\omega}{\tmode}}
  
   \inferrule
       {(\evalue{n}{\tptr{\omega}{\xi}})\in \sigma}
       {\Theta;\heap;\sigma \vdash_m n : \tptr{\omega}{\xi}}

   \inferrule
       {\tptr{\omega'}{\xi'} \sqsubseteq_{\Theta} \tptr{\omega}{\xi} 
            \\ \Theta;\heap;\sigma \vdash_m n : \tptr{\omega'}{\xi'}}
       {\Theta;\heap;\sigma \vdash_m n : \tptr{\omega}{\xi}}

   \inferrule
       { \xi \le m 
     \\\Xi(m,n)=\tau\;(\evalue{\overline{x'}}{\overline{\tau}})\;(\xi,e)
       \\  \overline{x} = \{x|(x:\tint) \in (\overline{x'}:\overline{\tau}) \}}
       {\Theta;\heap;\sigma \vdash_m n : \tptr{(\tfun{\overline{x}}{\overline{\tau}}{\tau})}{\xi}}
  
   \inferrule
       {\neg\funptr(\omega)\\ \xi \le m\\
        \forall i \in [0,\size(\omega)) \;.\;
            \Theta;\heap;(\sigma \cup \{(n:\tptr{\omega}{\xi})) \}\vdash_m \heap(m,n+i)}
       {\Theta;\heap;\sigma \vdash_m n : \tptr{\omega}{\xi}}
 \end{mathpar}
 }
{\footnotesize
\[
\begin{array}{l} 
\funptr(\tfun{\overline{x}}{\overline{\tau}}{\tau}) = \texttt{true}
\qquad
\funptr(\omega) = \texttt{false}\;\;{[\emph{owise}]}
\end{array}
\]
}
 \caption{Verification/Type Rules for Constants}
 \label{fig:const-type}
 \end{figure}
\end{DIFnomarkup}

\subsection{Compilation}\label{sec:compilation}

As we have shown in \Cref{fig:overview}, the \systemname compiler utilizes the sandbox mechanism \cite{rul2009towards} and the \checkedc compiler \cite{li22checkedc} to compile programs. Here, we introduce how \systemname compiles a program into these two components.

\begin{figure}[t!]
{\small
\hspace*{-0.5em}
\begin{tabular}{|c|c|c|c|}
\hline
& \cmode & \tmode & \umode \\
\hline
& \textsc{CBox} / \textsc{Core} & \textsc{CBox} / \textsc{Core} & \textsc{CBox} / \textsc{Core} \\
\hline
\cmode & $\estar{x}$ / $\getstar{\cmode}{x}$ 
 & $\texttt{sand\_get}(x)$ / $\getstar{\umode}{x}$ &  $\times$ \\
\hline
\umode & $\times$
 & $\estar{x}$ / $\getstar{\umode}{x}$ &  $\estar{x}$ / $\getstar{\umode}{x}$ \\
\hline
\end{tabular}

}
\caption{Compiled Targets for Dereference}
\label{fig:flagtable}
\end{figure}

In \systemname, context and pointer modes determine the particular heap/function store that a pointer points to,
i.e., $\cmode$ pointers point to checked regions, while $\tmode$ and $\umode$ pointers point to unchecked regions.
Unchecked regions are associated with a sandbox mechanism that permits exception handling of potential memory failures.
In the compiled LLVM code, pointer access operations have different syntaxes when the modes are different. 
\Cref{fig:flagtable} lists the different compiled syntaxes of a deference operation ($\estar{x}$) for the compiler implementation (\textsc{CBox}, stands for \systemname) and formalism (\textsc{Core}, stands for \lang). The columns represent different pointer modes and the rows represent context modes.
For example, when we have a $\tmode$-mode pointer in a $\cmode$-mode region, we compile a deference operation to the sandbox pointer access function ($\texttt{sand\_get}(x)$) accessing the data in the \systemname implementation. In \lang, we create a new deference data-structure on top of the existing $\estar{x}$ operation (in LLVM): $\getstar{m}{x}$. If the mode is $\cmode$, it accesses the checked heap/function store; otherwise, it accesses the unchecked one.

We now show how \lang deals with pointer modes, mode switching and function pointer compilations, 
with no loss of expressiveness
as the \checkedc contains the erase of annotations in \cite{li22checkedc} and \Cref{appx:comp1}.
For the compiler formalism, 
we present a compilation algorithm that converts from
\lang to \elang, an untyped language without metadata
annotations, which represents an intermediate layer we build on LLVM for simplifying compilation. 
In \elang, the syntax for deference, assignment, malloc, function calls are: $\getstar{m}{e}$, $\elassign{m}{e}{e}$, 
$\emalloc{m}{\omega}$, and $\elcall{m}{e}{\overline{e}}$.
The algorithm sheds
  light on how compilation can be implemented in the real Checked C
  compiler, while eschewing many vital details (\elang has many 
  differences with LLVM IR).

%This section shows how \systemname deals with 
%annotations can be safely erased: using static information a compiler
%can insert code to manage and check bounds metadata, with no loss of
%expressiveness. We present a compilation algorithm that converts from
%\lang to \elang, an untyped language without metadata
%annotations. The syntax and semantics \elang
  %closely mirrors that of \lang; it differs only in that literals lack
  %type annotations and its operational rules perform no
  %bounds and null checks, which are instead inserted during
  %compilation. Our compilation algorithm is evidence that \lang's
  %semantics, despite its apparent use of fat pointers, faithfully
  %represents Checked C's intended behavior. The algorithm also sheds
  %light on how compilation can be implemented in the real Checked C
  %compiler, while eschewing many important details (\elang has many 
  %differences with LLVM IR).

Compilation is defined by extending \lang's
typing judgment as follows:
\[\Gamma;\Theta;\rho \vdash_m e \gg \dot e:\tau\]
There is now a \elang output $\dot e$ and an input $\rho$, which maps
each (NT-)array pointer variable to its mode and
each variable \code{p} to a pair of \emph{shadow
  variables} that keep \code{p}'s up-to-date upper and lower bounds. 
These may differ from the bounds in \code{p}'s type due to bounds
widening.\footnote{Since lower bounds are never widened, the
  lower-bound shadow variable is unnecessary; we include it for uniformity.} 

% When $\Gamma$,$\Theta$ and $\rho$ are all empty, we write $e \gg \dot e$ rather than the
% complete judgment, implicitly assuming that $e$ is a well-typed and closed
% term.

We formalize rules for this judgment in PLT Redex~\cite{pltredex},
following and extending our Coq development for \lang. To give
confidence that compilation is correct, we use Redex's property-based
random testing support to show that compiled-to $\dot e $ simulates
$e$, for all $e$.

\myparagraph{Checked and Unchecked Blocks}
In the \systemname implementation,
$\euncheckedtext$ and $\echeckedtext$ blocks 
are compiled as context switching functions provided by the sandbox mechanism.
We compile $\eunchecked{\overline{x}}{e}$ to 
$\texttt{sandbox\_call}(\overline{x},e)$, where we call the sandbox 
to execute expression $e$ with the arguments $\overline{x}$.
$\echecked{\overline{x}}{e}$ is compiled to 
$\texttt{callback}(\overline{x},e)$, where we perform 
a \texttt{callback} to a checked block code $e$ inside a sandbox.
In \systemname, we adopt an aggressive execution scheme that
directly learns pointer addresses from compiled assembly to make the $\texttt{callback}$ happen.
In the formalism, we rely on the type system to 
guarantee the context switching without creating the extra function calls for simplicity.

%Fig.~\ref{fig:compilationexample} shows how an invocation of
%\code{strlen} on a null-terminated string is compiled into C
%code. Each dereference of a checked pointer requires a null check
%(See \textsc{S-DefNull} in Fig.~\ref{fig:semantics}), which the
%compiler makes explicit: Line~$3$ of the generated code has the null
%check on pointer \code{p} due to the \code{strlen},
%  and a similar check happens
%  at line~$8$ due to the pointer arithmetic on \code{p}.
%Dereferences also require bounds checks: line~$2$ checks \code{p} is
%in bounds before computing \code{strlen(p)}, while line~$10$ does
%likewise before computing \code{*(p+1)}.

\myparagraph{Function Pointers and Calls}
Function pointers are managed similarly to normal pointers,
but we insert checks to check if the pointer address is not null in 
the function store instead of heap, and whether or not the type is correctly represented, 
for both $\cmode$ and $\tmode$ mode pointers 
\footnote{$\cmode$-mode pointers are checked once in the beginning and $\tmode$-mode pointers are checked every time when use}.
For example, in compiling the \code{read_msg} function in \Cref{lst:humantaint},
we place a check \code{verify_fun(read_msg, not_null(c, p_lo, p_hi) && type_match)},
The compilation of function calls (compiling to $\elcall{m}{e}{\overline{e}}$) 
is similar to the manipulation of pointer access operations in \Cref{fig:flagtable}.
The other compilation rules are given in \Cref{appx:add-type-sem}.
}

%% file: implementation.tex
\section{Implementation} 
\label{sec:implementation}
As mentioned in~\Cref{sec:overview}, our implementation of~\systemname{} has two main components,~\ie source level program partitioner (\sourcerewriter) and compiler.
Given a well-typed~\systemname{} program and a sandbox configuration,~\sourcerewriter performs source-to-source transformation and splits the program into two parts (\ie two sets of source files) -- checked and tainted source files corresponding to~$\cmode$ and~\ucregion respectively.
Our compiler translates the checked source files and adds the necessary dynamic instruction.
The tainted source files need to be compiled with the target sandbox compiler.
Finally, these two sets of object files will be linked along with any sandbox library to produce the final executable that enforces our security guarantees.

We created the sandbox library once and for all for each sandbox.
This library abstracts the sandbox-specific details and exposes a uniform interface (header file) to be used in~\cregion.
For instance,~\code{_SBX_()} gets an opaque pointer to the target sandbox.
We will present our implementation using WebAssembly (WASM~\cite{bosamiya2020webassembly}) as our target sandbox.
We also formalize the implementation and show a simulation theorem in \Cref{appx:comp1}.

\subsection{\sourcerewriter} \label{subsec:checkmate}
The~\sourcerewriter is primarily implemented in C++ as a~\clang frontend tool (3K SLoc).
However, we use a small OCAML program (680 SLoc) to remove annotations and make the code compilable with the sandbox compiler (\sect{subsubsec:genucregion}).

\subsubsection{Additional Function Qualifiers}
\label{subsubsec:addfuncqual}
In addition to the~\code{_Tainted} qualifier that marks functions to be in~\ucregion, we provide a few other qualifiers that enable developers to provide additional information and ease the partitioning process.
Specifically, we provide three additional qualifiers:~\code{_Callback},~\code{_Mirror}, and~\code{_TLIB}.

\noindent\emph{\_Callback}:
Developers should use this qualifier to mark callback functions,~\ie functions in~\cregion, that can be called from the tainted region. 
The~\sourcerewriter inserts appropriate sandbox dependent mechanisms to enable this (\ref{subsub:gencregion}).

\noindent\emph{\_Mirror}: 
This qualifier permits copying the corresponding function into both ~\cregion and~\ucregion, which permits the handling of certain simple utility functions that are called from both regions. For example, ~\code{append_string} in our evaluation of parsons\_wasm has callers from both the regions. 
\begin{minted}[mathescape, escapeinside=||, fontsize=\footnotesize]{c}
_Mirror int append_string(_TPtr<char> buf,
const char* appendStr : itype(_Nt_array_ptr<const char>),
_TPtr<char> buf_start, size_t buf_len) {
/* Qualifier Rules:
1.) No access to global data NOT marked "const"
2.) Callees must be _Tainted or _Mirror
*/
...
}

\end{minted}
Qualifying ~\code{append_string} with~\code{_Mirror} duplicates the function in both regions, allowing calls to \code{append_string} with parameter to \code{appendStr} as an unchecked or checked pointer within $\umode$ and $\cmode$ regions, respectively. Consequently complexity from over-tainting is avoided as ~\code {appendStr} need not be tainted in ~\cregion and neither are callbacks required to access ~\code{append_string} from ~\ucregion.   
"\_Mirror" enforces control-flow and data-flow compile-time semantic rules to ensure all variable and function call dependencies of mirrored functions required for ~\ucregion's compilation are resolved. 

\noindent\emph{\_TLIB}:
This qualifier relaxes type-checking rules on library functions, allowing developers to use the function freely in ~\cregion.
\begin{minted}[mathescape, escapeinside=||, fontsize=\footnotesize]{c}
// First, manually check the memory is in tainted region.
// if yes, then call strncpy.
if (!is_mem_in_range(t_str, t_str + n, SBX_LOW(), SBX_HIGH()))
  handle_violation();
// our type checker ignores this because 
// the _TLIB annotation below.
strncat(dst, t_str, n);
\end{minted}

\begin{minted}[mathescape, escapeinside=||, fontsize=\footnotesize]{diff}
- extern char *strncat (char *__restrict __dest,
+ _TLIB extern char *strncat (char *__restrict __dest,
const char *__restrict __src, size_t __n); // In the header file
}
\end{minted}

Passing tainted pointer \code{t_str} to unqualified ~\code{strncat} above is disallowed without having additional ~\ucregion implementation for \code{strncat}. If a user ascertains that \code{t_str} has the right buffer size for \code{strncat}, she might label \code{strncat} with \code{\_TLIB}, so that \code{t_str} can be treated as an checked pointer parameter; such annotation relaxes type-checking for all the arguments to its calls. It is worth noting that \systemname does not enforce any semantics to ensure ~\code{\_TLIB} functions implemented in ~\cregion are non memory-modifying; therefore, using \code{\_TLIB} requires users' awareness of memory address leaks.

\subsubsection{Generating~\cregion Source Partition}
\label{subsub:gencregion}
We copy all non-tainted functions into~\cregion source files and make the following modifications to enable interaction with~\ucregion (\ie sandboxed code).
We created a library once and for all for each sandbox.
This library abstracts the sandbox-specific details and exposes a uniform interface (header file) to be used in~\cregion.
For instance,~\code{_SBX_()} gets an opaque pointer to the target sandbox.

\iffalse
\noindent\emph{Handling Tainted Buffers in~\cregion}:
All buffers that are marked as tainted in~\cregion should be manually allocated in the sandbox. For instance,~\code{buff} in function~\code{handle_request} of~\lst{lst:final}.
We perform source rewriting and make the buffer into a pointer variable and allocate the corresponding number of bytes using the sandbox-specific allocator. For instance, we perform the following rewriting for~\code{buff} in~\lst{lst:final} with WASM sandbox.
\begin{minted}[mathescape, escapeinside=||, fontsize=\footnotesize]{diff}
- char buff[MAX_MSG_SIZE] __Tainted;
+ _TPtr<char> buff = (_TPtr<char>)t_malloc(MAX_MSG_SIZE);  
\end{minted}
Note that we also add the necessary deallocator calls at function exit points.
\fi

\noindent\emph{Handling Calls to Tainted Functions}:
In~\cregion, we also need to modify calls to tainted functions as they execute inside the sandbox (separate address space) and thus cannot be invoked as regular functions.
However, modifying every call site of tainted functions is tedious and also requires precise pointer analysis~\cite{milanova2002precise} to handle indirect calls through function pointers.

We handle this by~\emph{indirection}: Instead of modifying the call sites, we modify the body of tainted functions to invoke the corresponding function in the sandbox.
For instance, we modify the body of tainted function~\code{process_req1} (\lst{lst:final}) in~\cregion as below:
\begin{minted}[mathescape, escapeinside=||, fontsize=\footnotesize]{diff}
int process_req1(char *msg, size_t m_l) _Tainted {
- int rc = -1, i;
- if (m_l > MIN_SIZE) {
- ...
+ return w2c_process_req1(msg, m_l);
}
\end{minted}
This ensures that all calls (even through function pointers) to the tainted function~\code{process_req1} are redirected to the sandbox.

\noindent\emph{Handling~\texttt{\_Callback} Qualifiers}:
As mentioned in~\sect{subsubsec:addfuncqual}, functions with these qualifiers can be called from~\ucregion.
Consider the following~\code{StringAuth} function that checks whether the provided user input ~\code{usertoken} is authenticated by accessing checked data. Since this needs to be invoked from ~\ucregion it is annotated as a ~\code{\_Callback}.
\begin{minted}[mathescape, escapeinside=||, fontsize=\footnotesize]{c}
_Callback _TPtr<char> StringAuth(
               _T_Array_Ptr<const char> usertoken : count(len),
               size_t len) {
...
// Checks whether usertoken is authenticated
/*
 These functions will be restricted to only accept
 tainted parameters.
*/
...
}
\end{minted}
These callback functions are only allowed to use tainted parameters as they will be called from a tainted region.

For each such function, we create a corresponding trampoline function that serves as the entry point for the callback function, as shown below:
\begin{minted}[mathescape, escapeinside=||, fontsize=\footnotesize]{diff}
+ unsigned int _T_StringAuth(void* sandbox,
+               unsigned int arg_1,
+               unsigned long int arg_2) {
+   // Perform necessary Type-conversion of arguments.
+   // uname <- conver arg_1
+   // len <- arg_2
+   ret = StringAuth(uname, len);
+   // ret_val <- ret
+   return ret_val;
+ }
\end{minted}

The trampoline function handles the invocations from sandbox (and hence the extra parameter~\code{sandbox}), performs necessary pointer argument conversion, and eventually invokes the callback.

We also add the code to register this trampoline function with the sandbox. The registration function for WASM sandbox is as shown below:
\begin{minted}[mathescape, escapeinside=||, fontsize=\tiny]{diff}
+ void registerCallback_StringAuth(void){
+ //callback function signature {ret <- int, arg_1 <- int, arg_2 <- long}
+ int ret_param_types[] = {0, 0, 1};
+ // 2 <- arg count, 1 <- ret count
+ __StringAuth__C_ = _SBXREG_((void*)_T_StringAuth,2,1, ret_param_types);
+ }
\end{minted}
This registration function creates an opaque handle for the trampoline function and enables~\ucregion to call the trampoline using the corresponding handle.

Lastly, we change the tainted function's body to include an indirect call to the sandbox's implementation of the tainted function. However, instead of passing the callback function pointer directly from the argument list, we pass the generated trampoline handle ~\code{__StringAuth__C_} as shown below:
\begin{minted}[mathescape, escapeinside=||, fontsize=\tiny]{diff}
_Tainted _TPtr<char> StringProc(_TPtr<_TPtr<const char>> user_input,
_TPtr<_TPtr<char>(_TPtr<const char> input, size_t len)>StringAuth) {
-  ...
-  //complex Function Body       
- return StringAuth(one_past_start, string_len);
+ return w2c_StringProc(_SBX_(), (unsigned int)string, __StringAuth__C_);

}
\end{minted}

\subsubsection{Generating~\ucregion Source Partition}
\label{subsubsec:genucregion}
We create~\ucregion partition by first copying all functions and variables marked as~\code{_Tained} and~\code{_Mirror}.
Next, we convert the partition into regular C by replacing all tainted types with the corresponding C types.
\newline
For instance:~\code{_T_Array_ptr<char>} will be replaced with~\code{char *}.

The~\code{read_msg} function signature in \lst{lst:final} will be modified in the ~\ucregion as shown below:
\begin{minted}[mathescape, escapeinside=||, fontsize=\tiny]{diff}
size_t read_msg(int sock_fd, char* msg,
-                             size_t sz) _Tainted {
+                             size_t sz) {
\end{minted}

% ** We Do NOT modify callback function in checked sources **
% Finally, we convert all calls to~\cregion callbacks with sandbox-specific invocations. For instance, for the~\code{StringAuth} call back (\sect{subsub:gencregion}), we modify~\code{...UNCHECKED FUNC..} as shown below:
% \begin{minted}[mathescape, escapeinside=||, fontsize=\tiny]{c}
% SAY HOW WE MODIFY UNCHECKED CODE
% TO INVOKE CALLBACK FUNCTIONS.
% \end{minted}
% \aravind{Arun, please fix the above example}

\subsection{Compiler}
\label{subsec:compilerimple}
The~\systemname compiler includes the type checker (\sect{sec:typechecking}) and necessary instrumentation to handle tainted pointers (\ie from sandbox) and perform runtime checks.
We implemented the compiler as a library (1.8K SLoc) on top of the existing~\checkedc compiler.
Our type checker verifies the constraints (\fig{fig:type-system-1}) and displays violations as errors, enabling developers to make additional annotations.
For instance, for the code in~\lst{lst:final}, our type checker emits an error saying that ``~\emph{buffer~\code{buff} is tainted, but it is allocated in~\cregion}''.
Consequently, the developer has to make the following changes to allocate~\code{buff} in the sandbox:
\begin{minted}[mathescape, escapeinside=||, fontsize=\tiny]{diff}
- char buff[MAX_MSG_SIZE] _Tainted;
+ _T_Array_ptr<char> buff: count(MAX_MSG_SIZE)
+    = sbx_allocate(MAX_MSG_SIZE);
...
+ sbx_free(buff)
return rc;
\end{minted}

For all tainted pointer access, our instrumentation adds a check (executed at runtime) to see we check if the address is within the valid range for the sandbox. An example of the check is shown below:
\begin{minted}[mathescape, escapeinside=||, fontsize=\tiny]{diff}
_TPtr<int> t = ...
+ if (!is_in_range(t, SBX_LOW(), SBX_HIGH()))
+  handle_violation();
*t = 1;
\end{minted}
We show the added check in C for clarify. However, our instrumentation works on LLVM IR.

The~\ucregion source files will be compiled with sandbox compiler,~\ie WASI-SDK (~\cite{wasi-clang}).
Given that WASM sandbox uses 32-bit addresses (v/s 64-bit for~
\cregion), our instrumentation also adds necessary pointer swizzling to have transparent access to composite data types,~\ie~\code{struct}'s with pointers. 

\subsection{Balance Assurance and Overhead}
\label{subsec:otherusecases}
\input{figures/newusecase}

Tainting all functions that are not checked may lead to a large number of transitions to the sandbox, which could add unnecessary overhead. In our example, in~\lst{lst:final}, marking~\code{read_msg} function as tainted would cause the system to transition to and from the sandbox every time a message is read, which could be inefficient. 

However, as mentioned in~\sect{subsec:checkedc}, checked pointers can be used alongside regular pointers, so the~\cregion{}~\emph{can} contain regular (unchecked) pointers as well.
This allows for flexibility in how the program is partitioned, such as only tainting a few high-risk unchecked functions rather than all of them. 
For our example in~\lst{lst:final}, we can make only the function \code{process_req1} as tainted, thereby transitioning into a sandbox only when the corresponding (high-risk) request is received.

In this way, the system can be used as a generic program partitioning technique that balances the trade-off between overhead and security assurance.
The general guidelines for this alternative use case are shown in~\fig{fig:mixedmodesupport}.
It is best to annotate all functions with checked types for security purposes. However, if it is difficult to convert all functions, it may be more practical to only mark certain high-risk functions with a high probability of vulnerabilities as tainted,
which maintains the assurance and overhead balance.
We add the additional constraint of not allowing casting from taint types to generic types to enable this generic use case without violating the safety guarantees of~\systemname.
Specifically, we disallow the following casting in~\cregion:
\begin{minted}[mathescape, escapeinside=||, fontsize=\tiny]{diff}
//_t_Ptr<int> y; int *z;
|\faRemove| z = (int *)y; // Not allowed.
\end{minted}

%% file: figures/newusecase.tex
\begin{figure}
\begin{center}
\begin{tikzpicture}[scale=0.75]
%\node[text width=3cm] at (1.5,-0.5) 
%    {some text spanning three};
%\draw[step=5cm,gray,very thin] (0,0) grid (5,5);
\fill[taintcolor] (2.5,2.5) rectangle (5,5);
\fill[green!10!white] (0,2.5) rectangle (2.5,5);
\fill[green!40!white] (2.5,0) rectangle (5,2.5);
\fill[green!40!white] (0,0) rectangle (2.5,2.5);

\Text[x=3.8cm,y=4.5cm,fontsize=\scriptsize]{\bf High Risk and}
\Text[x=3.8cm,y=4.2cm,fontsize=\scriptsize]{\bf Hard to Convert}
\Text[x=3.8cm,y=3.9cm,fontsize=\scriptsize]{\bf Functions.}
\Text[x=3.8cm,y=3.6cm,fontsize=\tiny]{(Isolated in}
\Text[x=3.8cm,y=3.3cm,fontsize=\tiny]{the Sandbox)}

\Text[x=1.2cm,y=4.5cm,fontsize=\scriptsize]{\bf Low Risk and}
\Text[x=1.2cm,y=4.2cm,fontsize=\scriptsize]{\bf Hard to Convert.}
\Text[x=1.2cm,y=3.9cm,fontsize=\scriptsize]{\bf Functions.}
\Text[x=1.2cm,y=3.6cm,fontsize=\tiny]{(Partially annotated}
\Text[x=1.3cm,y=3.3cm,fontsize=\tiny]{and execute in c-region)}
\Text[x=1.2cm,y=2.9cm,fontsize=\tiny]{}
%\Text[x=1.2cm,y=4.3cm,fontsize=\tiny]{(Will be partially converted)}

\Text[x=2.3cm,y=1.8cm,fontsize=\scriptsize]{\bf Easy to Convert Functions.}
\Text[x=2.3cm,y=1.5cm,fontsize=\tiny]{(Will be fully converted to Checked C}
\Text[x=2.3cm,y=1.2cm,fontsize=\tiny]{with all necessary annotations}
\Text[x=2.3cm,y=0.9cm,fontsize=\tiny]{ and execute in c-region)}

%\Text[x=3.8cm,y=2.1cm,fontsize=\tiny]{Retrofitting}
%\Text[x=3.8cm,y=1.8cm,fontsize=\tiny]{Techniques.}
%\Text[x=3.8cm,y=1.5cm,fontsize=\tiny]{(ASAN, CCured,}
%\Text[x=3.8cm,y=1.3cm,fontsize=\tiny]{etc)}

%\Text[x=0.9cm,y=0.6cm,fontsize=\tiny]{\bf \faFlagCheckered}

%\Text[x=2.0cm,y=1.4cm,fontsize=\tiny]{\bf \footnotesize iC3C}

%\draw [{Implies}-,double,line width=1pt] (1.0,0.8) -- (1.9,2.1);

\draw[-Stealth,thick] (0,0) -- (5.1,0) node[anchor=north west] {\bf \tiny high};
\Text[x=0cm,y=-0.2cm,fontsize=\tiny]{\bf least}
\Text[x=2.6cm,y=-0.2cm,fontsize=\footnotesize]{Functions' Risk Level}
\Text[x=-0.2,y=2.6,rotation=90,fontsize=\footnotesize]{Conversion effort}
\draw[-Stealth,thick] (0,0) -- (0,5.1) node[anchor=south east] {\bf \tiny high};
\end{tikzpicture}
\caption{Flexibile partioning using~\systemname{}.}
\label{fig:mixedmodesupport}
\end{center}
\end{figure}

%% file: evaluation.tex
\section{Evaluation}\label{sec:evaluation}

% \review{While I found the idea behind section V very interesting, the current version
%   of this section lacks some details that would help in better understanding (1) 
%   how the approach works, and (2) the overall scope of the approach. 
%   $\\$
%   For instance, the authors state that, following [19], they try to "exercise
%   interesting patterns" by adding "admissible but redundant typing rules" like
%   G-ASTR. There are a few points that are unclear here: (1) are these rules
%   discovered manually or automatically (starting from the Redex semantics)?, (2)
%   are there any guiding principles for coming up with rules that lead to
%   interesting cases?
%   $\\$
%   Later, the authors refer to "generation rules modified to be slightly more
%   permissive" to generate "a little" ill-typed terms. Again, are these rules
%   obtained automatically or defined manually? If the latter, did you follow any
%   methodology to derive such rules? Are these rules the same as the "admissible
%   but redundant typing rules" from above?}
% \liyi{Deena? Leo? }
We evaluate~\systemname in terms of the following three aspects:
\begin{itemize}
\item\textbf{Conversion Effort:} How much developer effort is needed to annotate parts of programs to run in~\systemname?
\item\textbf{Performance Overhead:} What is the performance overhead (both runtime and memory) in using~\systemname?
\item\textbf{Security Impact:} How effective is the isolation provided by~\systemname{} in preventing security vulnerabilities?
\end{itemize}

%Our evaluation of \systemname consists of a set of tests that can be classified into Micro-benchmarks and Program Benchmarks that evaluate \systemname on WebAssembly sandbox. Consequently, all further references to sandbox refer to unchecked code in the WebAssembly Sandbox. 
%MicroBenchmarking involves evaluating performance on fundamental operations involving tainted pointers, context switching between checked and sandboxed regions, and sandboxed execution of functions.
%We further go on to evaluate \systemname on six real-world programs pertaining to diversified domains to evaluate real-world run-time and memory performance.

\subsection{Dataset}
Network-facing programs such as servers directly interact with external input, often process complex data types, and are more susceptible to security issues.
We primarily focus on network servers as they can benefit most from our partitioning approach.
We use WebAssembly (WASM) as our target sandbox. Consequently, we also want any of the selected programs to be compilable with the WASM sandbox.
We selected network servers that we could (with minimal effort) compile with the WASM compiler.
We also selected a few standalone programs suggested by the ~\checkedc team~\cite{benchmarkcc}, which are good candidates to evaluate modifications to~\checkedc.
\tbl{table:dataset} shows the program selected as part of our evaluation dataset.

\subsection{Experimental Setup}
\label{experimentalsetup}
All experiments are performed on a 6-Core Intel i7-10700H machine with 40 GB of RAM, running Ubuntu 20.04.3 LTS.
We use WASM as our target sandbox and use a similar configuration as that of the recent work~\cite{rlbox-paper};
and Valgrind's "massif" memory profiler~\cite{seward2008valgrind} to measure the memory usage and consider the peak heap usage of an application as its memory consumption.
We measure runtime using the difference in elapsed clock cycles using~\code{clock()} API from POSIX's~\code{<time.h>} and linux's ~\code{time} command, and perform every measurement ten times and use the average as the final result.

\subsection{Conversion Effort}
The flexibility of~\systemname{} (\sect{subsec:otherusecases}) enables an application to be partitioned in several ways -- with varying levels of overhead and assurance. We explore the following three ways:

\noindent\emph{Checked C and Tainted Partitions ($CTP$):} This is the most comprehensive use case, where the~\cregion partition contains completely annotated Checked C code, and~\ucregion contains one or more tainted functions.
This provides the complete spatial safety of the code in~\cregion{} including isolation from~\ucregion{}.

\noindent\emph{Only Tainted Partition ($TP$):}
This is the general partitioning (\sect{subsec:otherusecases}) use case without Checked C.
This is similar to $CTP$, but~\cregion code does not use any Checked C annotations.
This provides only the isolation guarantee without spatial safety.

\noindent\emph{Only Tainted Pointers ($T_{Pr}$):}
In this use case, we only use tainted pointers, and all code lies in~\cregion.
This is needed for data isolation-only cases, where the developer might want to just isolate certain data (\eg user input) from the rest of the program.
As explained in~\sect{subsec:compilerimple}, all access through tainted pointers will be dynamically checked to ensure that every access is within the sandbox.
This provides partial spatial safety by ensuring that any spatial violation involving tainted pointer cannot affect~\cregion.

\input{tables/datasettable}
\subsubsection{Conversion Methodology}
We partitioned each program in our dataset using one of the above three methods.

Our goal is to isolate risky input processing routines from the application code.
We manually analyze each application's source code to identify which functions handle the input data and how.
We also look into previously reported vulnerabilities to identify risky input processing routines.
We pick one of the above methods to partition based on how the input data is processed.
Table ~\ref{table:conversioneffort} is a summary of our dataset partitioning.
For~\texttt{ProFTPD}, $T_{Pr}$ method is used and the input data is marked as tainted.
Consequently, five other pointers need to be marked as tainted according to the type rules. This results in a total of 6 pointer annotations. There is no code in~\ucregion as used $T_{Pr}$ method with only tainted pointers being used.
We follow the same approach for~\texttt{LibPNG} (\texttt{png2pnm} and~\texttt{pnm2png}). However, in this case, we have to annotate much more pointers (248) due to the complicated \texttt{libPNG}'s internal structures.

For~\texttt{MicroHTTPD} and~\texttt{UFTPD}, $TP$ method is used and we mark all of the direct input handled methods as tainted, which are consequently moved to the sandbox, with annotating several intermediate pointers as tainted.
For~\texttt{TinyBignum} and~\texttt{Parsons},
we follow $CTP$ and mark all input processing routines as tainted and place them in the sandbox.
The rest of the non-sandboxed code is annotated completely using~\checkedc types and placed in~\cregion.

We ensured that the partitioned programs retained their expected functionality by verifying using corresponding test suites.

\subsubsection{Conversion Effort}
The second last column shows the hour numbers for partitioning applications.
On average, it takes $\sim$ 3.5 hours for each partitioning. However, the exact time depends on the complexity of the application and the pointer usage.
Although the absolute time is high, partitioning is a one-time effort for each application.
We start by annotating functions and then iteratively fixing type-checker errors.
Most of the time (80\%) is spent on running the type-checker.
The type-checker stops at the first error without giving information about other variables that need to be fixed. 
For an instance, in the following code:
\begin{minted}[mathescape, escapeinside=||, fontsize=\footnotesize]{c}
_TPtr<int> y = ...; 
int *z;
int *x;
|\textcolor{red}{\faRemove}| z = y;
x = z;
\end{minted}
The type-checker displays an error only for the first assignment.
However, to correctly fix it, we need to annotate both~\code{x} and~\code{y}.
If $N$ pointers need to be annotated, then in the worst case, we might have to run the type-checker $N$ times, annotating one additional pointer in every run.
We plan to fix this in our future work by making the conversion procedure automatic.
%make the type-checker display all the affected pointer variables instead of just the type errors.

\subsection{Performance Overhead}
Recent work~\cite{jangda2019not} shows that code executed as part of WASM sandbox incurs significant runtime overhead,~\ie$\sim$200\%.
To better understand our runtime overhead, we first perform micro benchmarking of additional sandbox-related operations in~\systemname{}.

\subsubsection{Micro-Benchmarking}
\Cref{fig:microbenchmarks} shows our micro-benchmarking result.
We measure the following operations as part of this:

\noindent\emph{Memory access in WASM Sandbox ($SBX_{m}$):} 
All memory accesses in a sandbox need additional verification by the sandbox runtime, which results in runtime overhead.
We perform 100K memory accesses (read and write) in a loop, measure the time inside the sandbox, and compare it with the time executed as a regular program.
The results (~\fig{fig:microbenchmarks}) show that we incur 156.6\% overhead for memory accesses in the WASM sandbox compared to that in a normal program.
This is in line with the observations of the recent work~\cite{jangda2019not}.

\noindent\emph{Sandbox Roundtrip ($SBX_{RT}$):}
We measure the time to make a round trip between~\cregion and sandbox (\ucregion) compared to a regular function call and return.
We create a no-op function below:
\begin{minted}[mathescape, escapeinside=||, fontsize=\footnotesize]{c}
void noop() { return; }
\end{minted}
We place this~\code{noop} function in the sandbox and measure the time to call and return from it:
\begin{minted}[mathescape, escapeinside=||, fontsize=\footnotesize]{c}
s = clock(); sandbox_noop(); e = clock();
\end{minted}
We compare the time with a regular call when \code{noop} is in~\cregion.

As shown in ~\fig{fig:microbenchmarks}, we incur an overhead of $\sim 400\%$. This is also inline with the performance reported by prior works~\cite{jangda2019not, rlbox-paper}.
This is expected because transitions to/from sandbox require context switches which are more expensive than regular function calls (\ie\code{call} and~\code{ret} instructions).

\noindent\emph{Tainted Pointer Access in~\cregion ($TP_{c}$):}
As explained in~\sect{subsec:compilerimple}, we need to perform pointer swizzling to convert the sandbox-specific representation of tainted pointers to raw addresses.
In addition, our instrumentation also checks at runtime that all tainted pointers are within the sandbox address range before accessing them.
We measure this additional overhead by comparing tainted pointer accesses with regular pointer accesses.

As shown in~\fig{fig:microbenchmarks}, we incur 34\% overhead in accessing tainted pointers in~\cregion,
 due to additional validation checks, which require executing two compare instructions for every memory access.

\subsubsection{Overhead on Dataset}
\label{subsec:programoverhead}
\input{figures/runtimechart}

The first set of bars in~\Cref{fig:runtimeoverhead} shows the overhead of the partitioned programs.
%It is interesting to see a lot of variance in the runtime overhead.
The runtime overhead is proportional to the execution time 
spent in the sandbox and the number of transitions between~\cregion and the sandbox,
which coincides with the sandbox execution overhead observation in the micro-benchmarks experiments (\Cref{fig:microbenchmarks}).
\Cref{table:conversioneffort} shows the numbers of lines in the sandbox for our partitioned programs.
For~\texttt{ProFTPD}, we used only tainted pointers without code in the sandbox and transitions to/from the sandbox.
Consequently, the overhead is less than 4.2\%.
For~\texttt{UFTPD} and~\texttt{MicroHTTPD}, we sandboxed only request handlers (relatively small)
that parse messages and return the actions that need to be performed.
The server only invokes these handlers on particular requests resulting in less transitions with the sandbox.
As expected, the overhead is also less in these cases.
For~\texttt{TinyBigNum} and~\texttt{Parsons}, the overhead is high because of the relatively large amount of code in sandbox region.
In both applications, we place the frequently used parsing functions in the sandbox 
resulting in a lot of sandbox transitions with most of them in a loop.
The case is slightly different in~\texttt{pnm2png} and~\texttt{png2pnm},
where we made the entire~\code{png} structure tainted, 
which resulted in dynamic checks every time when the~\code{png} struct is accessed.
In summary, our results indicate that the runtime overhead largely depends on the sandbox.

\noindent\textbf{Overhead of only~\systemname{}.}
To verify the impact of the sandbox on the overall program runtime, we perform a NO-OP sandbox experiment.
Our goal is to measure the runtime overhead introduced ONLY by~\systemname{}.
We perform this experiment by skipping sandboxing.
Specifically, we run the completely annotated program as a regular application without any partitioning.
However, we compile the annotated program with~\systemname compiler, which will add the relevant runtime checks (\sect{subsec:compilerimple}).
We modify the instrumentation on~\code{taint}ed pointers to check for a valid pointer (instead of within sandbox bounds) -- this will add the same amount of checks as in sandboxing case, but the comparison values will be different.

On evaluating ~\systemname{} with NO-OP sandbox on our entire dataset, we observe significantly less overhead as compared to that of the WASM sandbox as shown in figure ~\ref{fig:runtimeoverhead}. Therefore, ~\systemname{} by itself contributes significantly less to the overhead as compared to the sandbox it uses. 
For~\texttt{TinyBigNum}, the overhead is higher at 54.7\%. Our analysis shows that this overhead is because we taint the main input buffer, which is processed in loops.
This leads to additional checking for every loop iteration resulting in higher overhead.
Another reason is that in the current implementation, our instrumentation is performed at the end after all optimization passes; thus, none of the instrumentation is optimized.
We plan to move our instrumentation before the optimization passes and exploit them to optimize further and decrease the runtime overhead.

\subsubsection{Memory Overhead}
All programs have a constant memory overhead ($\sim$81 KB) mainly for sandbox and a few variables related to creating sandbox and other helper functions.
However, similar to the original~\checkedc,~\systemname{} itself does not add any memory overhead,
because the compilation of \code{tainted} pointers do not come with any metadata.
%This is expected because, similar to~\checkedc types, ~\code{taint}ed pointers also do not have metadata and have the same runtime representation as regular pointers.

%ProFTPD's run-time overhead is expected as our conversion was small and strictly encapsulates CVE-2010-4221.
%Since UFTPD's test-suite lacks coverage for some of the \systemname changes, we manually write a script for 3 Tests that each trigger "quote CWD", "quote PORT", and FTP "get file" on a 4.0K file, following which, we record overhead as described in ~\ref{experimentalsetup}.
%Run-time overhead for UFTPD is expected as the sandboxed code was less performance intensive and FTP-protocol by itself overshadows \systemname's overhead. 

%Our evaluation of \systemname's performance results on the dataset yields the observations:
%\begin{itemize}
%  \item Run-time Overhead is proportional to the extent of 
%annotated pointers and sandboxed code.
%  \item Marshalling can be avoided if Taintedness of a pointer is propagated across its data-flow.
%  \item Marshalling might sometimes be relatively cost-effective instead of propagating the taintedness of a pointer throughout its data-flow. 
%\end{itemize}

\subsection{Security Impact}
\label{subsec:securityimpact}
The last column of Table ~\ref{table:conversioneffort} shows the list of all spatial safety vulnerabilities in the functions that have tainted types or are isolated in the~\ucregion{}.
We re-introduced these bugs in the annotated program and checked whether these bugs could be triggered by the corresponding crashing input or exploit (if available).
We also manually verified whether the bug can be triggered or prevented by~\systemname{}.
As expected,~\emph{all vulnerabilities} are prevented by~\systemname{}.

The symbols~\vulprevented{} and~\vulisolated{} indicate whether the vulnerability was detected by our dynamic instrumentation or isolated in the sandbox, respectively.
This shows that~\systemname{} provides an effective mechanism to prevent spatial safety vulnerabilities.

%% file: tables/datasettable.tex
\begin{table}[]
{\scriptsize
\begin{tabular}{c|c|c|r}
\toprule
\textbf{ID} & \textbf{Program} & \textbf{Description}          & \multicolumn{1}{c}{\textbf{\begin{tabular}[c]{@{}c@{}}Size\\ (SLoc)\end{tabular}}} \\ 
\midrule
\rowcolor{black!15} 1           & ProFTPD          & High performance FTP Server   & 556 K                                                                               \\ 
2           & MicroHTTPD       & Simple HTTPD Server           & 122 K                                                                                \\ 
\rowcolor{black!15} 3           & UFTPD            & UDP based FTP Server          & 3 K                                                                               \\ 
4           & \multicolumn{1}{c|}{\begin{tabular}[c]{@{}c@{}}LibPNG\\ (png2pnm and pnm2png)\end{tabular}}           & Program to convert between png and pnm & 76 K                                                                               \\ 
\rowcolor{black!15}5           & TinyBigNum          & Multiple-precision integer implementation & 1.6 K                                                                               \\ 
6           & Parsons          & JSON parsing library & 3.1 K                                                                               \\ 

\bottomrule
\end{tabular}
}
\caption{Evaluation Dataset.}
\label{table:dataset}

{\tiny
\begin{tabular}{c|c|c|c|r|c}
\toprule
\textbf{Program} & \textbf{\begin{tabular}[c]{@{}c@{}}Partition\\ Methodology\end{tabular}} & \textbf{\begin{tabular}[c]{@{}c@{}}Pointers\\ Annotated\end{tabular}} & \textbf{\begin{tabular}[c]{@{}c@{}}Lines in\\ \ucregion \\ (Sandbox)\end{tabular}} & \multicolumn{1}{c|}{\textbf{\begin{tabular}[c]{@{}c@{}}Time\\ Taken\\ (hours)\end{tabular}}} & \textbf{\begin{tabular}[c]{@{}c@{}}CVEs\\ Prevented\end{tabular}} \\ 
\midrule
\rowcolor{black!15} ProFTPD          &     $TP$                                                                     &       6                                                                &                                                                     N/A &                                                                                      1        & CVE-2010-4221 (\vulprevented{})                                                           \\ %\hline
png2pnm          & \multirow{2}{*}{$TP$}                                                        & \multirow{2}{*}{248}                                                     & \multirow{2}{*}{N/A}                                                    & \multirow{2}{*}{}                                                                            & \multirow{2}{*}{CVE-2018-144550 (\vulprevented{})}                                             \\ %\cline{1-1}
pnm2png          &                                                                          &                                                                       &                                                                      &                                                     8                                         &                                                                   \\ %\hline
\rowcolor{black!15} MicroHTTPD       &     $TP$                                                                     &          139                                                             &                                                                     450 &                                                                                       3       &                    N/A                                               \\ %\hline
UFTPD             &        $T_{Pr}$                                                                  &  146                                                                     &                                                                     90 &                                                           3                                   & \begin{tabular}[c]{@{}c@{}}CVE-2020-14149 (\vulprevented{}) \\ CVE-2020-5204 (\vulisolated{})\end{tabular}             \\ %\hline
\rowcolor{black!15} TinyBigNum       &      $CTP$                                                                    &   69                                                                    &                                                                     30 &                                                                                        2      &                             N/A                                      \\ %\hline
Parsons          &     $CTP$                                                                     &      364                                                                 &                                                                     800 &                                                      5                                        &                                                  N/A                 \\ 
\bottomrule
\end{tabular}
}
\caption{Summary of Conversion Efforts \& Security Impact.}
\label{table:conversioneffort}
\end{table}

%% file: figures/runtimechart.tex
\begin{figure}
\pgfplotsset{every x tick label/.append style={font=\tiny, rotate=30}}
\pgfplotsset{every y tick label/.append style={font=\tiny}}
\begin{tikzpicture}  
  
\begin{axis}  
[  
    ybar,  
    %ymode=log,
    enlargelimits=0.03,
    enlarge x limits=0.12,
    ylabel={Runtime Overhead (\%)}, % the ylabel must precede a # symbol.  
     axis x line*=bottom,
     axis y line*=left,
    symbolic x coords={Memory Access in SBX, SBX Roundtrip, Tainted-ptr Access}, % these are the specification of coordinates on the x-axis.  
    xtick=data,  
    bar width=8pt,
    width=6cm,
    x tick label/.append style={font=\tiny, rotate=30},
    y tick label/.append style={font=\tiny},
    axis background/.style={fill=gray!10},
    x label style={at={(axis description cs:0.5,-0.1)},anchor=north, font=\footnotesize},
    xlabel={},
    ylabel style={font=\footnotesize},
     nodes near coords, % this command is used to mention the y-axis points on the top of the particular bar.  
    nodes near coords align={vertical},  
    ]  
\addplot coordinates {(Memory Access in SBX,156) (SBX Roundtrip, 398) (Tainted-ptr Access, 34.16)};  
  
\end{axis}  
\end{tikzpicture} 
\vspace*{-1.5em}
\caption{\systemname Micro-Benchmarks.}
\label{fig:microbenchmarks}

\pgfplotsset{every x tick label/.append style={font=\tiny, rotate=30}}
\pgfplotsset{every y tick label/.append style={font=\tiny}}
\begin{tikzpicture}  
  
\begin{axis}  
[  
    ybar = 0.8,  
    enlargelimits=0.01,  
    enlarge x limits=0.12,
    ylabel={Runtime Overhead (\%)}, % the ylabel must precede a # symbol.  
     axis x line*=bottom,
     axis y line*=left,
    symbolic x coords={ProFTPD, TinyBigNum, UFTPD, MicroHTTPD, png2pnm, pnm2png, Parsons}, % these are the specification of coordinates on the x-axis.  
    xtick=data,  
    bar width=8pt,
    legend style={font=\tiny, at={(0.7,1.0)}},
    width=7cm,
    height=5cm,
    x tick label/.append style={font=\tiny, rotate=30},
    y tick label/.append style={font=\tiny},
    axis background/.style={fill=gray!10},
    x label style={at={(axis description cs:0.5,-0.1)},anchor=north, font=\tiny},
    xlabel={},
    ylabel style={font=\footnotesize},
    every node near coord/.append style={font=\tiny},
     nodes near coords, % this command is used to mention the y-axis points on the top of the particular bar.  
    nodes near coords align={vertical},  
    ]  
\addplot coordinates {(ProFTPD,4.2) (TinyBigNum,121.7) (UFTPD, 2.7) (MicroHTTPD,3.8) (png2pnm, 11.4) (pnm2png, 46.5) (Parsons, 267)};  
\addplot coordinates {(ProFTPD,2.0) (TinyBigNum,54.7) (UFTPD, 1.9) (MicroHTTPD,0.47) (png2pnm, 0.38) (pnm2png, 15.4) (Parsons, 1.9)};  
  \legend{WASM,NO-OP}
\end{axis}  
\end{tikzpicture} 
\vspace*{-1.5em}
\caption{Runtime Overhead of Partitioned Programs.}
\label{fig:runtimeoverhead}
\end{figure}

%% file: limitations.tex
\section{Limitations and Future Work}
\label{sec:limitations}
Despite the effectiveness of~\systemname{}, it has limitations:

\noindent\textbf{Sandbox Dependency:} \systemname{} assumes the availability of a sandbox and consequently inherits all the limitations of the corresponding sandbox.~\eg Programs should be compilable with the sandbox compiler.
Also, as shown in~\sect{subsec:programoverhead}, the performance of the partitioned applications mainly depends on the sandbox.
However, our implementation is not dependent on one specific sandbox and can be easily extended to other sandboxes. As a future work, we will extend our implementation to other sandboxes.

\noindent\textbf{Annotation Effort:} Currently, all~\code{taint} annotations have to be done manually -- such that these annotations satisfy our type checker rules (\sect{sec:typechecking}).
This could be tedious based on the complexity of the sandboxed function, its parameter complexity, and its dependency on other functions.
We plan to develop an automated annotation tool such that, given the initial annotations (\lst{lst:humantaint}), our tool will automatically add all the required annotations (\lst{lst:humanadjust}) according to our type rules.

%% file: comparison.tex
\section{Related Work}
\label{sec:related}

%Our work is most closely related to prior formalizations of C(-like)
%languages and program partitioning mechanism 
%that aim at enforcing memory safety.
A number of prior works have looked at formalizing the semantics of C,
including CompCert~\cite{Blazy2009,leroy:hal-00703441},
\citet{ellison-rosu-2012-popl}, \citet{Kang:2015:FCM:2813885.2738005},
and \citet{10.1145/2980983.2908081, Memarian:2019:ECS:3302515.3290380},
but they are not directly concerned with enforcing
spatial safety.

\myparagraph{Spatially Safe C Formalizations}
Several prior works~\cite{li22checkedc} formalize C-language transformations or C-language
dialects aiming to ensure spatial safety.
The difference between these works and \systemname is presented in \Cref{sec:intros,sec:overview}.
\citet{10.1145/2813885.2737979} extended the formalization
of \citet{ellison-rosu-2012-popl} to produce a semantics that detects
violations of spatial safety (and other forms of undefinedness) 
by focusing on bug finding, not compiling programs to use this semantics.

CCured~\cite{Necula2005} and Softbound~\cite{softbound} implement
spatially safe semantics for normal C via program transformation. Like
\lang, both systems' operational semantics annotate pointers with
their bounds. CCured's equivalent of array pointers are compiled to be
``fat,'' while SoftBound compiles bounds metadata to a separate
hashtable, thus retaining binary compatibility at higher checking
cost. \systemname uses static type information to enable bounds checks
without need of pointer-attached metadata.
Cyclone \cite{Jim2002,GrossmanMJHWC02} is a C dialect that aims to
ensure memory safety; its pointer types are similar to
CCured and its formalization~\cite{GrossmanMJHWC02} focuses on ensuring temporal safety. 
Deputy~\cite{Feng2006,Condit2007}
is another safe-C dialect that aims to avoid fat pointers,
with its formalization~\cite{Condit2007} defines its
semantics directly in terms of compilation.
None of the above work touches program partitioning and sandbox mechanism. 

%The most closely related work is the
%implementation and formalization of \checkedc done by \citet{checkedc}, \citet{machiry2022c}, %\citet{ruef18checkedc-incr} and \citet{li22checkedc}.
%The difference between these works and \systemname is listed in \Cref{sec:intros,sec:overview}.

\myparagraph{Program Partitioning Mechanism}
The unchecked and checked code region separation in \systemname represents an isolation mechanism to ensure that code executed as part of unchecked regions does not violate the safety guarantees in checked regions,
which is typically called program partitioning~\cite{rul2009towards}, and there has been considerable work~\cite{tan2017principles, brumley2004privtrans, bittau2008wedge, lind2017glamdring, liu2017ptrsplit} in the area. Most of these techniques are~\emph{data-centric}~\cite{lind2017glamdring, liu2017ptrsplit}, wherein program data drives the partitioning. E.g., Given sensitive data in a program, the goal is to partition functions into two parts or partitions based on whether a function can access the sensitive data.
The performance overhead of these approaches is dominated by marshaling costs and depends on the usage of sensitive data.
The overhead of state-of-the-art approaches~\cite{lind2017glamdring, liu2017ptrsplit} is prohibitive and varies from 37\%-163\%.
RLBox~\cite{rlbox-paper}, merged a type checker with a sandbox mechanism to better achieve the programming partitioning mechanism.
They allow tainted pointers to be shared among different code regions. 
However, the RLBox type system is primarily based on C++ templates, making it hard or impossible to apply directly to C programs.
Furthermore, their technique has no formal guarantees and can only isolate complete libraries instead of arbitrary functions.
\systemname is a generic formally verified system that enables partitioning arbitrary functions with stronger guarantees.

%% file: conclusion.tex
\section{Conclusions}
\label{sec:conclude}
We present~\systemname{}, a type-directed program partitioning mechanism with Checked C for incremental spatial memory safety.
Our system enables partitioning arbitrary functions of a program into a sandbox, so that these functions do not affect the rest of the program.
We formalize the type-system and semantics and prove various safety properties ensuring spatial memory safety.
We implemented~\systemname{} as an interactive pipeline enabling developers to partition applications interactively.
Our evaluation on~\numprog{} programs shows that~\systemname{} enables interactive partitioning of applications with relatively less effort.

%% file: appendix.tex
\section{Appendix}
\label{app:main}

\subsection{Converting C to \checkedc}
\label{app:convertctocc}
The safety guarantees of Checked C come with certain restrictions. For instance,
as shown below, Checked C programs cannot use address-taken variables in a
bounds expression as the bounds relations may not hold because of possible
modifications through pointers.
\begin{minted}[xleftmargin=30pt, mathescape, escapeinside=||, fontsize=\footnotesize]{c}
...
_Array_ptr<int> p : count (n) = NULL;
|\textcolor{red}{\faTimes}|..,&n,.
\end{minted}
Consequently, converting existing C programs to Checked C might require
refactoring,~\eg eliminate~\inlinecode{&n} from the program above without
changing its functionality.
This might require considerable effort~\cite{duanrefactoring} depending on the
program's complexity.
Recently, Machiry~\etal developed~\threec~\cite{machiry2022c} that tries to
automatically convert a program to Checked C by adding appropriate pointer
annotations.
However, as described in \threec, completely automated conversion
is \emph{infeasible}, and it requires the developer to convert some code regions
manually.

\subsection{Well-formedness and Subtype}
\label{app:le}

\begin{figure}[h]
{\small
  \begin{mathpar}

  \inferrule[]
  {}
  {m \vdash \tint}

  \inferrule[]
  {\xi \wedge m\vdash \tau \\ \xi \le m}
  {m \vdash \tptr{\tallarrayb{\bvar}{\tau}}{\xi}}

  \inferrule[]
  {\xi \wedge m \vdash \tau\\ \xi \le m}
  {m \vdash \tptr{\tau}{\xi}}

  \inferrule[]
  {\xi \wedge m \vdash \tau\\ \xi \le m \\\\ \fv(\overline{\tau})\cup\fv(\tau)\subseteq \overline{x}}
  {m \vdash \tptr{(\tfun{\overline{x}}{\overline{\tau}}{\tau}}{\xi})}
  \end{mathpar}
}
{\footnotesize
\[
\begin{array}{l} 
\tmode \wedge \cmode = \umode \qquad \xi \wedge \umode = \umode
\qquad \cmode \wedge m = m 
\qquad  m_1 \wedge m_2 = m_2 \wedge m_1
\end{array}
\]
}
 \caption{Well-formedness for Nested Pointers}
\label{fig:wftypes}
\end{figure}

\begin{figure}[h]
{\small
  \begin{mathpar}

    \inferrule[]
    {}
    {\Gamma \vdash n}

    \inferrule[]
    {x:\tint \in \Gamma}
    {\Gamma \vdash x + n}

    \inferrule[]
    {\Gamma \vdash b_l\\
    \Gamma \vdash b_h}
  {\Gamma \vdash (b_l,b_h)}

  \inferrule[]
  {}
  {\Gamma \vdash \tint}

  \inferrule[]
  {\Gamma \vdash \bvar \\
  \Gamma \vdash \tau}
  {\Gamma \vdash \tptr{\tallarrayb{\bvar}{\tau}}{m}}

  \inferrule[]
  {\Gamma \vdash \tau}
  {\Gamma \vdash \tptr{\tau}{m}}

  \inferrule[]
  {T \in D}
  {\Gamma \vdash \tptr{\tstruct{T}}{m}}

  \inferrule[]
  {\Gamma \vdash \tau}
  {\Gamma \vdash \tptr{\tau}{m}}

  \inferrule[]
  {\forall \tau_i\in \overline{\tau}\,.\,\Gamma[\forall x\in\overline{x}\,.\, x\mapsto \tint] \vdash \tau_i\\\\
   \Gamma[\forall x\in\overline{x}\,.\, x\mapsto \tint] \vdash \tau }
  {\Gamma \vdash \tfun{\overline{x}}{\overline{\tau}}{\tau}}
  \end{mathpar}
}
 \caption{Well-formedness for Types and Bounds}
\label{fig:wftypesandbounds}
\end{figure}

\Cref{fig:wftypes} defines the well-formedness for nested pointers, guaranteeing that no tainted ($\tmode$) pointer has a checked ($\cmode$) element field.
Mainly, in a nested pointer $\tptr{(... \tptr{\tau}{\xi_2} ...)}{\xi_1}$, $\xi_2\le \xi_1$.
\Cref{fig:wftypesandbounds} defines the well-formedness for bounds appearing in a type, which requires that any bound must be $\tint$ type and has instance in $\Gamma$.
All bound variables appearing in a type must be either an instance of the type environment $\Gamma$,
or a bound variable appearing in a function pointer.

\begin{DIFnomarkup}
\begin{figure}
{\small
\[\hspace*{-1.2em}
\begin{array}{l}
\textcolor{blue}{\text{Bound Inequality and Equality:}}\\[0.3em]
  \begin{array}{r@{~}c@{~}l@{~}c@{~}l}
     n \le n' &\Rightarrow& n &\le_{\Theta} & n'\\
     n \le n' &\Rightarrow& x+n &\le_{\Theta} & x+n'\\
     n \le n' \wedge \Theta(x)=\tgez &\Rightarrow& n &\le_{\Theta} & x+n'\\
     \Theta(x)=\teq{b} \wedge b+n\le_{\Theta}b'  &\Rightarrow& x+n & \le_{\Theta} & b'\\
     \Theta(x)=\teq{b}\wedge b'\le_{\Theta}b+n  &\Rightarrow& b' & \le_{\Theta} & x+n\\
     b \le_{\Theta} b' \wedge b' \le_{\Theta} b  &\Rightarrow& b & =_{\Theta} & b'
    \end{array}
  \\[1.5em]
\textcolor{blue}{\text{Type Equility:}}\\
  \begin{array}{r@{~}c@{~}l@{~}c@{~}l}
     && \tint & =_{\Theta} & \tint\\
     \omega =_{\Theta} \omega' &\Rightarrow& \tptr{\omega}{\xi} & =_{\Theta} & \tptr{\omega'}{\xi}\\
     \bvar =_{\Theta} \bvar' \wedge  \tau =_{\Theta} \tau'
             &\Rightarrow& \tallarrayb{\bvar}{\tau} & =_{\Theta} & \tallarrayb{\bvar'}{\tau'}\\

    \textit{cond}(\overline{x},\overline{\tau}\to\tau,\overline{y},\overline{\tau'}\to\tau')

 &\Rightarrow& \tfun{\overline{x}}{\overline{\tau}}{\tau} & 
                         =_{\Theta} & \tfun{\overline{y}}{\overline{\tau'}}{\tau'}\\
    \end{array}
  \\[0.7em]
\textcolor{blue}{\text{Subtype:}}\\[0.3em]

  \begin{array}{r@{~}c@{~}l@{~}c@{~}l}
    \tau =_{\Theta} \tau'&\Rightarrow&\tau &\sqsubseteq_{\Theta}& \tau'\\[0.2em]

    && \tptr{\tau}{\tmode}&\sqsubseteq_{\Theta}& \tptr{\tau}{\umode}\\[0.2em]

    0\le_{\Theta} b_l \wedge b_h \le_{\Theta} 1 &\Rightarrow& \tptr{\tau}{m}&\sqsubseteq_{\Theta}& \tarrayptr{b_l}{b_h}{\tau}{m}\\[0.2em]
    b_l \le_{\Theta} 0 \wedge 1 \le_{\Theta} b_h &\Rightarrow& \tarrayptr{b_l}{b_h}{\tau}{m} &\sqsubseteq_{\Theta}& \tptr{\tau}{m}\\[0.2em]
    b_l \le_{\Theta} 0 \wedge 1 \le_{\Theta} b_h &\Rightarrow& \tntarrayptr{b_l}{b_h}{\tau}{m} &\sqsubseteq_{\Theta}& \tptr{\tau}{m}\\[0.2em]
    %% b_l \le b_l' \wedge b_h' \le b_h &\Rightarrow&  \tarrayptr{b_l}{b_h}{\tau}{m} &\sqsubseteq&  \tarrayptr{b_l'}{b_h'}{\tau}{m}\\[0.6em]
    b_l \le_{\Theta} b_l' \wedge b_h' \le_{\Theta} b_h &\Rightarrow& \tntarrayptr{b_l}{b_h}{\tau}{m} &\sqsubseteq_{\Theta}& \tarrayptr{b_l'}{b_h'}{\tau}{m}\\[0.6em]
    b_l \le_{\Theta} b_l' \wedge b_h' \le_{\Theta} b_h &\Rightarrow& \tallarrayptr{b_l}{b_h}{\tau}{m} &\sqsubseteq_{\Theta}& \tallarrayptr{b_l'}{b_h'}{\tau}{m}
\\[0.2em]
\overline{\tau'}\sqsubseteq_{\Theta}\overline{\tau}\wedge \tau\sqsubseteq_{\Theta}\tau' &\Rightarrow& \tptr{\tfun{\overline{x}}{\overline{\tau}}{\tau}}{\xi} &\sqsubseteq_{\Theta}& \tptr{\tfun{\overline{x}}{\overline{\tau'}}{\tau'}}{\xi}

    \end{array}
\end{array}
  \]
}
{\footnotesize
\[
\begin{array}{l}
n'+n = add(n',n)
\qquad
(x+n')+n = x+add(n',n)\\
\textit{cond}(\overline{x},\tau,\overline{y},\tau')
=\exists\overline{z}\;.\;\overline{x}\cupdot\overline{z}
  \wedge \overline{y}\cupdot\overline{z}
  \wedge \size(\overline{x})=\size(\overline{y})=\size(\overline{z})
\\\qquad\qquad\qquad\qquad\qquad
  \wedge \tau[\overline{z}/\overline{x}]= \tau'[\overline{z}/\overline{x}]
\end{array}
\]
}
  \caption{Type Equality and Subtyping}
  \label{fig:checkc-subtype}
\end{figure}
\end{DIFnomarkup}

In \lang, type equality $\tau=_{\Theta}\tau'$
is a type construct equivalent relation defined by the bound equality ($=_{\Theta}$) in (NT-)array pointer types
and the alpha equivalence of two function types in \Cref{fig:checkc-subtype};
i.e., two (NT-)array pointer types $\tallarrayb{\bvar}{\tau} $ and $ \tallarrayb{\bvar'}{\tau'}$ are equivalent, if 
$\bvar =_{\Theta} \bvar'$ and $\tau=_{\Theta}\tau'$; two function types 
$\tfun{\overline{x}}{\overline{\tau}}{\tau} $ and $ \tfun{\overline{y}}{\overline{\tau'}}{\tau'}$
are equivalent, if we can find a same length (as $\overline{x}$ and $\overline{y}$) variable list $\overline{z}$ that is substituted for $\overline{x}$ and $\overline{y}$ in $\overline{\tau} \to {\tau}$ and $\overline{\tau'} \to {\tau'}$, resp.,
and the substitution results are equal.

 The \textsc{T-CastPtr} rule
permits casting from an expression of type $\tau'$ to a checked pointer when
$\tau' \sqsubseteq \tptr{\tau}{\cmode}$. This subtyping relation
$\sqsubseteq$ is given in Fig.~\ref{fig:checkc-subtype} and is built on the type equality
($\tau =_{\Theta} \tau'\Rightarrow\tau \sqsubseteq_{\Theta} \tau'$). The many
rules ensure the relation is transitive. Most of the rules manage
casting between array pointer types. The rule 
($0\le b_l \wedge b_h \le 1 \Rightarrow \tptr{\tau}{m}\sqsubseteq
\tarrayptr{b_l}{b_h}{\tau}{m}$) permits treating a singleton
pointer as an array pointer with $b_h\le 1$ and $0 \le b_l$.
Two function pointer types are subtyped ($\tptr{\tfun{\overline{x}}{\overline{\tau}}{\tau}}{\xi} \sqsubseteq_{\Theta} \tptr{\tfun{\overline{x}}{\overline{\tau'}}{\tau'}}{\xi}$), 
if the output type are subtyped ($\tau\sqsubseteq_{\Theta}\tau'$) and the argument types are reversely subtyped ($\overline{\tau'}\sqsubseteq_{\Theta}\overline{\tau}$).
%There is another casting rule in \Cref{app:main} stating that
% users are free to cast types in unchecked code regions, since unchecked regions can contain C code.

The subtyping relation given in Fig.~\ref{fig:checkc-subtype} involves
dependent bounds, i.e., bounds that may refer to variables. To decide
premises $b \leq_{\Theta} b'$ in \Cref{fig:checkc-subtype}, we need a decision procedure that accounts for
the possible values of these variables. This process considers
$\Theta$, tracked by the typing judgment, and $\varphi$, the current
stack snapshot (when performing subtyping as part of the type
preservation proof).

Since bounds expressions may
contain variables, determining assumptions like $b_l \leq_{\Theta} b_l'$
requires reasoning about the probable values of these variables'. The type
system uses $\Theta$ to make such reasoning more precise.
$\Theta$ is a map from variables $x$ to
equation predicates $P$, which have the form $P ::= \tgez \;|\; \teq{b}$.
It maps variables to equations that are recorded along the type checking procedure.
If $\Theta$ maps $x$ to $\tgez$, that means that $x \ge 0$;
$\teq{b}$ means that $x$ is equivalent to the bound value $b$ in the current context, 
such as in the type judgment for $e_2$ in Rule \textsc{T-LetInt} and \textsc{T-RetInt}.

$\sqsubseteq$ is parameterized by
$\Theta$, which provides the range of allowed values for a bound
variable; thus, more $\sqsubseteq$ relation is provable. For example,
rule \rulelab{T-LetInt} inserts a predicate $\teq{e_1}$ for variable $x$.
Assume that $e_1$ is equal to $0$,
when $x$ is used as a type variable in type $\tntarrayptr{0}{x}{\tint}{\cmode}$ in $e_2$,
the subtyping relation $\tntarrayptr{0}{x}{\tint}{\cmode} \sqsubseteq
\tntarrayptr{0}{0}{\tint}{\cmode}$ is provable when we know
\code{x}$\teq{0}$.

To capture bound variables in dependent types, the \checkedc subtyping
relation ($\sqsubseteq$) is parameterized by a restricted stack
snapshot $\varphi|_{\rho}$ and the predicate map $\Theta$, where
$\varphi$ is a stack and $\rho$ is a set of
variables. $\varphi|_{\rho}$ means to restrict the domain of $\varphi$
to the variable set $\rho$. Clearly, we have the relation:
$\varphi|_{\rho} \subseteq \varphi$. $\sqsubseteq$
being parameterized by $\varphi|_{\rho}$ refers to that when we
compare two bounds $b \le_{\Theta} b'$, we actually do
$\varphi|_{\rho}(b) \le_{\Theta} \varphi|_{\rho}(b')$ by interpreting the
variables in $b$ and $b'$ with possible values in $\varphi|_{\rho}$.
Let's define a subset relation $\preceq$ for two restricted stack
snapshot $\varphi|_{\rho}$ and $\varphi'|_{\rho}$:

\begin{DIFnomarkup}
\begin{figure*}[t]
{\small
  \begin{mathpar}
   \inferrule[T-ConstU]
       { \neg \cmode(\tau)}
       {\Gamma;\Theta\vdash_u \evalue{n}{\tau} : \tau}
\quad
   \inferrule[T-ConstC]
       {\Theta;\heap;\emptyset \vdash_c n : \tau}
       {\Gamma;\Theta\vdash_c \evalue{n}{\tau} : \tau}
\quad      
   \inferrule[T-Let]
    { x\not\in \fv(\tau') \\
        \Gamma;\Theta \vdash_m e_1 : \tau \\\\
          \Gamma[x\mapsto \tau];\Theta \vdash_m e_2 : \tau'
             }
    {\Gamma;\Theta \vdash_m \elet{x}{e_1}{e_2} : \tau'}
\quad
   \inferrule[T-LetInt]
    { x\in \fv(\tau') \Rightarrow e_1 \in \text{Bound} \\
        \Gamma;\Theta \vdash_m e_1 : \tint \\\\
           \Gamma[x\mapsto \tint];\Theta[x\mapsto \teq{e_1}] \vdash_m e_2 : \tau'
             }
    {\Gamma;\Theta \vdash_m \elet{x}{e_1}{e_2} : \tau'[e_1 / x]}
\quad
   \inferrule[T-RetInt]
    { \Gamma[x\mapsto \tint];\Theta[x\mapsto \teq{n}] \vdash_m e : \tau}
    {\Gamma;\Theta \vdash_m \eret{x}{\evalue{n}{\tint}}{e} : \tau}

    \inferrule[T-Mac]
              {\xi\le m}
              {\Gamma; \Theta \vdash_m \emalloc{\xi}{\omega} : \tptr{\omega}{\xi}}

    \inferrule[T-Add]
              {\Gamma; \Theta \vdash_m e_1 : \tint \\
                \Gamma; \Theta \vdash_m e_2 : \tint}
              {\Gamma; \Theta \vdash_m (e_1 \plus e_2) : \tint }

    \inferrule[T-Ind] 
              {\Gamma; \Theta \vdash_m e_1 : \tptr{\tallarrayb{\bvar}{\tau}}{\xi} \\
                \Gamma; \Theta \vdash_m e_2 : \tint \\
                \xi \leq m}              
              {\Gamma; \Theta \vdash_m \estar{(\ebinop{e_1}{e_2})} : \tau}

    \inferrule[T-Assign]
              {\Gamma; \Theta \vdash_m e_1 : \tptr{\tau}{\xi} \\
                \Gamma; \Theta \vdash_m e_2 : \tau' \\\\
                \tau'\sqsubseteq_{\Theta} \tau \\
                \xi \leq m}
              {\Gamma; \Theta \vdash_m \eassign{e_1}{e_2} : \tau}

    \inferrule[T-AssignArr]
              {\Gamma; \Theta \vdash_m e_1 : \tptr{\tallarrayb{\bvar}{\tau}}{\xi}\\\\
                \Gamma; \Theta \vdash_m e_2 : \tau' \\
                \tau'\sqsubseteq_{\Theta} \tau\\
                \xi \leq m}
              {\Gamma; \Theta \vdash_m \eassign{e_1}{e_2} : \tau}              

   \inferrule[T-IndAssign]
              {\Gamma; \Theta \vdash_m e_1 : \tptr{\tallarrayb{\bvar}{\tau}}{\xi}\\
                \Gamma; \Theta \vdash_m e_2 : \tint \\\\
                \Gamma; \Theta \vdash_m e_3 : \tau' \\
                \tau'\sqsubseteq_{\Theta} \tau \\
                \xi \leq m}
              {\Gamma; \defscope \vdash_m \eassign{(e_1 \plus e_2)}{e_3} : \tau}

  \end{mathpar}
}
% {\footnotesize
% \begin{center}
% $
% \begin{array}{l}
% \fm(e)\triangleq(\exists x\; n\; \tau. e=x+\evalue{n}{\tau}) \vee (\exists n\;\tau. e = \evalue{n}{\tau})
% \\[0.2em]
% \tau[\overline{e} / \overline{x}]\texttt{(with types }\evalue{\overline{x}}{\overline{\tau}}\texttt{)}\triangleq \forall e_i\in\overline{e}\;x_i\in\overline{x}\;\tau_i\in\overline{\tau}\;.\;\tau_i = \tint \wedge (x_i \in \fv(\tau) \Rightarrow \fm(e_i)) \Rightarrow \tau[e_i / x_i]
% \end{array}
% $
% \end{center}
% }
\caption{Remaining \lang Type Rules (extends Fig.~\ref{fig:type-system-1})}
\label{fig:rem-type-system}
\end{figure*}
\end{DIFnomarkup}

\begin{defi}[Subset of Stack Snapshots]
  Given two $\varphi|_{\rho}$ and $\varphi'|_{\rho}$,
  $\varphi|_{\rho} \preceq \varphi'|_{\rho}$, iff for $x\in\rho$ and
  $y$,
  $(x,y) \in \varphi|_{\rho} \Rightarrow (x,y) \in \varphi'|_{\rho}$.
\end{defi}

For every two restricted stack snapshots $\varphi|_{\rho}$ and
$\varphi'|_{\rho}$, such that
$\varphi|_{\rho} \preceq \varphi'|_{\rho}$, we have the following
theorem in \checkedc (proved in Coq):

\begin{thm}[Stack Snapshot Theorem]
  Given two types $\tau$ and $\tau'$, two restricted stack snapshots
  $\varphi|_{\rho}$ and $\varphi'|_{\rho}$, if
  $\varphi|_{\rho}\preceq \varphi'|_{\rho}$, and
  $\tau \sqsubseteq \tau'$ under the parameterization of
  $\varphi|_{\rho}$, then $\tau \sqsubseteq \tau'$ under the
  parameterization of $\varphi'|_{\rho}$.
\end{thm}

Clearly, for every $\varphi|_{\rho}$, we have
$\emptyset \preceq \varphi|_{\rho}$. The type checking stage is a
compile-time process, so $\varphi|_{\rho}$
is $\emptyset$ at the type checking stage. Stack snapshots are needed
for proving type preserving, as variables in bounds expressions are
evaluated away.

\subsection{Other Type Rules and Literal Validity Checks}\label{rem-type}

Here we show the type rules for other \lang operations in Fig.~\ref{fig:rem-type-system}.
Rule \textsc{T-Mac} deals with
$\emalloctext$ operations. There is a well-formedness check to require
that the possible bound variables in $\omega$ must be in the domain of
$\Gamma$ (see Fig.~\ref{fig:wftypesandbounds}). This is similar to the well-formedness assumption of the type environment (Definition~\ref{type-wellformed}). Rule \textsc{T-Add} deals with binary operations whose sub-terms are integer expressions.
The other rules are explained as follows.

\myparagraph{Pointer Access}
The \textsc{T-AssignArr} rule examines array assignment operations, returning the type of
pointed-to objects. Rules for pointers for other object types are
similar, such that Rule \textsc{T-Assign} assigns a value to a non-array pointer location.
The condition $\xi\le m$ ensures that checked and unchecked pointers 
can only be dereferenced in checked and unchecked regions, respectively;
The type rules do not attempt to reason whether the access is in bounds;
such check is deferred to the semantics.
Rule \textsc{T-Ind} serves the case for pointer arithmetic. For simplicity, in the \checkedc formalization, we do not allow arbitrary pointer arithmetic. The only pointer arithmetic operations allowed are the forms shown in rules \textsc{T-Ind} and \textsc{T-IndAssign} in Fig.~\ref{fig:rem-type-system}.  The predicate $\tau'\sqsubseteq_{\Theta} \tau$ requires that the value being assigned is a subtype of the pointer type.
The \textsc{T-IndAssign} rule is an extended assignment operation for handling assignments for array/NT-array pointers with pointer arithmetic.

% Subtyping and casting operations are briefly introduced in
% Sec.~\ref{sec:intros}~and~\ref{sec:overview}.  Subtyping is useful in
% static casting operations that allow users to view a pointer in one
% type as another, such as casting an NT-array pointer to an array
% one. \checkedc provides a set of safe static casting operations that
% have no cost in execution.  Moreover, subtyping acts as oracles for
% bound widening and dynamic casting operations; thus, \checkedc is
% different from a complete static array pointer bound system.  For
% example, if $e$ has type $\tau'$ and $\varphi$ is the current stack
% snapshot, the semantics of $\edyncast{\tau}{e}$ does not transition to
% an error state when $\varphi(\tau')\sqsubseteq\varphi(\tau)$.  In a
% function call, for every argument, \lang permits users to input a
% subtype entity and we prove that this does not affect the correctness
% of the program.

\begin{DIFnomarkup}
 \begin{figure}[t]
 {\small

 \begin{mathpar}
   \inferrule
       {}
       {\Theta;\heap;\sigma \vdash_m n : \tint}

   \inferrule
       {}
       {\Theta;\heap;\sigma \vdash_m 0 : \tptr{\omega}{\xi}}

   \inferrule
       {(m = \cmode \Rightarrow \xi \neq \cmode) \\\\ (m=\umode \Rightarrow \xi = \umode)}
       {\Theta;\heap;\sigma \vdash_{\cmode} n : \tptr{\omega}{\tmode}}
  
   \inferrule
       {(\evalue{n}{\tptr{\omega}{\xi}})\in \sigma}
       {\Theta;\heap;\sigma \vdash_m n : \tptr{\omega}{\xi}}

   \inferrule
       {\tptr{\omega'}{\xi'} \sqsubseteq_{\Theta} \tptr{\omega}{\xi} 
            \\ \Theta;\heap;\sigma \vdash_m n : \tptr{\omega'}{\xi'}}
       {\Theta;\heap;\sigma \vdash_m n : \tptr{\omega}{\xi}}

   \inferrule
       { \xi \le m 
     \\\Xi(m,n)=\tau\;(\evalue{\overline{x'}}{\overline{\tau}})\;(\xi,e)
       \\  \overline{x} = \{x|(x:\tint) \in (\overline{x'}:\overline{\tau}) \}}
       {\Theta;\heap;\sigma \vdash_m n : \tptr{(\tfun{\overline{x}}{\overline{\tau}}{\tau})}{\xi}}
  
   \inferrule
       {\neg\funptr(\omega)\\ \xi \le m\\
        \forall i \in [0,\size(\omega)) \;.\;
            \Theta;\heap;(\sigma \cup \{(n:\tptr{\omega}{\xi})) \}\vdash_m \heap(m,n+i)}
       {\Theta;\heap;\sigma \vdash_m n : \tptr{\omega}{\xi}}
 \end{mathpar}
 }
{\footnotesize
\[
\begin{array}{l} 
\funptr(\tfun{\overline{x}}{\overline{\tau}}{\tau}) = \texttt{true}
\qquad
\funptr(\omega) = \texttt{false}\;\;{[\emph{owise}]}
\end{array}
\]
}
 \caption{Verification/Type Rules for Constants}
 \label{fig:const-type}
 \end{figure}
\end{DIFnomarkup}

\myparagraph{Literal Constant Validity}
Rules \textsc{T-ConstU} and \textsc{T-ConstC}
describes type assumptions for literals appearing in a program.
$\cmode(\tau)$ judges that a literal pointer 
in an unchecked region cannot be of a checked type,
which represents an assumption that programmers 
cannot guess a checked pointer address and utilize it in an unchecked region in \systemname.
In rule \textsc{T-ConstC}, we requires a static 
verification procedure for validating a literal pointer, 
which is similar to the dynamic verification process in \Cref{sec:typechecking}. 

The verification process $\Theta;\heap;\sigma \vdash_m n : \tau$ checks (\Cref{fig:const-type})
validate the literal $\evalue{n}{\tau}$, 
where $\heap(m)$ is the initial heap that the literal resides on and
$\sigma$ is a set of literals assumed to be checked.
A global function store $\Xi(m)$ is also required to check the validity of a function pointer.
A valid function pointer should appear in the right store region ($\cmode$ or $\umode$)
and the address stores a function with the right type.
The last rule in \Cref{fig:const-type} describes the validity check for a non-function pointer, 
where every element in the pointer range ($[0,\size(\omega))$) should be well
typed.

\begin{figure*}[t]
{\small
\begin{mathpar}

\inferrule [S-Var]{} {(\varphi,\heap,x)\longrightarrow (\varphi,\heap,\varphi(x))}

    \inferrule[S-DefArrayC]{\heap(\cmode,n)=\evalue{n_a}{\tau_a} \\ 0 \in [n_l,n_h)}
    {(\varphi,\heap,\estar{\evalue{n}{\tntarrayptr{n_l}{n_h}{\tau}{\cmode}}}) \longrightarrow (\varphi,\heap,\evalue{n_a}{\tau})}

    \inferrule[S-DefArrayT]{\heap(\umode,n)=\evalue{n_a}{\tau_a} \\ 0 \in [n_l,n_h) 
               \\  \emptyset;\heap ; \emptyset \vdash_{\umode}\evalue{n_a}{\tau}}
    {(\varphi,\heap,\estar{\evalue{n}{\tntarrayptr{n_l}{n_h}{\tau}{\tmode}}}) \longrightarrow (\varphi,\heap,\evalue{n_a}{\tau})}

    \inferrule[S-DefArrayBound]{0 \not\in [n_l,n_h)}
     { (\varphi,\heap,\estar{\evalue{n}{\tallarrayptr{n_l}{n_h}{\tau}{\xi}}}) \longrightarrow (\varphi,\heap,\ebounds)}
\quad
    \inferrule[S-DefNTArrayBound]{0 \notin [n_l,n_h]}
    {(\varphi,\heap,\estar{\evalue{n}{\tntarrayptr{n_l}{n_h}{\tau}{\xi}}}) \longrightarrow (\varphi,\heap,\ebounds)}
\quad
    \inferrule[S-Checked]{}{(\varphi,\heap,\echecked{\overline{x}}{\evalue{n}{\tau}}) \longrightarrow (\varphi,\heap,\evalue{n}{\tau})}

    \inferrule[S-Ret]{}{(\varphi,\heap,\ret{x}{\mu}{\evalue{n}{\tau}}) \longrightarrow (\varphi[x\mapsto \mu],\heap,\evalue{n}{\tau})}

        \inferrule[S-Let]{}{(\varphi,\heap,\elet{x}{\evalue{n}{\tau}}{e}) \longrightarrow (\varphi[x\mapsto \evalue{n}{\tau}],\heap,\ret{x}{\varphi(x)}{e})}

    \inferrule[S-AssignNull]{}
      {(\varphi,\heap,\eassign{\evalue{0}{\tptr{\omega}{\xi}}}{\evalue{n_1}{\tau_1}}) \longrightarrow (\varphi,\heap,\enull)}

    \inferrule[S-AssignArrC]{\heap(\cmode,n)=\evalue{n_a}{\tau_a}\\ 0 \in [n_l,n_h) }
      {(\varphi,\heap,\eassign{\evalue{n}{\tallarrayptr{n_l}{n_h}{\tau}{\cmode}}}{\evalue{n_1}{\tau_1}}) \longrightarrow (\varphi,\heapup{\cmode}{n}{\evalue{n_1}{\tau_a}},\evalue{n_1}{\tau})}

    \inferrule[S-AssignArrT]{\heap(\umode,n)=\evalue{n_a}{\tau_a}\\ 0 \in [n_l,n_h) 
               \\ \emptyset;\heap ; \emptyset \vdash_{\umode}\evalue{n_a}{\tau}}
      {(\varphi,\heap,\eassign{\evalue{n}{\tallarrayptr{n_l}{n_h}{\tau}{\tmode}}}{\evalue{n_1}{\tau_1}}) \longrightarrow (\varphi,\heapup{\umode}{n}{\evalue{n_1}{\tau_a}},\evalue{n_1}{\tau})}

    \inferrule[S-AssignArrBound]{0 \not\in [n_l,n_h) }
      {(\varphi,\heap,\eassign{\evalue{n}{\tallarrayptr{n_l}{n_h}{\tau}{\cmode}}}{\evalue{n_1}{\tau_1}}) \longrightarrow (\varphi,\heap,\ebounds)}

    \inferrule[S-AssignC]{\heap(\cmode,n)=\evalue{n_a}{\tau_a} }
      {(\varphi,\heap,\eassign{\evalue{n}{\tptr{\tau}{\cmode}}}{\evalue{n_1}{\tau_1}}) \longrightarrow (\varphi,\heap[n \mapsto \evalue{n_1}{\tau}],\evalue{n_1}{\tau})}

  \inferrule[S-Malloc]{\varphi(\omega)=\omega_a \\ \mathtt{alloc}(\heap,\xi,\omega_a)=(n,\heap')}
   { (\varphi,\heap,\emalloc{\xi}{\omega}) \longrightarrow (\varphi,\heap',\evalue{n}{\tptr{\omega_a}{\xi}})}

  \inferrule[S-MallocBound]{\varphi(\omega)=\tallarray{n_l}{n_h}{\tau}\\ (n_l \neq 0 \vee n_h \le 0)}
    { (\varphi,\heap,\emalloc{\omega}) \longrightarrow (\varphi,\heap',\ebounds)}

  \inferrule[S-IfNTNotC]{\varphi(x)=\evalue{n}{\tntarrayptr{n_l}{n_h}{\tau}{\cmode}} \\ \heap(\cmode,n)\neq 0\\ 0 < n_h}
             {(\varphi,\heap,\eif{\estar{x}}{e_1}{e_2}) \longrightarrow (\varphi,\heap,e_1)}

    \inferrule[S-IfT]{n \neq 0 }
    {(\varphi,\heap,\eif{\evalue{n}{\tau}}{e_1}{e_2}) \longrightarrow (\varphi,\heap,e_1)}

    \inferrule[S-IfF]{}
    {(\varphi,\heap,\eif{\evalue{0}{\tau}}{e_1}{e_2}) \longrightarrow (\varphi,\heap,e_2)}

    \inferrule[S-Add]{n = n_1 + n_2}
    {(\varphi,\heap,\evalue{n_1}{\tint} \plus \evalue{n_2}{\tint}) \longrightarrow (\varphi,\heap, n)}

    \inferrule[S-AddArr]{n = n_1 + n_2\\ n_l' = n_l - n_2 \\ n_h' = n_h - n_2}
    {(\varphi,\heap,\evalue{n_1}{\tallarrayptr{n_l}{n_h}{\tau}{\xi}} \plus \evalue{n_2}{\tint}) \longrightarrow (\varphi,\heap, \evalue{n}{\tallarrayptr{n_l'}{n_h'}{\tau}{\xi}})}

n    \inferrule[S-AddArrNull]{}
    {(\varphi,\heap,\evalue{0}{\tallarrayptr{n_l}{n_h}{\tau}{\xi}} \plus \evalue{n_2}{\tint}) \longrightarrow (\varphi,\heap, \enull)}

\end{mathpar}

}
\caption{Remaining \lang Semantics Rules (extends Fig.~\ref{fig:type-system-1})}
\label{fig:rem-semantics}
\end{figure*}

A checked pointer checks validity in type step as rule \textsc{T-ConstC},
while a tainted/unchecked pointer does not check for such during the type checking.
Tainted pointers are verified through the validity check in dynamic execution as we mentioned above.

If the literal's type is an integer, an unchecked pointer, or a null
pointer, it is well typed, as shown by the top three rules in
Fig.~\ref{fig:const-type}. However, if it is a checked pointer
$\tptr{\omega}{\cmode}$, we need to ensure that what it points to in
the heap is of the appropriate pointed-to type ($\omega$), and also
recursively ensure that any literal pointers reachable this way are
also well-typed. This is captured by the bottom rule in the figure,
which states that for every location $n+i$ in the pointers' range
$[n, n+\size(\omega))$, where $\size$ yields the size of its argument,
  then the value at the location $\heap(n+i)$ is also well-typed.
  However, as heap snapshots can contain cyclic structures (which
  would lead to infinite typing deriviations), we use a scope $\sigma$
  to assume that the original pointer is well-typed when checking the
  types of what it points to. The middle rule then accesses the scope
  to tie the knot and keep the derivation finite, just like in
  \citet{ruef18checkedc-incr}.

\myparagraph{Let Bindings}
Rules \textsc{T-Let} and \textsc{T-LetInt} type a $\elettext$ expression, which also admits
type dependency. 
In particular, the result of evaluating a $\elettext$
may have a type that refers to one of its bound variables (e.g., if
the result is a checked pointer with a variable-defined bound). If so, we must substitute away this variable once it goes out of scope (\textsc{T-LetInt}). 
Note that we restrict the expression $e_1$ to syntactically match the
structure of a Bounds expression $b$ (see Fig.~\ref{fig:checkc-syn}).
Rule \textsc{T-RetInt} types a $\erettext$ expression when $x$ is of type $\tint$.
$\erettext$ does not appear in source programs but is introduced by the semantics when
evaluating a let binding (rule \textsc{S-Let} in
Fig.~\ref{fig:rem-semantics}). 

\subsection{Other Semantic Rules}\label{sec:rem-semantics}

Fig.~\ref{fig:rem-semantics} shows the remaining semantic rules for
$\lang$. We explain a selected few rules in this subsection.
% other few low-level semantic rules for variable and dereference and $\emalloctext$ operations in \checkedc. Other operations are defined in the same manner.

\myparagraph{Checked and Tainted Pointer Operations}
Rule \textsc{S-Var} loads the value for $x$ in stack $\varphi$.
Rules \textsc{S-DefArrayC} and \textsc{S-DefArrayT} dereference an $\cmode$ and $\tmode$ array pointer, respectively.
In \lang, the difference between array and NT-array dereference is that the range of $0$ is at $[n_l,n_h)$ not $[n_l,n_h]$, meaning that one cannot dereference the upper-bound position in an array.
Rules \textsc{DefArrayBound} and \textsc{DefNTArrayBound} describe an error case for a dereference operation.
If we are dereferencing an array/NT-array pointer and the mode is $\cmode$ (or $\tmode$), $0$ must be in the range from $n_l$ to $n_h$ (meaning that the dereference is in-bound); if not, the system results in a $\ebounds$ error. Obviously, the dereference of an array/NT-array pointer also experiences a $\enull$ state transition if $n\le 0$.

Rules \textsc{S-Malloc} and \textsc{S-MallocBound} describe the $\emalloctext$ semantics. Given a valid type $\omega_a$ that contains no free variables, $\mathtt{alloc}$ function returns an address pointing at the first position of an allocated space whose size is equal to the size of $\omega_a$ for a specific mode $\xi$, 
and a new heap snapshot $\heap'$ that marks the allocated space for the new allocation. 
If $\xi$ is $\cmode$, the new $\heap$ allocation is in the $\cmode$ region, while $\tmode$ and $\umode$ mode allocation creates new pieces in $\umode$ region.
The $\emalloctext$ is transitioned to the address $n$ with the type ${\tptr{\omega_a}{\xi}}$ and new updated heap. It is possible for $\emalloctext$ to transition to a $\ebounds$ error if the $\omega_a$ is an array/NT-array type $\tallarray{n_l}{n_h}{\tau}$, and either $n_l \neq 0$ or $n_h \le 0$. This can happen when the bound variable is evaluated to a bound constant that is not desired. 

Rules \textsc{S-AssignArrC} and \textsc{S-AssignArrT} are for assignment operations.
\textsc{S-AssignArrC} assigns to an array as long as 0 (the point of
dereference) is within the bounds designated by the pointer's annotation
and strictly less than the upper bound. 
Rule \textsc{S-AssignArrT} is similar to \textsc{S-AssignArrC} for tainted pointers.
Any dynamic heap access of a tainted pointer requires a \textit{verification}.
Performing such a verification equates to performing a literal type check for a pointer constant in \Cref{fig:const-type}.

\myparagraph{Let Bindings}
The semantics manages variable scopes using the special $\erettext$
form. \textsc{S-Let} evaluates to a configuration whose expression is
$\ret{x}{\varphi(x)}{e})$.
It keeps the old value of $x$ in the stack of $\texttt{ret}$ 
and updates $x$'s value in $\varphi$ with the new value of the let binding.
After $e$ is evaluated to a value $\evalue{n}{\tau}$,
we restore $\varphi$'s value back to the one stored in the stack of $\texttt{ret}$ as $\mu$,
which is explain in rule \rulelab{S-Ret}.

\subsection{Compilation Formalism}\label{appx:comp1}

The main subtlety of compiling \checkedc to Clang/LLVM is to capture the annotations on pointer literals
that track array bound information, which is used in premises
of rules like \textsc{S-DefArrayC} and
  \textsc{S-AssignArrC} to prevent spatial safety violations.
The \checkedc compiler \cite{li22checkedc} inserted additional pointer checks 
for verifying pointers are not null and the bounds are within their limits.
The latter is done by introducing additional shadow variables for storing (NT-)array pointer bound information.

\begin{figure}[t!]
{\small
\hspace*{-0.5em}
\begin{tabular}{|c|c|c|c|}
\hline
& \cmode & \tmode & \umode \\
\hline
& \textsc{CBox} / \textsc{Core} & \textsc{CBox} / \textsc{Core} & \textsc{CBox} / \textsc{Core} \\
\hline
\cmode & $\estar{x}$ / $\getstar{\cmode}{x}$ 
 & $\texttt{sand\_get}(x)$ / $\getstar{\umode}{x}$ &  $\times$ \\
\hline
\umode & $\times$
 & $\estar{x}$ / $\getstar{\umode}{x}$ &  $\estar{x}$ / $\getstar{\umode}{x}$ \\
\hline
\end{tabular}

}
\caption{Compiled Targets for Dereference}
\label{fig:flagtable}
\end{figure}

In \systemname, context and pointer modes determine the particular heap/function store that a pointer points to,
i.e., $\cmode$ pointers point to checked regions, while $\tmode$ and $\umode$ pointers point to unchecked regions. 
Unchecked regions are associated with a sandbox mechanism that permits exception handling of potential memory failures.
In the compiled LLVM code, pointer access operations have different syntaxes when the modes are different.
\Cref{fig:flagtable} lists the different compiled syntaxes of a deference operation ($\estar{x}$) for the compiler implementation (\textsc{CBox}, stands for \systemname) and formalism (\textsc{Core}, stands for \lang). The columns represent different pointer modes and the rows represent context modes.
For example, when we have a $\tmode$-mode pointer in a $\cmode$-mode region, we compile a deference operation to the sandbox pointer access function ($\texttt{sand\_get}(x)$) accessing the data in the \systemname implementation. In \lang, we create a new deference data-structure on top of the existing $\estar{x}$ operation (in LLVM): $\getstar{m}{x}$. If the mode is $\cmode$, it accesses the checked heap/function store; otherwise, it accesses the unchecked one.

This section shows how \lang deals with pointer modes, mode switching and function pointer compilations, 
with no loss of expressiveness
as the \checkedc contains the erase of annotations in \cite{li22checkedc}.
For the compiler formalism, 
we present a compilation algorithm that converts from
\lang to \elang, an untyped language without metadata
annotations, which represents an intermediate layer we build on LLVM for simplifying compilation. 
In \elang, the syntax for deference, assignment, malloc, function calls are: $\getstar{m}{e}$, $\elassign{m}{e}{e}$, 
$\emalloc{m}{\omega}$, and $\ecall{e}{\overline{e}}$.
The algorithm sheds light on how compilation can be implemented in the real \systemname
  compiler, while eschewing many vital details (\elang has many differences with LLVM IR).

Compilation is defined by extending \lang's
typing judgment as follows:
\[\Gamma;\Theta;\rho \vdash_m e \gg \dot e:\tau\]
There is now a \elang output $\dot e$ and an input $\rho$, which maps
each (NT-)array pointer variable to its mode and
each variable \code{p} to a pair of \emph{shadow
  variables} that keep \code{p}'s up-to-date upper and lower bounds. 
These may differ from the bounds in \code{p}'s type due to bounds
widening.\footnote{Since lower bounds are never widened, the
  lower-bound shadow variable is unnecessary; we include it for uniformity.} 

% When $\Gamma$,$\Theta$ and $\rho$ are all empty, we write $e \gg \dot e$ rather than the
% complete judgment, implicitly assuming that $e$ is a well-typed and closed
% term.

We formalize rules for this judgment in PLT Redex~\cite{pltredex},
following and extending our Coq development for \lang. To give
confidence that compilation is correct, we use Redex's property-based
random testing support to show that compiled-to $\dot e $ simulates
$e$, for all $e$.

% We developed a \checkedc compiler to compile a \checkedc program to a C program.
% \mwh{We formalized compilation from CoreChkC to a version of CoreChkC
%   but with the metadata removed, right? This is not a Checked C
%   compiler. You go on to see stuff about CompCert, CLight,
%   etc. This is confusing. We should be talking about how this relates
%   to what was just presented. Pick definitive names for things. }
% Given a \checkedc program $e$, we build a compilation process ($\gg$), such that $e \gg \dot e$, where $\dot e$ is the corresponding C program for $e$ in A-normal form (ANF). 
% The compilation process ($\gg$) relies on the type checking step $\Gamma;\Theta\vdash_m e:\tau$. 
% Especially, it relies on $\Gamma$ to provide the type information for variables in $e$. 
% We utilize CompCert/CLight syntax and semantics \cite{Leroy:2009:FVC:1666192.1666216,Blazy2009} as our translation target language.
%  Besides, we defined two data structures in the CompCert format for representing $\enull$ and $\ebounds$ states.
% We write $\xrightarrow{c}$ for the semantics of CLight. \mwh{Of our
%   target language, which I presume does not have all the features that
%   CLight has. Should mention up front that we did all of this in PLT Redex.}

\myparagraph{Approach}
Here, we explain the rules for compilation by
examples, using a C-like syntax; the complete rules are given in
\cite{checkedc-tech-report}.
Each rule performs up to three tasks: (a) conversion of $e$ to
A-normal form; (b) insertion of dynamic checks and bound widening expressions; 
and (c) generate right pointer accessing expressions based on modes.
A-normal form conversion is straightforward: compound expressions are managed by storing results of subexpressions into temporary variables,
as in the following example.

{\vspace*{-0.5em}
{\small
\begin{center}
$
\begin{array}{l}
$\code{let y=(x+1)+(6+1)}$
\;
\begin{frame}

\tikz\draw[-Latex,line width=2pt,color=orange] (0,0) -- (1,0);

\end{frame}
\;
\begin{array}{l}
$\code{let a=x+1;}$\\
$\code{let b=6+1;}$\\
$\code{let y=a+b}$\\
\end{array}
\end{array}
$
\end{center}
}
}

This simplifies the management of effects from subexpressions. The
next two steps of compilation are more interesting.
We state them based on different \lang operations.

\begin{figure}[t!]
  \begin{small}
\begin{lstlisting}[mathescape,xleftmargin=4 mm]
int deref_array(n : int,
     p :  $\color{green!40!black}\tntarrayptr{0}{n}{\tint}{\cmode}$,
     q : $\color{green!40!black}\tntarrayptr{0}{n}{\tint}{\tmode}$) {
  /* $\color{purple!40!black}\rho$(p) = p_lo,p_hi,p_m */
  /* $\color{purple!40!black}\rho$(q) = q_lo,q_hi,q_m */
    * p;
    * q = 1;
}
...
/* p0 : $\color{purple!40!black}\tntarrayptr{0}{5}{\tint}{\cmode}$ */
/* q0 : $\color{purple!40!black}\tntarrayptr{0}{5}{\tint}{\tmode}$ */
deref_array(5, p0, q0);
    \end{lstlisting}
\begin{frame}

\tikz\draw[-Latex,line width=2pt,color=orange] (0,0) -- (1,0);

\end{frame}
\begin{lstlisting}[mathescape,xleftmargin=4 mm]
deref_array(int n, int* p, int * q) {
  //m is the current context mode
  let p_lo = 0; let p_hi = n; 
  let q_lo = 0; let q_hi = n; 
  /* runtime checks */
  assert(p_lo <= 0 && 0 <= p_hi);
  assert(p != 0);
  *(mode(p) $\wedge$ m,p);
  verify(q, not_null(m, q_lo, q_hi) 
             && q_lo <= 0 && 0 <= q_hi);
  *(mode(q) $\wedge$ m,q)=1;
}
...
deref_array(5, p0, q0);
    \end{lstlisting}
\end{small}
    \caption{Compilation Example for Dependent Functions}
\label{fig:compilationexample1}
\end{figure}

% \review{Fig 9: if this is actual C code, then your null-check at line 6 will be
%   eliminated by the compiler. At line 3, you performed a pointer addition, which
%   is only defined when `p` is non-null. So, either `p` is non-null, and the
%   NULL-check can be eliminated; or, `p` is NULL, but line 3 was undefined
%   behavior, meaning the compiler is allowed to do anything, notably eliminate
%   the NULL-check. This is where I am super confused, and either:
%   - CoreC is not really the C language, and has different semantics...? but is
%     this well-defined in the context of LLVM?
%   - there is a problem that was not caught by the PLT-Redex-based testing.
% \yiyun{We have clarified at the start of IV that CoreC is an untyped
%   variant of CoreChkC, and does not aim to represent C per se, or LLVM
%   IR. We aimed to avoid confusion by rewriting the examples in a way that is more closely
%     related to the syntax presented in Fig 3. Pointer arithmetic between 0 and a non-zero
%     index is always valid because CoreC there is technically only
%     integer arithmetic.}}

\myparagraph{Pointer Accesses and Modes}
In every declaration of a pointer,
if the poniter is an (NT-)array,
we first allocate two \emph{shadow variables}
to track the lower and upper bounds which are potentially changed for pointer arithmetic and NT-array bound widening.
Each $\cmode$-mode NT-array pointer variable is associated with its type information in a store.
Additionally, we place bounds and null-pointer checks, such as the line 6 and 7 in \Cref{fig:compilationexample1}.
In addition, in the formalism, before every use of a tainted pointer (\Cref{fig:compilationexample1} line 9 and 10), 
there is an inserted verification step similar to \Cref{fig:const-type},
which checks if a pointer is well defined in the heap (\code{not_null}) and the spatial safety.
Predicate \code{not_null} checks that every element in the pointer's range (\code{p_lo} and \code{p_hi}) is well defined in the heap.  
The modes in compiled deference (\code{*(mode(p) }$\wedge$\code{ m,p)})
 and assignment (\code{*(mode(q) }$\wedge$\code{ m,q)=1}) operations 
are computed based on the meet 
operation ($\wedge$) of the pointer mode (e.g. \code{mode(p)}) and the current context mode (\code{m}).

\myparagraph{Checked and Unchecked Blocks}
In the \systemname implementation,
$\euncheckedtext$ and $\echeckedtext$ blocks 
are compiled as context switching functions provided by sandbox.
$\eunchecked{\overline{x}}{e}$ is compiled to 
$\texttt{sandbox\_call}(\overline{x},e)$, where we call the sandbox 
to execute expression $e$ with the arguments $\overline{x}$.
$\echecked{\overline{x}}{e}$ is compiled to 
$\texttt{callback}(\overline{x},e)$, where we perform 
a \texttt{callback} to a checked block code $e$ inside a sandbox.
In \systemname, we adopt an aggressive execution scheme that
directly learns pointer addresses from compiled assembly to make the $\code{_Callback}$ happen.
In the formalism, we rely on the type system to 
guarantee the context switching without creating the extra function calls for simplicity.

%Fig.~\ref{fig:compilationexample} shows how an invocation of
%\code{strlen} on a null-terminated string is compiled into C
%code. Each dereference of a checked pointer requires a null check
%(See \textsc{S-DefNull} in Fig.~\ref{fig:semantics}), which the
%compiler makes explicit: Line~$3$ of the generated code has the null
%check on pointer \code{p} due to the \code{strlen},
%  and a similar check happens
%  at line~$8$ due to the pointer arithmetic on \code{p}.
%Dereferences also require bounds checks: line~$2$ checks \code{p} is
%in bounds before computing \code{strlen(p)}, while line~$10$ does
%likewise before computing \code{*(p+1)}.

\myparagraph{Function Pointers and Calls}
Function pointers are managed similarly to normal pointers,
but we insert checks to check if the pointer address is not null in 
the function store instead of heap, and whether or not the type is correctly represented, 
for both $\cmode$ and $\tmode$ mode pointers 
\footnote{$\cmode$-mode pointers are checked once in the beginning and $\tmode$-mode pointers are checked every time when use}.
The compilation of function calls (compiling to $\elcall{m}{e}{\overline{e}}$) 
is similar to the manipulation of pointer access operations in \Cref{fig:flagtable}.
For compiling dependent function calls,
\Cref{fig:compilationexample1} provides a hint.
Notice that the bounds for the array pointer \code{p} are not passed as
arguments. Instead, they are initialized according to \code{p}'s
type---see line~4 of the original \lang program at the top of the figure.
Line~$3$ of the generated code
sets the lower bound  to \code{0} and the
upper bound to \code{n}.

\subsection{Constraints and Metatheory}
\label{sec:meta}

Here, we first show some Well-formedness and consistency definitions that are required in \Cref{sec:theorem}, and then show the simulation theorem for the \lang compiler.
Type soundness relies on several \emph{well-formedness}:

\begin{definition}[Type Environment Well-formedness]\label{type-wellformed}
A type environment $\Gamma$ is well-formed if every variable mentioned as type bounds in $\Gamma$ are bounded by $\tint$ typed variables in $\Gamma$.
\end{definition}

\begin{definition}[Heap Well-formedness]
For every $m$, A heap $\heap$ is well-formed if (i) $\heap(m,0)$ is undefined, and
(ii) for all $\evalue{n}{\tau}$ in the range of $\heap(m)$, type $\tau$
contains no free variables. 
\end{definition}

\begin{definition}[Stack Well-formedness]
A stack snapshot $\varphi$ is well-formed if
for all $\evalue{n}{\tau}$ in the range of $\varphi$, type $\tau$
contains no free variables. 
\end{definition}

We also need to introduce a notion of
\emph{consistency}, relating heap environments before and after a
reduction step, and type environments, predicate sets, and stack
snapshots together.

\begin{definition}[Stack Consistency]
A type environment $\Gamma$, variable predicate set $\Theta$, and
stack snapshot $\varphi$ are consistent---written $\Gamma;\Theta\vdash
\varphi$---if for every variable $x$, $\Theta(x)$ is defined implies
$\Gamma(x) = \tau$ for some $\tau$ and 
$\varphi(x) =\evalue{n}{\tau'}$ for some $n,\tau'$ where $\tau' \sqsubseteq_{\Theta} \tau$. 
\end{definition}

\begin{definition}[Checked Stack-Heap Consistency]
A stack snapshot $\varphi$ is consistent with heap $\heap$---written $\heap \vdash \varphi$---if
for every variable $x$, $\varphi(x)= \evalue{n}{\tau}$ with $\mode(\tau)=\cmode$ implies $\emptyset;\heap(\cmode);\emptyset \vdash_{\cmode} n:\tau$.
\end{definition}

\begin{definition}[Checked Heap-Heap Consistency]
A heap $\heap'$ is consistent with $\heap$---written $\heap \triangleright \heap'$---if
for every constant $n$, $\emptyset;\heap;\emptyset \vdash_{\cmode} n:\tau$ implies $\emptyset;\heap';\emptyset \vdash_{\cmode} n:\tau$.
\end{definition}

We formalize both the compilation procedure and the simulation
theorem in the PLT Redex model we developed for \lang (see Sec.~\ref{sec:syntax}),
and then attempt to falsify it via Redex's support for random
testing. Redex allows us
  to specify compilation as logical rules (an extension
  of typing), but then execute it algorithmically to
  automatically test whether simulation holds. This process revealed
  several bugs in compilation and the theorem statement.
%
  % us gain confidence that our original pen and paper proof of
  % simulation remains true with the addition of variable bounds. }
We ultimately plan to prove simulation in the Coq model.

%Turning to the simulation theorem: We first introduce notation
%used to specify the theorem.
We use the notation $\gg$ to
indicate the \emph{erasure} of stack and heap---the rhs is the same as
the lhs but with type annotations removed:
\begin{equation*}
  \begin{split}
    \heap  \gg & \dot \heap \\
    \varphi \gg & \dot \varphi
  \end{split}
\end{equation*}
In addition, when $\Gamma;\emptyset\vdash
\varphi$ and $\varphi$ is well-formed, we write $(\varphi,\heap,e) \gg_m (\dot \varphi, \dot \heap,
\dot e)$ to denote $\varphi \gg \dot \varphi$, $\heap \gg \dot \heap$
and $\Gamma;\Theta;\emptyset \vdash_m e \gg \dot e : \tau$ for some $\tau$ respectively. $\Gamma$ is omitted from the notation since the well-formedness of $\varphi$ and its consistency with respect to $\Gamma$ imply that $e$ must be closed under $\varphi$, allowing us to recover $\Gamma$ from $\varphi$.
Finally, we use $\xrightarrow{\cdot}^*$ to denote the transitive closure of the
reduction relation of $\elang$. Unlike the $\lang$, the semantics of
$\elang$ does not distinguish checked and unchecked regions.

\begin{figure}[t]
{\small
\[
\begin{array}{c}
\begin{tikzpicture}[
            > = stealth, % arrow head style
            shorten > = 1pt, % don't touch arrow head to node
            auto,
            node distance = 3cm
        ]

\begin{scope}[every node/.style={draw}]
    \node (A) at (0,1.5) {$\varphi_0,\heap_0, e_0$};
    \node (B) at (4,1.5) {$\varphi_1, \heap_1 ,e_1$};
    \node (C) at (0,0) {$\dot \varphi_0, \dot \heap_0 ,\dot e_0$};
    \node (D) at (4,0) {$\dot \varphi_1, \dot \heap_1, \dot e_1$};
    \node (E) at (2,-1.5) {$\dot \varphi,\dot \heap ,\dot e$};
\end{scope}
\begin{scope}[every edge/.style={draw=black}]

    \path [->] (A) edge node {$\longrightarrow_{\cmode}$} (B);
    \path [<->] (A) edge node {$\gg$} (C);
    \path [<->] (B) edge node {$\gg$} (D);
    \path [dashed,<->] (C) edge node {$\sim$} (D);
    \path [dashed,->] (C) edge node {$\xrightarrow{\cdot}^*$} (E);
    \path [dashed,->] (D) edge node[above] {$\xrightarrow{\cdot}^*$} (E);
\end{scope}

\end{tikzpicture}
\end{array}
\]
}
\caption{Simulation between \lang and \elang }
\label{fig:checkedc-simulation-ref}
\end{figure}

Fig.~\ref{fig:checkedc-simulation-ref} gives an overview of 
the simulation theorem.\footnote{We ellide the  possibility of $\dot e_1$ evaluating to $\ebounds$ or $\enull$ in the diagram for readability.} The simulation theorem is specified in a way
that is similar to the one by~\citet{merigoux2021catala}.

An ordinary simulation property would
replace the middle and bottom parts of the figure with the
following: \[(\dot \varphi_0, \dot \heap_0, \dot e_0) 
  \xrightarrow{\cdot}^* (\dot \varphi_1, \dot \heap_1, \dot e_1)\]
Instead, we relate two erased configurations using the relation $\sim$,
which only requires that the two configurations will eventually reduce
to the same state.

% The two theorems are translation preservation and simulation. We donate $\xrightarrow{c}$ as the transition semantics of CLight.
\begin{thm}[Simulation ($\sim$)]\label{simulation-thm}
For \lang expressions $e_0$, stacks $\varphi_0$, $\varphi_1$, and heap snapshots $\heap_0$, $\heap_1$, 
if $\heap_0 \vdash \varphi_0$, $(\varphi_0,\heap_0,e_0)\gg_c (\dot \varphi_0,\dot \heap_0, \dot e_0)$,
and if there exists some $r_1$ such that $(\varphi_0, \heap_0, e_0)
\rightarrow_c (\varphi_1, \heap_1, r_1)$, then the following facts hold:

\begin{itemize}

\item if there exists $e_1$ such that $r=e_1$ and $(\varphi_1, \heap_1, e_1) \gg (\dot \varphi_1, \dot \heap_1, \dot e_1)$, then there exists some $\dot \varphi$,$\dot \heap$, $\dot e$, such that
$(\dot \varphi_0, \dot \heap_0,\dot e_0) \xrightarrow{\cdot}^* (\dot
\varphi,\dot \heap,\dot e)$ and $(\dot
\varphi_1,\dot \heap_1,\dot e_1) \xrightarrow{\cdot}^* (\dot \varphi,
\dot \heap,\dot e)$.

\item if $r_1 = \ebounds$ or $\enull$, then we have $(\dot \varphi_0, \dot \heap_0,\dot e_0) \xrightarrow{\cdot}^* (\dot
\dot \varphi_1,\dot \heap_1, r_1)$ where $\varphi_1 \gg \dot
\varphi_1$, $\heap_1 \gg \dot \heap_1$.

\end{itemize}
\end{thm}

As our random generator never generates
$\euncheckedtext$ expressions (whose behavior could be undefined), we can only test a the simulation theorem 
as it relates to checked code. This limitation makes it
unnecessary to state the other direction of the simulation theorem
where $e_0$ is stuck, because Theorem~\ref{thm:progress} guarantees
that $e_0$ will never enter a stuck state if it is well-typed in
checked mode.

The current version of the Redex model has been tested against $21500$
expressions with depth less than $12$. Each expression can
reduce multiple steps, and we test simulation between every two
adjacent steps to cover a wider range of programs, particularly the
ones that have a non-empty heap.

\subsection{Additional Program evaluations}\label{appx:add-prog-eval}

Here, we provide the description of additional program evaluations.

\myparagraph{parsons}
Parsons is annotated comprehensively in two variants parsons\_wasm and parsons\_tainted. parsons\_wasm has most of its input parsing functions moved into the sandbox, whilst having all its pointers marked as tainted. These sandboxed functions interact with the checked region by making indirect calls through RLBOX's callback mechanism. However, with parsons\_tainted, we do not move any of the functions to the sandbox but still mark all the pointers as tainted. The test suite itself consists of 328 tests comprehensively testing the JSON parser's functionality. Benchmarks for both of these forks are recorded using the mean difference between the \systemname and generic-C/checked-C variants when executing 10 consecutive iterations of the test suite. parsons\_wasm expectedly shows 200/266\% runtime overhead when evaluated against checked-c and generic-c respectively due to the performance limitation of WebAssembly. However, evaluating parsons\_tainted against checked-c shows \systemname to be faster because \systemname by itself performs lighter run-time-instrumentation on tainted pointers as compared to the run-time bounds checking performed on checked pointers by checked-c. Furthermore, we only see an average peak memory of 9.5 KiB as compared to the anticipated 82 KiB overhead as Valgrind does not consider the WASM Shadow memory allocated to the tainted pointers.

\myparagraph{LibPNG}
\systemname changes for libPNG is narrow in scope and begins with the encapsulation CVE-2018-144550 and a buffer overflow in compare\_read(). However, we also annotate sections of Lib-png that involve reading, writing, and image processing (interlace, intrapixel, etc) on user-input image data as tainted. That is, rows of image bytes are read into tainted pointers and the taintedness for the row\_bytes is propagated throughout the program. All our changes extend to the png2pnm and pnm2png executables. To evaluate png2pnm, we take the mean of 10 iterations of a test script that runs png2pnm on 52 png files located within the libpng's pngsuite. To test pnm2png, we take the mean of 10 iterations of pnm2png in converting a 52MB 5184x3456 pixels large pnm image file to png. Valgrind's reported lower Heap space consumption for \systemname converted code is due to the discounted space consumed on the heap by the Sandbox's shadow memory. Consequently, when evaluating pnm2png, \systemname's heap consumption was 52 MB lower as the entire image was loaded onto the shadow memory.  

\myparagraph{MicroHTTPD}
MicroHTTPD demonstrates the practical difficulties in converting a program to \systemname. Our conversion for this program was aimed at sandboxing memory vulnerabilities CVE-2021-3466 and CVE-2013-7039. CVE-2021-3466 is described as a vulnerability from a buffer overflow that occurs in an unguarded "memcpy" which copies data into a structure pointer (struct MHD\_PostProcessor pp) which is type-casted to a char buffer (char *kbuf = (char *) \&pp[1]). Our changes would require making the "memcpy" safe by marking this pointer as tainted. However, this would either require marshaling the data pointed by this structure (and its sub-structure pointer members) pointer or would require marking every reference to this structure pointer as tainted, which in turn requires every pointer member of this structure to be tainted. Marshalling data between structure pointers is not easy and demands substantial marshaling code due to the spatial non-linearity of its pointer members unlike a char*. This did not align with our conversion goals which were aimed at making minimal changes. Consequently, the above CVE stands un-handled by \systemname.  Our changes for CVE-2013-7039 involve marking the user input data arguments of this function as tainted pointers and in the interests of seeking minimal conversion changes, we do not propagate the tainted-ness on these functions. Following up on the chronological impossibility of sandboxing bugs before they are discovered and the general programmer intuiting, we moved many of the core internal functions (like MHD\_str\_pct\_decode\_strict\_() and MHD\_http\_unescape()) into the sandbox. 

\myparagraph{Tiny-bignum}
Due to its small size and simplicity, \systemname changes for Tiny-bignum was chosen to be comprehensive. Furthermore, bignum\_to\_string() was moved to the sandbox due to a memory-unsafe use of sprintf(). 

\myparagraph{\textbf{ProFTPD}:}
\systemname changes for ProFTPD were lim-
ited was small and exactly encapsulates CVE-2010-4221 by marking the user input to the unsafe function
"pr\_netio\_telnet\_gets()" as tainted. Although we propagate the taintedness of the above function's argument, run-time overhead measured by following the methodology in ~\ref{experimentalsetup} was minimal as the data-flow graph for the pointer was small and thereby, required less pointers to be annotated.

\myparagraph{\textbf{UFTPD}}
\systemname changes for UFTPD were aimed at sandboxing CVE-2020-14149 and CVE-2020-5204. CVE-2020-14149 was recorded as a NULL pointer dereference in the handle\_CWD() which could have led to a DoS in versions before 2.12, thereby, requiring us to sandbox this function. CVE-2020-5204 was recorded as a buffer overflow vulnerability in the handle\_PORT() due to sprintf() which also required us to sandbox this function. Although we could have chosen to only mark the faulty pointers as tainted, we intended to keep our changes more generic.